# Hierarchical Bayesian Uncertainty Quantification of Finite Element Models using Modal Statistical Information

Omid Sedehi[1,*], Costas Papadimitriou[2], and Lambros S. Katafygiotis[3]

**Abstract**

This paper develops a Hierarchical Bayesian Modeling (HBM) framework for uncertainty quantification of Finite Element (FE) models based on modal information. This framework uses an existing Fast Fourier Transform (FFT) approach to identify experimental modal parameters from time-history data and employs a class of maximum-entropy probability distributions to account for the mismatch between the modal parameters. It also considers a parameterized probability distribution for capturing the variability of structural parameters across multiple data sets. In this framework, the computation is addressed through Expectation-Maximization (EM) strategies, empowered by Laplace approximations. As a result, a new rationale is introduced for assigning optimal weights to the modal properties when updating structural parameters. According to this framework, the modal features' weights are equal to the inverse of the aggregate uncertainty, comprised of the identification and prediction uncertainties. The proposed framework is coherent in modeling the entire process of inferring structural parameters from response-only measurements and is comprehensive in accounting for different sources of uncertainty, including the variability of both modal and structural parameters over multiple data sets, as well as their identification uncertainties. Numerical and experimental examples are employed to demonstrate the HBM framework, wherein the environmental and operational conditions are almost constant. It is observed that the variability of parameters across data sets remains the dominant source of uncertainty while being much larger than the identification uncertainties.

[1*] Postdoctoral Fellow, Department of Civil and Environmental Engineering, The Hong Kong University of Science and Technology, Hong Kong, China, Email: osedehi@connect.ust.hk **(Corresponding Author)**.
[2] Professor, Department of Mechanical Engineering, University of Thessaly, Volos, Greece, Email: costasp@uth.gr
[3] Professor, Department of Civil and Environmental Engineering, The Hong Kong University of Science and Technology, Hong Kong, China, Email: katafygiotis.lambros@gmail.com.

**Nomenclature**

$Conv$ : convergence metric

$D_s$ : $s$th data set

$\mathbb{E}[.]$ : expected value

$G_c(\phi_{i,s})$ : unit-norm constrain function

$J(.)$ or $L(.)$ : loss function

$\mathbb{M}(\boldsymbol{\theta})$ : FE model class parameterized by $\boldsymbol{\theta}$

$\mathbb{M}_p(\Theta)$ : probabilistic model class parameterized by $\Theta$

$\mathcal{M}$ : a class of FE models

$\mathcal{M}_p$ : a class of probabilistic models

$N_{DOF}$ : number of degrees of freedom of the FE model

$N_f$ : number of FFT data points

$N_m$ : number of dynamical modes of interest

$N_o$ : number of observed response quantities

$N_s$ : number of substructures

$N_{sa}$ : total number of samples

$N_T$ : total number of parameters

$N_\theta$ : number of unknown structural parameters

$p(.)$ : probability density function

$P(.)$ : probability of a model class

$S_{ei,s}$ : prediction error variance of $i$th mode and $s$th dataset

$\Delta t_s$ : sampling interval of the data set $D_s$

$Tol$ : tolerance used for checking the convergence

$\nabla.$ : gradient operator

$\hat{}$ : over-hat denotes the most probable values of parameters

$.|_0$ : initial values of parameters

$\|.\|$ : Euclidean norm

$\partial$ : partial differential operator

$\beta$ : a set of data-set-specific parameters

$\chi_{i,s}$ : a coefficient for matching the length and direction of modes

$\delta_{i,s}$ : Lagrange multiplier

$\varphi$ : a set of hyper-parameters

$\eta$ : a set of nuisance parameters

$\varpi_{i,s}^2$ : $i$th analytical modal frequency

$\omega_{i,s}^2$ : $i$th experimental modal frequency

$\omega_{k,s} = 2\pi k / (N_s \Delta t_s)$ : discrete angular frequency

$\rho$ : correlation coefficient

$\sigma_{\phi_i}^2$ : isotropic variance of the $i$th mode shape

$\tau_{\omega_i^2}^2$ : dimensionless variance of the $i$th modal frequency

$\xi_{i,s}$ : $i$th modal damping ratio

$\mathbf{D} = \{D_s, s = 1,...,N_D\}$ : full data set

$\mathbf{E}_k(\boldsymbol{\omega}_s^2, \boldsymbol{\Phi}_s, \boldsymbol{\eta}_s) \in \mathbb{R}^{N_o N_m \times N_o N_m}$ : theoretical PSD matrix

$\hat{\mathbf{F}}_{k,s} \in \mathbb{C}^{N_o}$ : FFT of the measured response

$\mathbf{H}$ : Hessian matrix of appropriate dimension

$\mathbf{K}(\boldsymbol{\theta}) \in \mathbb{R}^{N_{DOF} \times N_{DOF}}$ : global stiffness matrix

$\mathbf{K}_0 \in \mathbb{R}^{N_{DOF} \times N_{DOF}}$ : known part of the global stiffness matrix

$\mathbf{K}_p \in \mathbb{R}^{N_{DOF} \times N_{DOF}}$ : initial stiffness matrix of the $p$th substructure

$\mathbf{M}(\boldsymbol{\theta}) \in \mathbb{R}^{N_{DOF} \times N_{DOF}}$ : global mass matrix

$\mathbf{M}_0 \in \mathbb{R}^{N_{DOF} \times N_{DOF}}$ : known part of the global mass matrix

$\mathbf{M}_p \in \mathbb{R}^{N_{DOF} \times N_{DOF}}$ : initial mass matrix of the *p*th substructure

$\mathbf{S}_s \in \mathbb{R}^{N_o \times N_o}$ : modal force power spectral density matrix

$\mathbf{y}_{j,s} \in \mathbb{R}^{N_o}$ : measured vibration response $D_s$

$\mathbf{X} \in \mathbb{R}^N$ : a random vector

$\boldsymbol{\eta}_s \in \mathbb{R}^{2N_m + N_m(N_m+1)/2}$ : a vector of nuisance parameters

$\boldsymbol{\varepsilon}_{\boldsymbol{\omega}_s^2} \in \mathbb{R}^{N_m}$ : discrepancy between modal frequencies

$\boldsymbol{\varepsilon}_{\boldsymbol{\Phi}_s} \in \mathbb{R}^{N_m N_o}$ : discrepancy between mode shapes

$\boldsymbol{\Phi}_s \in \mathbb{R}^{N_m N_o}$ : a vector of experimental mode shapes

$\boldsymbol{\phi}_{i,s} \in \mathbb{R}^{N_o}$ : *i*th experimental mode shape

$\boldsymbol{\Psi}(\boldsymbol{\theta}_s) \in \mathbb{R}^{N_m N_{DOF}}$ : a vector of analytical mode shapes

$\boldsymbol{\psi}_{i,s} \in \mathbb{R}^{N_{DOF}}$ : *i*th analytical mode shape

$\boldsymbol{\Gamma}_s \in \mathbb{R}^{N_m N_o \times N_m N_{DOF}}$ : selection matrix

$\boldsymbol{\gamma}_{i,s} \in \mathbb{R}^{N_o \times N_{DOF}}$ : matching matrix

$\boldsymbol{\mu} \in \mathbb{R}^N$ : mean vector

$\boldsymbol{\mu}_0 \in \mathbb{R}^N$ : prior mean vector

$\boldsymbol{\mu}_{\boldsymbol{\theta}} \in \mathbb{R}^{N_\theta}$ : hyper mean vector of structural parameters

$\boldsymbol{\varpi}^2(\boldsymbol{\theta}_s) \in \mathbb{R}^{N_m}$ : a vector of analytical modal frequencies

$\boldsymbol{\theta} \in \mathbb{R}^{N_\theta}$ : structural parameters

$\boldsymbol{\omega}_s^2 \in \mathbb{R}^{N_m}$ : a vector of experimental modal frequencies

$\boldsymbol{\Sigma}_{\boldsymbol{\omega}^2} \in \mathbb{R}^{N_m \times N_m}$ : prediction error covariance matrix of modal frequencies

$\boldsymbol{\Sigma}_{\boldsymbol{\Phi}} \in \mathbb{R}^{N_o N_m \times N_o N_m}$ : prediction error covariance matrix of mode shapes

$\hat{\boldsymbol{\Sigma}}_{\boldsymbol{\omega}_s^2} \in \mathbb{R}^{N_m \times N_m}$ : identification uncertainty of modal frequencies

$\hat{\boldsymbol{\Sigma}}_{\boldsymbol{\Phi}_s} \in \mathbb{R}^{N_o N_m \times N_o N_m}$ : identification uncertainty of mode shapes

$\hat{\boldsymbol{\Sigma}}_{\boldsymbol{\eta}_s} \in \mathbb{R}^{(N_m^2/2 + 5/2 N_m) \times (N_m^2/2 + 5/2 N_m)}$ : identification uncertainty of nuisance parameters

$\boldsymbol{\Sigma}_{\boldsymbol{\theta}} \in \mathbb{R}^{N_\theta \times N_\theta}$ : hyper covariance matrix of structural parameters

$\boldsymbol{\Sigma}_s \in \mathbb{R}^{N_o N_m \times N_o N_m}$ : prediction error covariance matrix

$\boldsymbol{\Sigma} \in \mathbb{R}^{N \times N}$ : covariance matrix

$\boldsymbol{\Sigma}_0 \in \mathbb{R}^{N \times N}$ : prior covariance matrix

## 1. Introduction

FE models have revolutionized simulation-based design and analysis of structural and mechanical systems. However, the success of these approaches hinges upon the proper choice of model parameters. In recent years, model updating methods have emerged to identify structural parameters based on data and produce validated and data-informed FE models. This methodology also finds considerable merits and applications to vibration-based Structural Health Monitoring (SHM), where vibration measurements are used for updating FE models and detecting abrupt elemental stiffness drops [1–7].

The most traditional model updating perspective is deterministic techniques, which combine least-squares regression with regularization tools [3,8]. However, the presence of uncertainties has necessitated developing probabilistic methods, such as Maximum Likelihood (ML) [9] and Bayesian approaches [4,10–12]. While probabilistic methods remain the most popular choice, non-probabilistic techniques, interval-based analysis approach, and Fuzzy model updating constitute other types of methods, progressed considerably in recent years, e.g., [13–18]. Aside from the methodology, the model updating problem can be formulated in numerous forms, depending on the type of information, such as input-output, output-only, time-history data, frequency response data, and modal data [15,19,20].

The present study focuses on a particular class of Bayesian methods, wherein structural models are updated based on modal data. Note that the modal data can be acquired through numerous tools, such as subspace modal identification, peak-picking approach, and Bayesian methods [20–24]. In this regard, pioneering works include identifying stiffness parameters based on the modal information to detect structural damage [25–30]. This approach has gradually been refined and progressed through significant research works. In the original formulation, one-to-one matching of the analytical and experimental modal properties was required, but the new formulation in Yuen et al. relaxes this unnecessary premise by considering a new prediction error model [31]. Christodoulou and Papadimitriou have demonstrated that the weighting of modal features plays a crucial role in the reliable calculation of the most probable values (MPV) of structural parameters and have proposed a novel approach for identifying the weights [32]. Goller et al. have applied such model updating schemes to aerospace structures [33]. Later, they have introduced an evidence-based method for the calibration of the weights [34]. In this work, however, the extent of weighting modal features was limited to finding only a scalar, governing the relative confidence placed in the modal frequencies and shapes. Yan and Katafygiotis [35] have developed a Bayesian method for model updating based on the modal information that is obtained by applying the Bayesian Power Spectral Density (BPSD) approach [24,36,37]. This work legitimately weights modal features based on their identification uncertainty such that, when identifying the unknown structural parameters, more uncertain features will be given smaller weights and vice versa. However, imposing a perfect matching between the analytical and experimental modal frequencies makes this approach less attractive since in practice this condition is often difficult to satisfy.

Au and Zhang [38,39] have developed a two-stage method, which identifies the modal parameters and their uncertainties through the Bayesian FFT-based approach and then updates the stiffness/mass parameters [20,40,41]. Like [35], the modal parameters are weighed based on the identification uncertainty obtained from the Bayesian modal identification so that more uncertain modal parameters are assigned smaller weights for updating the stiffness/mass parameters. Notable applications of this methodology can be found in [42,43]; however, the original formulation in [38,39] postulates a direct and perfect matching between the modal parameters, which rarely occurs in practice. A refined version of this approach has appeared recently in [44], which considers a Gaussian discrepancy model to account for the mismatch between the analytical and experimental modal parameters. Nevertheless, as pointed out in [44], the computation in this new formulation is challenging due to the heuristics involved in satisfying unit-norm constraints of mode shapes, as well as positive-definiteness conditions of non-diagonal covariance matrices.

Updating FE models based on complex modal features and non-classical damping models has recently been studied in [45]. Although optimization-based model updating methods have a long history of usage, applications of sampling techniques to model updating have gained more popularity in recent years [46,47]. In this regard, a sampling-based transfer learning approach to model updating has appeared in [48]. In these recent studies, the weighting of modal features is ignored entirely despite its concerning effects on the identification results when modeling errors and sensor misalignment potentially exist.

The modal parameters can be identified only up to a certain precision. This precision, commonly referred to as the identification uncertainty, represents how well the mathematical model describes the data and how much noise is introduced by the sensing devices. This type of uncertainty can be reduced when the quality/quantity of measured data is enhanced [49–51]. On the other hand, the modal and structural parameters considerably vary depending on environmental effects, ambient conditions, and service loadings when they are identified from different tests (or data sets) [52–54]. This test-to-test (dataset-to-dataset) variability of the parameters can be attributed to the model's inadequacy to describe the physics of the problem, and it may predominantly exceed the identification uncertainty calculated by conventional Bayesian methods [49,50]. A two-level probabilistic hierarchical model can account for the variability promoted due to modeling error [49,55–62]. The key idea is to release the model parameters to vary over data sets and describe their test-to-test variability through a parameterized probability distribution, whose parameters, e.g., the mean vector and covariance matrix, should be updated by fusing multiple data sets. Once the variability is characterized, it is possible to reduce it using non-parametric probabilistic models trained on environmental and operational conditions, as demonstrated in [63–65]. Even if the information about the environmental conditions is missing, it has been indicated that the regression can be performed by considering additional latent parameters [66].

Despite recent advances, updating structural models based on modal data still requires further elaboration in terms of methodology and computation to handle specifically: 1) optimal and rational weighting of modal parameters in a concise manner 2) quantification of both aleatory and epistemic uncertainties 3) satisfying unit-norm constraint of modal parameters and considering its impact on the covariance matrix of prediction errors. These research gaps are addressed in this paper by developing hierarchical Bayesian modeling (HBM) framework for the uncertainty quantification of structural models using modal statistical data. In this study, the modal statistical information is first obtained using an FFT-based formulation of the likelihood function [20,23], and next, the modal information is employed to update the parameters of the embedded physics-based model. While the present study extends the HBM to deal with the problem of inferring structural parameters from multiple vibration data sets using modal statistical information, the HBM presented in [50,60] deals with the model inference in the time domain; and the framework in [56] addresses only operational modal analysis. The proposed framework differs from [49,65] in the formulation since it incorporates the identification uncertainty of experimental modal parameters. Moreover, it substitutes the underlying assumptions with less stringent ones and relaxes the diagonal choices of covariance matrices postulated in [49,65]. In terms of computation, this study is novel as it combines EM algorithms with Laplace approximations, aiming to enhance computational efficiency and provide theoretical insights on the uncertainty quantification and the weighting of modal features. In contrast, contributions [49,65] mainly use Gibbs sampling approaches, by which theoretical properties of the HBM framework cannot be studied to the extent explored herein. Overall, specific contributions of the paper can be outlined as follows:

- Develop a new Bayesian formulation and implement new computational algorithms
- Consider both identification uncertainty and variability of structural or modal parameters

- Propose a new rationale to attribute optimal weights to the modal properties based on their aggregate uncertainty, comprising the sum of identification uncertainty and ensemble variability
- Demonstrate and verify the proposed framework using both numerical and experimental examples

This paper continues with Section 2, describing the HBM framework and its mathematical formulation. In Section 3, the mathematical derivation of the computational algorithm is discussed. Section 4 compares it with existing methods and provides insights on the uncertainty quantification aspects. Section 5 exhibits one numerical and two experimental examples to test and validate the framework, and Section 6 summarizes conclusions.

## 2. Hierarchical Bayesian framework

### 2.1. Structural model class

Let the FE model $\mathbb{M}(\boldsymbol{\theta}) \in \mathcal{M}$ be selected from a class of models $\mathcal{M}$, described as a function of free parameters $\boldsymbol{\theta}$ used for characterizing the stiffness and mass matrices. A convenient approach is to express the stiffness and mass matrices as a linear function of substructures' stiffness/mass matrices, which gives:

$$\mathbf{K}(\boldsymbol{\theta}) = \mathbf{K}_0 + \sum_{p=1}^{N_s} \theta_p \mathbf{K}_p \quad ; \quad \mathbf{M}(\boldsymbol{\theta}) = \mathbf{M}_0 + \sum_{p=1}^{N_s} \theta_{p+N_s} \mathbf{M}_p \tag{1}$$

where $\mathbf{K}(\boldsymbol{\theta})$ and $\mathbf{M}(\boldsymbol{\theta})$ are the global stiffness and mass matrices, respectively; $\mathbf{K}_0$ and $\mathbf{M}_0$ are the known parts of the global stiffness and mass matrices, respectively; $\mathbf{K}_p$ and $\mathbf{M}_p$ are the stiffness and mass matrices corresponding to the $p$th substructure, whose contribution to the global stiffness and mass matrices is determined as the multiplication of $\mathbf{K}_p$ and $\mathbf{M}_p$ by the unknown parameters $\theta_p$ and $\theta_{p+N_s}$, respectively. By this parameterization, the FE model consists of $N_{DOF}$ degrees-of-freedom (DOFs) and $N_s$ substructures. The unknown parameters are collected into $\boldsymbol{\theta}$, which is intended to be updated based on vibration data.

The modal parameters of the FE model include $\boldsymbol{\varpi}^2(\boldsymbol{\theta}) = [\varpi_1^2 \ ... \ \varpi_{N_m}^2]^T$ and $\boldsymbol{\Psi}(\boldsymbol{\theta}) = [\boldsymbol{\psi}_1^T \ ... \ \boldsymbol{\psi}_{N_m}^T]^T$, where each pair $\{\varpi_i^2, \boldsymbol{\psi}_i\}$ represents the square of the angular modal frequency and the mode shape vector, corresponding to the $i$th dynamical mode when $N_m$ modes are considered. Each pair of the analytical modal parameters satisfies the following equation:

$$\mathbf{K}(\boldsymbol{\theta})\boldsymbol{\psi}_i = \varpi_i^2 \mathbf{M}(\boldsymbol{\theta})\boldsymbol{\psi}_i \quad , \quad \forall i = 1, 2, ..., N_m \tag{2}$$

The primary goal is to update the structural parameters ($\boldsymbol{\theta}$) and their uncertainties by fusing multiple sets of vibration measurements. Let $\mathbf{D} = \{D_s, s = 1, ..., N_D\}$ denote the full data set containing $N_D$ independent vibration data sets, where $D_s = \{\mathbf{y}_{j,s} \triangleq \mathbf{y}_s(j\Delta t_s), j = 0, 1, ..., N_s - 1\}$ is a sequence of response-only measurements obtained from $N_o$ DOFs and at $\Delta t_s$ intervals. In practice, updating dynamical models directly based on time-history data, as proposed in [4,67], can be non-trivial and difficult from a practical standpoint as it would often require the knowledge of input forces and initial conditions, as well as the direct matching of measured and modeled responses. Therefore, it is commonly preferred to update models by minimizing the mismatch between the experimental and analytical modal parameters. This paper also follows the same strategy, proposing a new HBM framework.

### 2.2. Probability models

Let $\boldsymbol{\omega}^2 = \begin{bmatrix} \omega_1^2 \ ... \ \omega_{N_m}^2 \end{bmatrix}^T$ and $\boldsymbol{\Phi} = \left[ \left( \|\boldsymbol{\phi}_1\|^{-1} \boldsymbol{\phi}_1^T \right) \ ... \ \left( \|\boldsymbol{\phi}_{N_m}\|^{-1} \boldsymbol{\phi}_{N_m}^T \right) \right]^T$ be two vectors comprising the squares of the modal frequencies (angular) and the mode shape vectors, respectively, where each pair $\{\omega_i^2, \boldsymbol{\phi}_i\}$ corresponds to the $i$th dynamical mode of the structure when $N_m$ modes are observed. It is assumed that these modal parameters can be inferred directly from each time-history data set ($D_s$) using existing modal identification methods. In practice, this assumption can easily be satisfied when each data set consists of a sufficiently large number of data points, and the dynamical modes of interest are excited adequately. Therefore, each data set $D_s$ would provide an individual realization for the experimental modal parameters that can be different from other data sets. This issue creates variability in the parameters of the FE model $\{\varpi^2(\boldsymbol{\theta}), \boldsymbol{\Psi}(\boldsymbol{\theta}), \boldsymbol{\theta}\}$ as well. To formalize this variability, the parameters $\{\boldsymbol{\omega}_s^2, \boldsymbol{\Phi}_s, \varpi^2(\boldsymbol{\theta}_s), \boldsymbol{\Psi}(\boldsymbol{\theta}_s), \boldsymbol{\theta}_s\}$ are introduced, where the subscript $s$ corresponds to $D_s$. The discrepancy between the experimental and analytical modal parameters is the basis for updating the structural parameters, described by

$$\begin{bmatrix} \boldsymbol{\omega}_s^2 \\ \boldsymbol{\Phi}_s \end{bmatrix} = \begin{bmatrix} \varpi^2(\boldsymbol{\theta}_s) \\ \boldsymbol{\Gamma}_s \boldsymbol{\Psi}(\boldsymbol{\theta}_s) \end{bmatrix} + \begin{bmatrix} \boldsymbol{\varepsilon}_{\boldsymbol{\omega}_s^2} \\ \boldsymbol{\varepsilon}_{\boldsymbol{\Phi}_s} \end{bmatrix} \tag{3}$$

where $\boldsymbol{\Gamma}_s$ is a matrix used for specifying the observed DOFs while matching and normalizing the analytical and experimental modes. It should be highlighted that the selection matrix ($\boldsymbol{\Gamma}_s$) is data-set-specific to accommodate dissimilar sensor configurations. The prediction error vectors $[\boldsymbol{\varepsilon}_{\boldsymbol{\omega}_s^2}^T \quad \boldsymbol{\varepsilon}_{\boldsymbol{\Phi}_s}^T]^T$ are considered statistically independent and identically distributed (*i.i.d.*), described as

$$\boldsymbol{\varepsilon}_{\boldsymbol{\omega}_s^2} \underset{i.i.d.}{\sim} N(\mathbf{0}, \boldsymbol{\Sigma}_{\boldsymbol{\omega}^2}) \quad ; \quad \boldsymbol{\varepsilon}_{\boldsymbol{\Phi}_s} \underset{i.i.d.}{\sim} N(\mathbf{0}, \boldsymbol{\Sigma}_{\boldsymbol{\Phi}}) \tag{4}$$

where $\boldsymbol{\Sigma}_{\boldsymbol{\omega}^2}$ and $\boldsymbol{\Sigma}_{\boldsymbol{\Phi}}$ are the prediction error covariance matrices corresponding to the modal frequencies and mode shape vectors, respectively. These unknown parameters represent the matching accuracy of the modal parameters. They are considered to be invariable over data sets, which should be inferred based on all data sets. Note that the *i.i.d.* assumption helps simplify the formulation and reduce computational costs up to a large extent. Moreover, its simplicity intends to satisfy the principle of model parsimony [11], favoring simple models over more complicated ones. This assumption has been used extensively in the literature within the context of model updating, e.g., [11,27,38,55], and it holds when the modal parameters indeed follow such a distribution. Although this assumption might not be physically correct, it aims to avoid excessive complexity and potential overfitting in the probabilistic model.

Given these assumptions, the probability distribution of the experimental modal parameters conditional on the analytical ones can be written as

$$p(\boldsymbol{\omega}_s^2, \boldsymbol{\Phi}_s \mid \boldsymbol{\varpi}^2(\boldsymbol{\theta}_s), \boldsymbol{\Psi}(\boldsymbol{\theta}_s), \boldsymbol{\Sigma}_{\boldsymbol{\omega}^2}, \boldsymbol{\Sigma}_{\boldsymbol{\Phi}}) = N\left(\boldsymbol{\omega}_s^2 \mid \boldsymbol{\varpi}^2(\boldsymbol{\theta}_s), \boldsymbol{\Sigma}_{\boldsymbol{\omega}^2}\right) N\left(\boldsymbol{\Phi}_s \mid \boldsymbol{\Gamma}_s \boldsymbol{\Psi}(\boldsymbol{\theta}_s), \boldsymbol{\Sigma}_{\boldsymbol{\Phi}}\right) \qquad (5)$$

It should be noted that the modal frequencies and mode shapes can be correlated as they are governed by the same set of equations. However, this cross-correlation is ignored in this study for the sake of simplicity and ease of computation. Additionally, to avoid excessive computational overheads, the cross-correlation of the modal parameters of different modes is neglected, giving:

$$p(\boldsymbol{\omega}_s^2, \boldsymbol{\Phi}_s \mid \boldsymbol{\varpi}^2(\boldsymbol{\theta}_s), \boldsymbol{\Psi}(\boldsymbol{\theta}_s), \boldsymbol{\Sigma}_{\boldsymbol{\omega}^2}, \boldsymbol{\Sigma}_{\boldsymbol{\Phi}}) = \prod_{i=1}^{N_m} N(\omega_{i,s}^2 \mid \varpi_{i,s}^2, \sigma_{\omega_i^2}^2) N(\|\boldsymbol{\phi}_{i,s}\|^{-1} \boldsymbol{\phi}_{i,s} \mid \chi_{i,s} \boldsymbol{\gamma}_{i,s} \boldsymbol{\psi}_{i,s}, \boldsymbol{\Sigma}_{\boldsymbol{\phi}_i}) \qquad (6)$$

where $\|.\|$ denotes the Euclidean norm of a vector, $\boldsymbol{\gamma}_{i,s}$ is a known matrix matching the analytical modal quantities with the observed ones, and $\chi_{i,s} = \text{sign}(\boldsymbol{\phi}_{i,s}^T \boldsymbol{\gamma}_{i,s} \boldsymbol{\psi}_{i,s}) / \|\boldsymbol{\gamma}_{i,s} \boldsymbol{\psi}_{i,s}\|$ is a scalar considered to normalize both $\boldsymbol{\phi}_{i,s}$ and $\boldsymbol{\gamma}_{i,s} \boldsymbol{\psi}_{i,s}$ in order to have a consistent sign and norm. Due to this parameter synthesis, it is clear that $\boldsymbol{\Sigma}_{\boldsymbol{\omega}^2} = \text{diag}\left[\sigma_{\omega_1^2}^2, ..., \sigma_{\omega_{N_m}^2}^2\right]$, $\boldsymbol{\Sigma}_{\boldsymbol{\Phi}} = \text{block-diag}\left[\boldsymbol{\Sigma}_{\boldsymbol{\phi}_1}, ..., \boldsymbol{\Sigma}_{\boldsymbol{\phi}_{N_m}}\right]$, and $\boldsymbol{\Gamma}_s = \text{block-diag}\left[\chi_{1,s} \boldsymbol{\gamma}_{1,s}, ..., \chi_{N_m,s} \boldsymbol{\gamma}_{N_m,s}\right]$.

Due to Eq. (2), there is a deterministic relationship between $\boldsymbol{\varpi}^2(\boldsymbol{\theta}_s), \boldsymbol{\Psi}(\boldsymbol{\theta}_s)$ and $\boldsymbol{\theta}_s$. Thus, Eq. (6) also characterizes the conditional relationship between the experimental modal parameters and the structural ones. This relationship also implies that the structural parameters are variable over data sets, whose variability can be modeled using a parameterized Gaussian distribution given by

$$p(\boldsymbol{\theta}_s \mid \boldsymbol{\mu}_{\boldsymbol{\theta}}, \boldsymbol{\Sigma}_{\boldsymbol{\theta}}) = N(\boldsymbol{\theta}_s \mid \boldsymbol{\mu}_{\boldsymbol{\theta}}, \boldsymbol{\Sigma}_{\boldsymbol{\theta}}) \qquad (7)$$

where $\boldsymbol{\mu}_{\boldsymbol{\theta}}$ and $\boldsymbol{\Sigma}_{\boldsymbol{\theta}}$ are the hyper mean vector and covariance matrix, respectively. These hyper-parameters characterize the second-moment statistical information of the structural parameters, which should be identified based on all data sets. The use of Gaussian distributions is justifiable in the maximum entropy sense. In effect, when the entropy is maximized subjected to the second-moment information, multivariate Gaussian distributions are obtained. It is possible to replace the Gaussian distribution in Eq. (7) with other distributions. However, the formulation and computational algorithms has to change accordingly.

## 2.3. Likelihood function

In this section, the likelihood function is formulated using an existing FFT-based approach [20,23], which will appear later in the Bayesian formulation. The readers are referred to [20] for further information about its validity. This approach describes the Fourier transform of the measured vibration response by a mathematical model of the Frequency Response Function (FRF), characterizing $N_m$ dynamical modes of interest. This model is parameterized by the experimental modal features ($\boldsymbol{\omega}_s^2$ and $\boldsymbol{\Phi}_s$), as well as some auxiliary parameters subsumed into $\boldsymbol{\eta}_s$. As proposed in [20], the likelihood function of each data set ($D_s$) can be constructed as

$$p(D_s | \boldsymbol{\omega}_s^2, \boldsymbol{\Phi}_s, \boldsymbol{\eta}_s) = \prod_{k=1}^{N_f} p(\hat{\mathbf{F}}_{k,s} | \boldsymbol{\omega}_s^2, \boldsymbol{\Phi}_s, \boldsymbol{\eta}_s)$$
$$= \prod_{k=1}^{N_f} \left[ \pi^{-n} \left| \mathbf{E}_k(\boldsymbol{\omega}_s^2, \boldsymbol{\Phi}_s, \boldsymbol{\eta}_s) \right|^{-1} \exp\left( -\hat{\mathbf{F}}_{k,s}^* \left[ \mathbf{E}_k(\boldsymbol{\omega}_s^2, \boldsymbol{\Phi}_s, \boldsymbol{\eta}_s) \right]^{-1} \hat{\mathbf{F}}_{k,s} \right) \right] \qquad (8)$$

where $\hat{\mathbf{F}}_{k,s}$ is a complex vector representing the FFT response at the discrete angular frequency $\omega_{k,s} = 2\pi k / (N_s \Delta t_s)$; the subscript $k = \{1, 2, ..., N_f\}$ characterizes the discrete frequencies up to the Nyquist frequency, satisfying $k < N_f = \text{int}[N/2]$, where int[.] returns the integer part of real numbers; $\mathbf{E}_k(\boldsymbol{\omega}_s^2, \boldsymbol{\Phi}_s, \boldsymbol{\eta}_s)$ is the theoretical PSD matrix given as

$$\mathbf{E}_k(\boldsymbol{\omega}_s^2, \boldsymbol{\Phi}_s, \boldsymbol{\eta}_s) = \boldsymbol{\Phi}_s \mathbf{H}_k \boldsymbol{\Phi}_s^T + \boldsymbol{\Sigma}_s$$
$$\mathbf{H}_k = diag(\mathbf{h}_k) \mathbf{S}_s diag(\mathbf{h}_k^*) \quad ; \quad \boldsymbol{\Sigma}_s = block - diag\left[ S_{e1,s} \mathbf{I}_{N_o}, ..., S_{eN_m,s} \mathbf{I}_{N_o} \right] \qquad (9)$$
$$\mathbf{h}_k = \begin{bmatrix} h_{1,k} & \cdots & h_{N_m,k} \end{bmatrix}^T \quad ; \quad h_{i,k} = \frac{(\omega_{k,s} \mathbf{i})^{-q} p_{ik,s}}{1 - (\omega_{i,s} / \omega_{k,s})^2 - 2\mathbf{i}\xi_{i,s}(\omega_{i,s} / \omega_{k,s})}$$

where $\xi_{i,s}$ is the modal damping ratio of the $i$th mode inferred from the $s$th data set; $\mathbf{S}_s$ is the modal force PSD matrix; and $S_{ei,s}$ is the prediction errors PSD corresponding to $i$th mode and the $s$th data set. The proposed setting also implies that the vector of auxiliary parameters comprises $\boldsymbol{\eta}_s = [\xi_{1,s} \ ... \ \xi_{N_m,s} \ S_{e1,s} \ ... \ S_{eN_m,s} \ (\text{vec}(\mathbf{S}_s))^T \ ]^T$.

This likelihood function has a sharp peak around the MPV, so it can well be approximated using an asymptotic approximation [20,23]. Due to this technique, the likelihood function can be replaced by a Gaussian distribution, having a mean equal to the MPV and a covariance matrix equal to the Hessian inverse evaluated at the MPV. The Laplace approximation is expected to be more accurate as the number of data points grows because the likelihood function becomes more and more peaked around the MPV [68]. By applying this approximation, the likelihood function of Eq. (8) can be simplified into

$$p(D_s | \omega_s^2, \mathbf{\Phi}_s, \mathbf{\eta}_s) \propto N\left(\hat{\omega}_s^2 | \omega_s^2, \hat{\mathbf{\Sigma}}_{\omega_s^2}\right) N\left(\hat{\mathbf{\Phi}}_s | \mathbf{\Phi}_s, \hat{\mathbf{\Sigma}}_{\mathbf{\Phi}_s}\right) N\left(\hat{\mathbf{\eta}}_s | \mathbf{\eta}_s, \hat{\mathbf{\Sigma}}_{\mathbf{\eta}_s}\right) \qquad (10)$$

where $\hat{\omega}_s^2$, $\hat{\mathbf{\Phi}}_s$, and $\hat{\mathbf{\eta}}_s$ denote the MPV of the parameters, calculated by minimizing the negative logarithm of $p(D_s | \omega_s^2, \mathbf{\Phi}_s, \mathbf{\eta}_s)$ with respect to the underlying parameters:

$$(\hat{\omega}_s^2, \hat{\mathbf{\Phi}}_s, \hat{\mathbf{\eta}}_s) = \underset{\omega_s^2, \mathbf{\Phi}_s, \mathbf{\eta}_s}{\text{Argmin}} \left[-\ln p(D_s | \omega_s^2, \mathbf{\Phi}_s, \mathbf{\eta}_s)\right] \qquad (11)$$

In addition, the covariance matrices $\hat{\mathbf{\Sigma}}_{\omega_s^2}$, $\hat{\mathbf{\Sigma}}_{\mathbf{\Phi}_s}$, and $\hat{\mathbf{\Sigma}}_{\mathbf{\eta}_s}$ constitute the block-diagonal elements of the Hessian inverse of the negative log-likelihood function, $\nabla\nabla[-\ln p(D_s | \omega_s^2, \mathbf{\Phi}_s, \mathbf{\eta}_s)]$, evaluated at MPV. In the literature, extensive efforts were devoted to calculating this approximation in a reliable manner using explicit derivations. Interested readers may refer to [20] for further details.

In Eq. (10), the cross-correlation between modal frequencies and mode shape vectors is neglected for the sake of simplicity. Later, in the second illustrative example, we will show that this simplifying assumption is indeed reasonable.

## 2.4. Posterior distributions

In this section, a hierarchical Bayesian formulation is developed to infer the structural parameters from modal statistical information. Fig. 1 shows a graphical representation of the proposed probabilistic model. The convention followed for creating this graph is the plate notation described in [69]. In this figure, the large rectangular area indicates $N_D$ independent data sets; the experimental modal parameters should be inferred from $N_f$ data points; the arrows indicate the conditional dependence of the parameters; the white circles show the unknown parameters; the gray circle indicates the data points. In this graphical model, the experimental modal parameters $(\boldsymbol{\omega}_s^2, \boldsymbol{\Phi}_s)$ should first be inferred from each data set ($D_s$) according to Section 2.3. Subsequently, the structural parameters ($\boldsymbol{\theta}_s$) should be updated while considering the discrepancy between the analytical and experimental modal parameters. Based on the realizations of $\boldsymbol{\theta}_s$'s, the hyper-parameters $\{\boldsymbol{\mu}_{\boldsymbol{\theta}}, \boldsymbol{\Sigma}_{\boldsymbol{\theta}}, \boldsymbol{\Sigma}_{\omega^2}, \boldsymbol{\Sigma}_{\boldsymbol{\Phi}}\}$ should be updated. The parameters $\boldsymbol{\eta}_s$'s are inferred in the modal identification stage, but later they will be marginalized out from the posterior distribution as nuisance parameters.

For notation convenience, the data-set-specific modal parameters and the hyper-parameters are respectively collected into $\beta = \{\boldsymbol{\omega}_s^2, \boldsymbol{\Phi}_s, \boldsymbol{\varpi}^2(\boldsymbol{\theta}_s), \boldsymbol{\Psi}(\boldsymbol{\theta}_s), \boldsymbol{\theta}_s\}_{s=1}^{N_D}$ and $\varphi = \{\boldsymbol{\mu}_{\boldsymbol{\theta}}, \boldsymbol{\Sigma}_{\boldsymbol{\theta}}, \boldsymbol{\Sigma}_{\omega^2}, \boldsymbol{\Sigma}_{\boldsymbol{\Phi}}\}$. The parameters collected in $\beta$ are mutually independent and variable over data sets, whereas the parameters in $\varphi$ are invariable over different data sets. Additionally, the nuisance parameters are collected into the set $\eta = \{\boldsymbol{\eta}_s\}_{s=1}^{N_D}$.

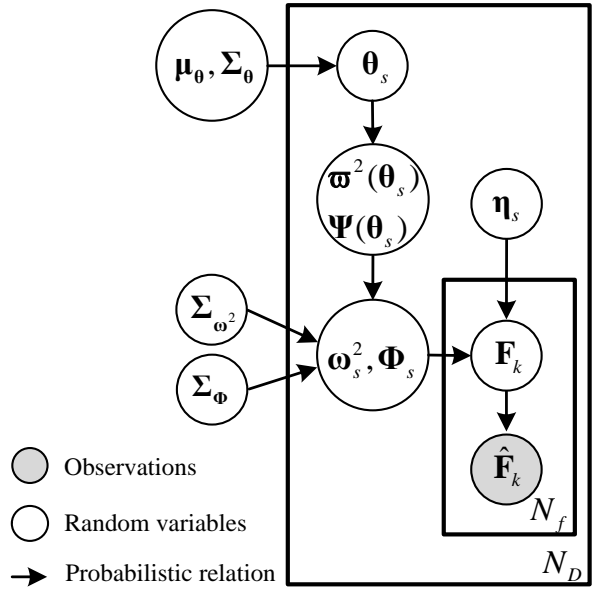

**Fig. 1**, A graphical representation of the proposed hierarchical model (Indices *s* and *k* indicate the *s*th data set and the *k*th data point, respectively)

The combination of the structural model class ($\mathbb{M}(\boldsymbol{\theta}) \in \mathcal{M}$), the probability models adopted earlier, and the dynamical model used for describing the experimental modal parameters constitute a hierarchical probabilistic model class $\mathbb{M}_p(\Theta) \in \mathcal{M}_p$. This model class belongs to a class of candidates $\mathbb{M}_p$ and is described via the parameters $\Theta = \{\beta, \varphi, \eta\}$. Given this explanation, the joint prior distribution of all parameters can be expressed as

$$p(\beta,\varphi,\eta \mid \mathbb{M}_p) = p(\boldsymbol{\mu_\theta}, \boldsymbol{\Sigma_\theta}, \boldsymbol{\Sigma_{\omega^2}}, \boldsymbol{\Sigma_\Phi} \mid \mathbb{M}_p)$$
$$\times \prod_{s=1}^{N_D} \Big[ p(\boldsymbol{\omega}_s^2, \boldsymbol{\Phi}_s \mid \boldsymbol{\varpi}^2(\boldsymbol{\theta}_s), \boldsymbol{\Psi}(\boldsymbol{\theta}_s), \boldsymbol{\Sigma_{\omega^2}}, \boldsymbol{\Sigma_\Phi}, \mathbb{M}_p) p(\boldsymbol{\theta}_s \mid \boldsymbol{\mu_\theta}, \boldsymbol{\Sigma_\theta}, \mathbb{M}_p) p(\boldsymbol{\eta}_s \mid \mathbb{M}_p) \Big] \qquad (12)$$

where $p(\boldsymbol{\mu_\theta}, \boldsymbol{\Sigma_\theta}, \boldsymbol{\Sigma_{\omega^2}}, \boldsymbol{\Sigma_\Phi} \mid \mathbb{M}_p)$ is the prior distribution of the hyper-parameters, and $p(\boldsymbol{\eta}_s \mid \mathbb{M}_p)$ is the prior distribution of the nuisance parameters. These distributions are considered uniform; all other distributions were introduced earlier through Eqs. (3-7). Uniform distributions are preferred herein to avoid introducing additional parameters and variables whose identification can require extra effort. Another reason is our willingness to put more confidence in the data rather than the prior knowledge by using diffuse prior knowledge. Based on these assumptions, the joint prior distribution can be simplified into

$$p(\beta,\varphi,\eta \mid \mathbb{M}_p) \propto \prod_{s=1}^{N_D} N(\boldsymbol{\omega}_s^2 \mid \boldsymbol{\varpi}^2(\boldsymbol{\theta}_s), \boldsymbol{\Sigma_{\omega^2}}) N(\boldsymbol{\Phi}_s \mid \boldsymbol{\Gamma}_s \boldsymbol{\Psi}(\boldsymbol{\theta}_s), \boldsymbol{\Sigma_\Phi}) N(\boldsymbol{\theta}_s \mid \boldsymbol{\mu_\theta}, \boldsymbol{\Sigma_\theta}) \qquad (13)$$

Due to the Bayes' rule, the joint posterior distribution of all parameters is proportional to the multiplication of the likelihood function and the joint prior distribution, yielding:

$$p(\beta,\varphi,\eta \mid \mathbf{D}, \mathbb{M}_p) \propto p(\mathbf{D} \mid \beta,\varphi,\eta, \mathbb{M}_p) p(\beta,\varphi,\eta \mid \mathbb{M}_p) \qquad (14)$$

where $p(\mathbf{D} \mid \beta,\varphi,\eta, \mathbb{M}_p)$ is the full data set likelihood, and $p(\beta,\varphi,\eta \mid \mathbb{M}_p)$ is the joint prior distribution given by Eq. (13). As explained earlier, the time-history data sets are statistically independent and directly depend only on the parameters involved in the modal identification stage, implying:

$$p(\mathbf{D} \mid \beta,\varphi,\eta, \mathbb{M}_p) = \prod_{s=1}^{N_D} p(D_s \mid \boldsymbol{\omega}_s^2, \boldsymbol{\Phi}_s, \boldsymbol{\eta}_s) \qquad (15)$$

where $p(D_s \mid \boldsymbol{\omega}_s^2, \boldsymbol{\Phi}_s, \boldsymbol{\eta}_s)$ is the likelihood function of $D_s$ given by Eq. (10). Substituting the likelihood of each data set from Eq. (10) and combining Eqs. (13-15) give the joint posterior distribution as

$$p(\beta,\varphi,\eta|\mathbf{D},\mathbb{M}_p) \propto$$
$$\prod_{s=1}^{N_D} N(\hat{\boldsymbol{\omega}}_s^2|\boldsymbol{\omega}_s^2,\hat{\boldsymbol{\Sigma}}_{\boldsymbol{\omega}_s^2})N(\hat{\boldsymbol{\Phi}}_s|\boldsymbol{\Phi}_s,\hat{\boldsymbol{\Sigma}}_{\boldsymbol{\Phi}_s})N(\hat{\boldsymbol{\eta}}_s|\boldsymbol{\eta}_s,\hat{\boldsymbol{\Sigma}}_{\boldsymbol{\eta}_s})N(\boldsymbol{\omega}_s^2|\boldsymbol{\varpi}^2(\boldsymbol{\theta}_s),\boldsymbol{\Sigma}_{\boldsymbol{\omega}^2})N(\boldsymbol{\Phi}_s|\boldsymbol{\Gamma}_s\boldsymbol{\Psi}(\boldsymbol{\theta}_s),\boldsymbol{\Sigma}_{\boldsymbol{\Phi}})N(\boldsymbol{\theta}_s|\boldsymbol{\mu}_{\boldsymbol{\theta}},\boldsymbol{\Sigma}_{\boldsymbol{\theta}})$$

(16)

Immediately from this equation, the nuisance parameters ($\eta$) can be marginalized out explicitly, which leads to

$$p(\beta,\varphi|\mathbf{D},\mathbb{M}_p) \propto$$
$$\prod_{s=1}^{N_D} N(\hat{\boldsymbol{\omega}}_s^2|\boldsymbol{\omega}_s^2,\hat{\boldsymbol{\Sigma}}_{\boldsymbol{\omega}_s^2})N(\hat{\boldsymbol{\Phi}}_s|\boldsymbol{\Phi}_s,\hat{\boldsymbol{\Sigma}}_{\boldsymbol{\Phi}_s})N(\boldsymbol{\omega}_s^2|\boldsymbol{\varpi}^2(\boldsymbol{\theta}_s),\boldsymbol{\Sigma}_{\boldsymbol{\omega}^2})N(\boldsymbol{\Phi}_s|\boldsymbol{\Gamma}_s\boldsymbol{\Psi}(\boldsymbol{\theta}_s),\boldsymbol{\Sigma}_{\boldsymbol{\Phi}})N(\boldsymbol{\theta}_s|\boldsymbol{\mu}_{\boldsymbol{\theta}},\boldsymbol{\Sigma}_{\boldsymbol{\theta}})$$

(17)

Although this equation is general and can be used to obtain the MPV or draw samples from the posterior distribution, the computation can be prohibitively expensive since the number of data-set-specific parameters linearly grows with the number of data sets. Moreover, the main interest lies in updating the hyper-parameters that encapsulate common information about the parameters. Therefore, it is beneficial to reduce the computational cost as much as possible, as will be performed in the next sections.

**Remark 1.** The probability distributions are written conditional on the data ($\mathbf{D}$) and the probabilistic model ($\mathbb{M}_p$). This dependence emphasizes that the Bayesian interpretation of probability applies throughout the text, describing the notion of probability as the relative plausibility of an event or outcome conditional on the data and modeling assumptions [10].

## 2.5. Marginal posterior distribution

This section provides a marginal posterior distribution, which will appear beneficial for deriving the computation algorithms and studying theoretical properties. To this end, the posterior distribution $p(\beta,\varphi|\mathbf{D},\mathbb{M}_p)$ is marginalized over $\boldsymbol{\omega}_s^2$'s and $\boldsymbol{\Phi}_s$'s in a closed-form manner, leading to:

$$p(\{\boldsymbol{\theta}_s,\boldsymbol{\varpi}^2(\boldsymbol{\theta}_s),\boldsymbol{\Psi}(\boldsymbol{\theta}_s)\}_{s=1}^{N_D},\varphi|\mathbf{D},\mathbb{M}_p)$$
$$\propto \prod_{s=1}^{N_D}\left[N(\hat{\boldsymbol{\omega}}_s^2|\boldsymbol{\varpi}^2(\boldsymbol{\theta}_s),\hat{\boldsymbol{\Sigma}}_{\boldsymbol{\omega}_s^2}+\boldsymbol{\Sigma}_{\boldsymbol{\omega}^2})N(\hat{\boldsymbol{\Phi}}_s|\boldsymbol{\Gamma}_s\boldsymbol{\Psi}(\boldsymbol{\theta}_s),\hat{\boldsymbol{\Sigma}}_{\boldsymbol{\Phi}_s}+\boldsymbol{\Sigma}_{\boldsymbol{\Phi}})N(\boldsymbol{\theta}_s|\boldsymbol{\mu}_{\boldsymbol{\theta}},\boldsymbol{\Sigma}_{\boldsymbol{\theta}})\right]$$

(18)

This result is an immediate consequence of Theorem 1 presented in Appendix (A), which gives an explicit formula for the integral of the multiplication of two Gaussian distributions, as would be required when integrating Eq. (17) over $\omega_s^2$'s and $\mathbf{\Phi}_s$'s.

Although Eq. (18) helps provide theoretical insight on uncertainty quantification aspects of the proposed framework (see Section 4), the calculation of the MPV based on this equation still encounters computational difficulties. For example, when $\mathbf{\Sigma_\Phi}$ is non-diagonal, there will be no explicit formulation for the MPV, and the optimization may trap into undesirable local optimum. Additionally, due to the scaling of the mode shape vectors by the Euclidean norm, $\mathbf{\Sigma_\Phi}$ is anticipated to be singular. For these reasons, we base the computation method upon the posterior distribution given by Eq. (17) in the next section, where explicit formulations can be derived for updating the covariance matrices while considering singularity in $\mathbf{\Sigma_\Phi}$.

## 3. Computational algorithms

At first glance, it might deem possible to follow a full sampling strategy like Algorithm 1 and draw samples directly from the joint posterior distribution, given by Eq. (17). However, based on the dimensions of the parameters, the total number of parameters ($N_T$) turns out to be excessively large, as given by

$$N_T = N_D \left[ \underbrace{N_m}_{\omega_s^2} + \underbrace{N_m N_o}_{\mathbf{\Phi}_s} + \underbrace{N_m}_{\varpi^2(\boldsymbol{\theta}_s)} + \underbrace{N_m N_o}_{\boldsymbol{\Gamma}_s \boldsymbol{\Psi}(\boldsymbol{\theta}_s)} + \underbrace{N_\theta}_{\boldsymbol{\theta}_s} \right] + \underbrace{N_m}_{\Sigma_{\omega^2}} + \underbrace{\frac{N_m N_o (N_o+1)}{2}}_{\Sigma_\Phi} + \underbrace{N_\theta}_{\mu_\theta} + \underbrace{\frac{N_\theta (N_\theta+1)}{2}}_{\Sigma_\theta} \quad (19)$$

While this large number of parameters is well beyond the power of our multi-core processors, it also elucidates the essence of establishing new computational tools, which can overcome computational challenges. This perspective also justifies the simplifications and approximations made in this section, which help boost up computational efficiency. Thus, a two-stage strategy is proposed herein, which approximates the hyper-parameters ($\varphi$) with their MPV and allows quantifying the uncertainty in the data-set-specific parameters ($\beta$). For this purpose, Eq. (17) is rewritten as

$$p(\beta,\varphi|\mathbf{D},\mathbb{M}_p) \propto p(\mathbf{D}|\beta,\Sigma_{\omega^2},\Sigma_{\Phi},\mathbb{M}_p)\prod_{s=1}^{N_D} N(\theta_s|\mu_\theta,\Sigma_\theta) \quad (20)$$

where $p(\mathbf{D}|\beta,\Sigma_{\omega^2},\Sigma_{\Phi},\mathbb{M}_p)$ is given by

$$p(\mathbf{D}|\beta,\Sigma_{\omega^2},\Sigma_{\Phi},\mathbb{M}_p) = \prod_{s=1}^{N_D} N(\hat{\omega}_s^2|\omega_s^2,\hat{\Sigma}_{\omega_s^2})N(\hat{\Phi}_s|\Phi_s,\hat{\Sigma}_{\Phi_s})N(\omega_s^2|\varpi^2(\theta_s),\Sigma_{\omega^2})N(\Phi_s|\Gamma_s\Psi(\theta_s),\Sigma_{\Phi})$$
(21)

This rearrangement of expressions indicates that inferring $\{\beta,\Sigma_{\omega^2},\Sigma_{\Phi}\}$ can separately be performed from $\{\mu_\theta,\Sigma_\theta\}$. Thus, in the first stage, the MPV of $\{\beta,\Sigma_{\omega^2},\Sigma_{\Phi}\}$ are calculated through an Expectation-Conditional-Maximization (ECM) strategy, involving the following steps:

$$\begin{cases} E\text{-}Step: Calculate\ L(\{\varpi^2(\theta_s),\Psi(\theta_s),\theta_s\}_{s=1}^{N_D},\Sigma_{\omega^2},\Sigma_{\Phi}) = \mathbb{E}_{\{\omega_s^2,\Phi_s\}_{s=1}^{N_D}}\left[\ln p(\mathbf{D}|\beta,\Sigma_{\omega^2},\Sigma_{\Phi},\mathbb{M}_p)\right] \\ M\text{-}Step\ 1: Maximize\ L(\{\varpi^2(\theta_s),\Psi(\theta_s),\theta_s\}_{s=1}^{N_D},\Sigma_{\omega^2},\Sigma_{\Phi})\ for\ \{\theta_s\}_{s=1}^{N_D} \\ M\text{-}Step\ 2: Maximize\ L(\{\varpi^2(\theta_s),\Psi(\theta_s),\theta_s\}_{s=1}^{N_D},\Sigma_{\omega^2},\Sigma_{\Phi})\ for\ \{\Sigma_{\omega^2},\Sigma_{\Phi}\} \end{cases} \quad (22)$$

where the expectation $\mathbb{E}_{\{\omega_s^2,\Phi_s\}_{s=1}^{N_D}}[.]$ is taken with respect to the parameters in the set $\{\omega_s^2,\Phi_s\}_{s=1}^{N_D}$, and the mathematical expression of $L(\{\varpi^2(\theta_s),\Psi(\theta_s),\theta_s\}_{s=1}^{N_D},\Sigma_{\omega^2},\Sigma_{\Phi})$ is given as

$$L(\{\varpi^2(\theta_s),\Psi(\theta_s),\theta_s\}_{s=1}^{N_D},\Sigma_{\omega^2},\Sigma_{\Phi}) \doteq -\frac{N_D}{2}\ln|\Sigma_{\omega^2}| - \frac{N_D}{2}\ln|\Sigma_{\Phi}|$$
$$-\frac{1}{2}\sum_{s=1}^{N_D}tr\left[\hat{\Sigma}_{\omega_s^2}^{-1}\mathbb{E}_{\omega_s^2}\left[(\hat{\omega}_s^2-\omega_s^2)(\hat{\omega}_s^2-\omega_s^2)^T\right] + \Sigma_{\omega^2}^{-1}\mathbb{E}_{\omega_s^2}\left[(\omega_s^2-\varpi^2(\theta_s))(\omega_s^2-\varpi^2(\theta_s))^T\right]\right] \quad (23)$$
$$-\frac{1}{2}\sum_{s=1}^{N_D}tr\left[\hat{\Sigma}_{\Phi_s}^{-1}\mathbb{E}_{\Phi_s}\left[(\hat{\Phi}_s-\Phi_s)(\hat{\Phi}_s-\Phi_s)^T\right] + \Sigma_{\Phi}^{-1}\mathbb{E}_{\Phi_s}\left[(\Phi_s-\Gamma_s\Psi(\theta_s))(\Phi_s-\Gamma_s\Psi(\theta_s))^T\right]\right] + cte.$$

In the context of the ECM algorithm, the "E-Step" stands for "Expectation," and the "M-Step" refers to "Maximization." As the number of iterations of the algorithm grows, the ECM is guaranteed to converge to a local optimum [70]. Additionally, when the initial estimations of the parameters are within a close neighborhood of the global optimum, the ECM is expected to converge to the global optimum. For this reason, we will later prescribe approximate solutions that may fall adequately close to the desired optima. Note that another strategy for checking the global optimality of the estimates is to change the initial point and investigate whether a different optimum can be attained by re-running the algorithm.

The mathematical derivation of this ECM algorithm is explained in Appendix (B), along with the prescribed initial estimates of the hyper-parameters. Algorithm 1 implements the proposed ECM algorithm under two main phases; In Phase I, the FFT-based modal identification approach is employed for calculating the experimental modal frequencies and mode shapes from time-history data sets. This process should be repeated for each data set independently based on Section 2.3, which provides the MPV and identification uncertainty of the experimental modal parameters. Subsequently, in Phase II, the acquired modal information is used to infer the unknown model parameters. In this algorithm, initial estimates of $\Sigma_{\omega^2}$ and $\Sigma_{\Phi}$ are prescribed to avoid undesirable optima. Once this Phase is completed, the estimates of the modal parameters, the structural parameters, and the covariance matrices of the discrepancy model will be acquired.

In Algorithm 1, the convergence criterion is an important aspect of the iterative ECM algorithm. In this paper, the convergence criterion is considered as

$$Conv = \left\| \hat{\Xi}^{(k)} - \hat{\Xi}^{(k-1)} \right\|^2 / \left\| \hat{\Xi}^{(k-1)} \right\|^2 < Tol \qquad (24)$$

where $\hat{\Xi}^{(k)}$ denotes a vector of parameters estimated at the $k$th iteration of the ECM or EM algorithm. The convergence metric ($Conv$) is next checked against a predefined tolerance ($Tol$), set by the user and considered to be a small number, e.g., $10^{-6}$.

Sometimes, it is computationally beneficial to ignore the identification uncertainty of the experimental modal parameters. In this case, the proposed ECM algorithm is still applicable but requires small modifications. In particular, the E-Step will no longer be required as $\omega_s^2$ and $\Phi_s$ must be replaced by their optimal values $\hat{\omega}_s^2$ and $\hat{\Phi}_s$, respectively. Then, the M-Steps 1 and 2 should be performed similar to the procedure provided in Algorithm 1 while substituting $\mathbb{E}[\omega_{i,s}^2]$, $\mathbb{E}[\omega_{i,s}^4]$, $\mathbb{E}\left[ \left\| \phi_{i,s} \right\|^{-1} \phi_{i,s} \right]$, and $\mathbb{E}\left[ \left\| \phi_{i,s} \right\|^{-2} \phi_{i,s} \phi_{i,s}^T \right]$ by $\hat{\omega}_{i,s}^2$, $\hat{\omega}_{i,s}^4$, $\hat{\phi}_{i,s}$ and $\hat{\phi}_{i,s} \hat{\phi}_{i,s}^T$, respectively. Additionally, the variances $\hat{\sigma}_{\omega_{i,s}^2}^2$'s and the covariance matrices $\hat{\Sigma}_{\phi_{i,s}}$'s should be set to zero.

**Algorithm 1.** The ECM algorithm proposed for inferring $\beta$, $\Sigma_{\omega^2}$, and $\Sigma_{\Phi}$

---
*Phase I: FFT-based modal identification*

1: **For** each data set $D_s$, $s = 1:N_D$ {

2: **Minimize** $-\ln p(D_s | \omega_s^2, \Phi_s, \eta_s)$, given by Eqs. (8-11), and **calculate** the MPV of the experimental modal parameters, i.e., $\hat{\omega}_s^2$ and $\hat{\Phi}_s$

3: **Calculate** the Hessian matrix inverse of $-\ln p(D_s | \omega_s^2, \Phi_s, \eta_s)$ at the MPV using Eqs. (8-11), and **select** block matrices that embed the covariance matrices $\hat{\Sigma}_{\omega_s^2}$ and $\hat{\Sigma}_{\Phi_s}$

4: **} End For**

*Phase II: ECM algorithm* (See Appendix (B) for the derivation)

5: **Set** the initial values $\hat{\theta}_s$ (May use least-squares solutions)

6: **Set** the initial values of $\varpi^2(\hat{\theta}_s)$ and $\Psi(\hat{\theta}_s)$ by solving the eigenvalue problem in Eq. (2)

7: **For** $i = 1:N_m$ {**Calculate** the elements of $\Sigma_{\omega^2}$ and $\Sigma_{\Phi}$ from Eqs. (B33-B34):

8: $\quad \hat{\tau}_{\omega_i^2}^2 = \frac{1}{N_D}\sum_{s=1}^{N_D}\hat{\omega}_{i,s}^{-4}\left[(\varpi_{i,s}^2 - \hat{\omega}_{i,s}^2)^2\right] + \frac{1}{N_D}\sum_{s=1}^{N_D}\hat{\omega}_{i,s}^{-4}\hat{\sigma}_{\omega_{i,s}^2}^2$

9: $\quad \hat{\Sigma}_{\phi_i} = \frac{1}{N_D}\sum_{s=1}^{N_D}\left[(\hat{\chi}_{i,s}\gamma_{i,s}\hat{\psi}_{i,s} - \hat{\phi}_{i,s})(\hat{\chi}_{i,s}\gamma_{i,s}\hat{\psi}_{i,s} - \hat{\phi}_{i,s})^T + \hat{\Sigma}_{\phi_{i,s}}\right]$

10: **} End For**

11: **Set** the initial values of *Conv* and *Tol* (Example: *Conv* =1 and *Tol* = 1e-6)

12: **While** *Conv* > *Tol* {

13: **For** $s = 1:N_D$ {

14: $\quad$ **For** $i = 1:N_m$ {**(E-Step)**

15: $\quad\quad$ **Calculate** the first moment of $\omega_{i,s}^2$ from Eq. (B6): $\mathbb{E}[\omega_{i,s}^2] = (\hat{\sigma}_{\omega_{i,s}^2}^{-2} + \hat{\omega}_{i,s}^{-4}\tau_{\omega_i^2}^{-2})^{-1}(\hat{\sigma}_{\omega_{i,s}^2}^{-2}\hat{\omega}_{i,s}^2 + \hat{\omega}_{i,s}^{-4}\tau_{\omega_i^2}^{-2}\varpi_{i,s}^2)$

16: $\quad\quad$ **Calculate** the second moment of $\omega_{i,s}^2$ from Eq. (B7): $\mathbb{E}[\omega_{i,s}^4] = (\hat{\sigma}_{\omega_{i,s}^2}^{-2} + \hat{\omega}_{i,s}^{-4}\tau_{\omega_i^2}^{-2})^{-1} + (\mathbb{E}[\omega_{i,s}^2])^2$

17: $\quad\quad$ **Solve** the following eigenvalue problem, and and eigenvector

$$\begin{bmatrix} \hat{\Sigma}_{\phi_{i,s}}^{-1} + \Sigma_{\phi}^{-1} & \mathbf{b}_{i,s}\mathbf{b}_{i,s}^T \\ \mathbf{I}_{N_o} & \hat{\Sigma}_{\phi_{i,s}}^{-1} + \Sigma_{\phi}^{-1} \end{bmatrix}\begin{bmatrix} \phi_{i,s} \\ \mathbf{z}_{i,s} \end{bmatrix} = \delta_{i,s}\begin{bmatrix} \phi_{i,s} \\ \mathbf{z}_{i,s} \end{bmatrix}$$

18: $\quad\quad$ **Calculate** the smallest eigenvalue ($\tilde{\delta}_{i,s}$) and the correspond eigenvector ($\tilde{\phi}_{i,s}$) from Eq. (B18)

19: $\quad\quad$ **Calculate** the Hessian inverse based on Eqs. (B28-B29): $\mathbb{E}[\phi_{i,s}\phi_{i,s}^T] = \mathbf{H}^{-1}(\phi_{i,s})|_{\phi_{i,s}=\tilde{\phi}_{i,s}} + \tilde{\phi}_{i,s}\tilde{\phi}_{i,s}^T$

20: $\quad$ **} End For**

21: **} End For**

22: **For** $s = 1:N_D$ {**M-Step 1:**

23: $\quad$ **Minimize** $L(\theta_s)$ given by Eq. (B30) and **calculate** the MPV ($\hat{\theta}_s$) (Derivatives are supplied in Appendix (D))

24: $\quad$ **Calculate** $\varpi^2(\hat{\theta}_s)$ and $\Psi(\hat{\theta}_s)$ by Eq. (2)

25: **} End For**

26: **M-Step 2: Update** the elements of $\Sigma_{\omega^2}$ and $\Sigma_{\Phi}$ from Eqs. (B31-B32):

27: $\hat{\tau}_{\omega_i^2}^2 = \frac{1}{N_D}\sum_{s=1}^{N_D}\hat{\omega}_{i,s}^{-4}\left[\mathbb{E}[\omega_{i,s}^4] - 2\varpi_{i,s}^2\mathbb{E}[\omega_{i,s}^2] + \varpi_{i,s}^4\right]$

28: $\hat{\Sigma}_{\phi_i} = \frac{1}{N_D}\sum_{s=1}^{N_D}\left[\mathbb{E}\left[\|\phi_{i,s}\|^{-2}\phi_{i,s}\phi_{i,s}^T\right] + \chi_{i,s}^2\gamma_{i,s}\psi_{i,s}\psi_{i,s}^T\gamma_{i,s}^T - \chi_{i,s}\mathbb{E}\left[\|\phi_{i,s}\|^{-1}\phi_{i,s}\right]\psi_{i,s}^T\gamma_{i,s}^T + \chi_{i,s}\gamma_{i,s}\psi_{i,s}\mathbb{E}\left[\|\phi_{i,s}\|^{-1}\phi_{i,s}^T\right]\right]$

29: **Update** *Conv* based on Eq. (24)

30: **} End While**

---

In the second stage of the proposed computational approach, the marginal distribution provided in Eq. (18) is used. From this equation, the hyper-parameters $\Sigma_{\omega^2}$ and $\Sigma_{\Phi}$ are marginalized out approximately by substituting them with their MPV provided by the ECM algorithm. Doing so results in

$$p(\{\boldsymbol{\varpi}^2(\boldsymbol{\theta}_s), \boldsymbol{\Psi}(\boldsymbol{\theta}_s), \boldsymbol{\theta}_s\}_{s=1}^{N_D}, \boldsymbol{\mu_\theta}, \boldsymbol{\Sigma_\theta} | \mathbf{D}, \mathbb{M}_p)$$
$$\propto \prod_{s=1}^{N_D} \left[ N(\hat{\boldsymbol{\omega}}_s^2 | \boldsymbol{\varpi}^2(\boldsymbol{\theta}_s), \hat{\boldsymbol{\Sigma}}_{\boldsymbol{\omega}_s^2} + \hat{\boldsymbol{\Sigma}}_{\boldsymbol{\omega}^2}) N(\hat{\boldsymbol{\Phi}}_s | \boldsymbol{\Gamma}_s \boldsymbol{\Psi}(\boldsymbol{\theta}_s), \hat{\boldsymbol{\Sigma}}_{\boldsymbol{\Phi}_s} + \hat{\boldsymbol{\Sigma}}_{\boldsymbol{\Phi}}) N(\boldsymbol{\theta}_s | \boldsymbol{\mu_\theta}, \boldsymbol{\Sigma_\theta}) \right] \quad (25)$$

where $\hat{\boldsymbol{\Sigma}}_{\boldsymbol{\omega}^2}$ and $\hat{\boldsymbol{\Sigma}}_{\boldsymbol{\Phi}}$ respectively denote the MPV of $\boldsymbol{\Sigma}_{\boldsymbol{\omega}^2}$ and $\boldsymbol{\Sigma}_{\boldsymbol{\Phi}}$, obtained by Algorithm 1. The deterministic relationship between the analytical modal parameters and the structural parameters, given by Eq. (2), allows simplifying the latest equation as

$$p(\{\boldsymbol{\theta}_s\}_{s=1}^{N_D}, \boldsymbol{\mu_\theta}, \boldsymbol{\Sigma_\theta} | \mathbf{D}, \mathbb{M}_p) \propto \prod_{s=1}^{N_D} \left[ \exp(-J(\boldsymbol{\theta}_s)) N(\boldsymbol{\theta}_s | \boldsymbol{\mu_\theta}, \boldsymbol{\Sigma_\theta}) \right] \quad (26)$$

where $J(\boldsymbol{\theta}_s)$ describes the fitting accuracy of the experimental and analytical modal parameters, given by

$$J(\boldsymbol{\theta}_s) = \frac{1}{2} tr \left[ \left( \hat{\boldsymbol{\Sigma}}_{\boldsymbol{\omega}_s^2} + \hat{\boldsymbol{\Sigma}}_{\boldsymbol{\omega}^2} \right)^{-1} \left( \hat{\boldsymbol{\omega}}_s^2 - \boldsymbol{\varpi}^2(\boldsymbol{\theta}_s) \right) \left( \hat{\boldsymbol{\omega}}_s^2 - \boldsymbol{\varpi}^2(\boldsymbol{\theta}_s) \right)^T \right]$$
$$+ \frac{1}{2} tr \left[ \left( \hat{\boldsymbol{\Sigma}}_{\boldsymbol{\Phi}_s} + \hat{\boldsymbol{\Sigma}}_{\boldsymbol{\Phi}} \right)^{-1} \left( \hat{\boldsymbol{\Phi}}_s - \boldsymbol{\Gamma}_s \boldsymbol{\Psi}(\boldsymbol{\theta}_s) \right) \left( \hat{\boldsymbol{\Phi}}_s - \boldsymbol{\Gamma}_s \boldsymbol{\Psi}(\boldsymbol{\theta}_s) \right)^T \right] \quad (27)$$

In Eqs. (26-27), the dependence on the analytical modal parameters is dropped, as expressed explicitly via the structural parameters ($\boldsymbol{\theta}_s$). Moreover, the identifiability of $\boldsymbol{\theta}_s$ can be characterized based on the number of minima of the loss function $J(\boldsymbol{\theta}_s)$ in the sense Katafygiotis and Beck [5] suggest. In practice, global identifiability is highly desired, reflecting that the collected data is sufficiently informative for inferring the structural parameters. Otherwise, we might have to improve the sensor configuration or the FE model to attain identifiability. Global identifiability is also important from a computational standpoint since it puts more flexible options ahead of the users. Specifically, when $\boldsymbol{\theta}_s$ is globally identifiable, the Laplace asymptotic approximation aids to substitute $\exp(-J(\boldsymbol{\theta}_s))$ with an approximate Gaussian distribution, leading to

$$\exp(-J(\boldsymbol{\theta}_s)) \propto N(\hat{\boldsymbol{\theta}}_s | \boldsymbol{\theta}_s, \hat{\boldsymbol{\Sigma}}_{\boldsymbol{\theta}_s}) \quad (28)$$

where $\hat{\boldsymbol{\theta}}_s$ is the desired minimum acquired by the foregoing ECM algorithm and $\hat{\boldsymbol{\Sigma}}_{\boldsymbol{\theta}_s} = \left[ \nabla^T \nabla J(\boldsymbol{\theta}_s) \right]^{-1}_{\boldsymbol{\theta}_s = \hat{\boldsymbol{\theta}}_s}$ is the Hessian inverse evaluated at $\hat{\boldsymbol{\theta}}_s$. As the MPV of $\boldsymbol{\theta}_s$ are available, applying this approximation requires only the calculation of the Hessian matrix and its inverse. This process can be performed based on Appendix (D) using analytical derivatives of $J(\boldsymbol{\theta}_s)$. Then, Eq. (28) can be rewritten as

$$p(\{\boldsymbol{\theta}_s\}_{s=1}^{N_D}, \boldsymbol{\mu}_{\boldsymbol{\theta}}, \boldsymbol{\Sigma}_{\boldsymbol{\theta}} | \mathbf{D}, \mathbb{M}_p) \propto \prod_{s=1}^{N_D} \left[ N(\hat{\boldsymbol{\theta}}_s | \boldsymbol{\theta}_s, \hat{\boldsymbol{\Sigma}}_{\boldsymbol{\theta}_s}) N(\boldsymbol{\theta}_s | \boldsymbol{\mu}_{\boldsymbol{\theta}}, \boldsymbol{\Sigma}_{\boldsymbol{\theta}}) \right] \tag{29}$$

This equation is the basis to calculate the MPV of the remaining parameters through the following EM algorithm:

$$\begin{cases} E\text{-}Step: Calculate\ L(\boldsymbol{\mu}_{\boldsymbol{\theta}}, \boldsymbol{\Sigma}_{\boldsymbol{\theta}}) = \mathbb{E}_{\{\boldsymbol{\theta}_s\}_{s=1}^{N_D}} \left[ \ln p(\{\boldsymbol{\theta}_s\}_{s=1}^{N_D}, \boldsymbol{\mu}_{\boldsymbol{\theta}}, \boldsymbol{\Sigma}_{\boldsymbol{\theta}} | \mathbf{D}, \mathbb{M}_p) \right] \\ M\text{-}Step: Maximize\ L(\boldsymbol{\mu}_{\boldsymbol{\theta}}, \boldsymbol{\Sigma}_{\boldsymbol{\theta}})\ for\ (\boldsymbol{\mu}_{\boldsymbol{\theta}}, \boldsymbol{\Sigma}_{\boldsymbol{\theta}}) \end{cases} \tag{30}$$

where the objective function $L(\boldsymbol{\mu}_{\boldsymbol{\theta}}, \boldsymbol{\Sigma}_{\boldsymbol{\theta}})$ is computed as

$$L(\boldsymbol{\mu}_{\boldsymbol{\theta}}, \boldsymbol{\Sigma}_{\boldsymbol{\theta}}) = -\frac{N_D}{2} \ln |\boldsymbol{\Sigma}_{\boldsymbol{\theta}}| - \frac{1}{2} \sum_{s=1}^{N_D} tr \left[ \boldsymbol{\Sigma}_{\boldsymbol{\theta}}^{-1} \left[ \mathbb{E}_{\boldsymbol{\theta}_s}[\boldsymbol{\theta}_s \boldsymbol{\theta}_s^T] + \boldsymbol{\mu}_{\boldsymbol{\theta}} \boldsymbol{\mu}_{\boldsymbol{\theta}}^T - \boldsymbol{\mu}_{\boldsymbol{\theta}} \mathbb{E}_{\boldsymbol{\theta}_s}[\boldsymbol{\theta}_s^T] - \mathbb{E}_{\boldsymbol{\theta}_s}[\boldsymbol{\theta}_s] \boldsymbol{\mu}_{\boldsymbol{\theta}}^T \right] \right] + cte. \tag{31}$$

The proposed EM algorithm leads to explicit formulations for calculating the MPV of the hyper-parameters $(\boldsymbol{\mu}_{\boldsymbol{\theta}}, \boldsymbol{\Sigma}_{\boldsymbol{\theta}})$. The mathematical derivation of this algorithm is provided in Appendix (C), and Algorithm 2 summarizes the computational procedure proposed for estimating the structural parameters. It starts with the computation of the identification uncertainty of the structural parameters for each data set. Then, it uses a "while loop" to implement the EM algorithm of Eq. (30). This process starts with calculating the initial estimations of the hyper-parameters $\boldsymbol{\mu}_{\boldsymbol{\theta}}$ and $\boldsymbol{\Sigma}_{\boldsymbol{\theta}}$ based on the approximate solutions derived in Appendix (C), aiming to guide the EM algorithm toward a good neighborhood of the solution. Then, the "while loop" will continue until convergence.

**Algorithm 2.** Proposed method for inferring the hyper-parameters of the structural parameters

1: **For** each data set, $s = 1 : N_D$ {
2:  **Calculate** the identification covariance matrix by calculating the Hessian inverse of $J(\boldsymbol{\theta}_s)$ at the optimal values,
3:  $\hat{\boldsymbol{\Sigma}}_{\boldsymbol{\theta}_s} = \left[ \nabla^T \nabla J(\boldsymbol{\theta}_s) \right]^{-1}_{\boldsymbol{\theta}_s = \hat{\boldsymbol{\theta}}_s}$   Note: Analytical derivatives of $J(\boldsymbol{\theta}_s)$ are supplied for use in Appendix (D).
4: } **End For**

*EM algorithm* (See Appendix (C) for the derivation):

5: **Set** initial estimations of $\boldsymbol{\mu}_{\boldsymbol{\theta}}$ and $\boldsymbol{\Sigma}_{\boldsymbol{\theta}}$ from Eqs. (C7-C8):
6:  $\hat{\boldsymbol{\mu}}_{\boldsymbol{\theta}} = \dfrac{1}{N_D} \sum_{s=1}^{N_D} \hat{\boldsymbol{\theta}}_s$
7:  $\hat{\boldsymbol{\Sigma}}_{\boldsymbol{\theta}} = \dfrac{1}{N_D} \sum_{s=1}^{N_D} \left[ (\hat{\boldsymbol{\theta}}_s - \hat{\boldsymbol{\mu}}_{\boldsymbol{\theta}})(\hat{\boldsymbol{\theta}}_s - \hat{\boldsymbol{\mu}}_{\boldsymbol{\theta}})^T + \hat{\boldsymbol{\Sigma}}_{\boldsymbol{\theta}_s} \right]$
8: **Set** the initial values of the *Conv* and *Tol*. (example: *Conv* =1 and *Tol* = 1e-6)
9: **While** *Conv* > *Tol* {
10: **For** $s = 1 : N_D$ {**E-Step: calculate** second-moment information from Eqs. (C3-C4)
11:  $\mathbb{E}[\boldsymbol{\theta}_s] = (\hat{\boldsymbol{\Sigma}}_{\boldsymbol{\theta}_s}^{-1} + \boldsymbol{\Sigma}_{\boldsymbol{\theta}}^{-1})^{-1} (\hat{\boldsymbol{\Sigma}}_{\boldsymbol{\theta}_s}^{-1} \hat{\boldsymbol{\theta}}_s + \boldsymbol{\Sigma}_{\boldsymbol{\theta}}^{-1} \boldsymbol{\mu}_{\boldsymbol{\theta}})$
12:  $\mathbb{E}[\boldsymbol{\theta}_s \boldsymbol{\theta}_s^T] = (\boldsymbol{\Sigma}_{\boldsymbol{\theta}}^{-1} + \hat{\boldsymbol{\Sigma}}_{\boldsymbol{\theta}_s}^{-1})^{-1} + \mathbb{E}[\boldsymbol{\theta}_s] \mathbb{E}[\boldsymbol{\theta}_s]^T$
13: } **End For**
14: **M-Step: update** the hyper-parameters $\boldsymbol{\mu}_{\boldsymbol{\theta}}$ and $\boldsymbol{\Sigma}_{\boldsymbol{\theta}}$ from Eqs. (C5-C6)
15:  $\hat{\boldsymbol{\mu}}_{\boldsymbol{\theta}} = \dfrac{1}{N_D} \sum_{s=1}^{N_D} \mathbb{E}[\boldsymbol{\theta}_s]$
16:  $\hat{\boldsymbol{\Sigma}}_{\boldsymbol{\theta}} = \dfrac{1}{N_D} \sum_{s=1}^{N_D} \left[ \mathbb{E}[\boldsymbol{\theta}_s \boldsymbol{\theta}_s^T] + \hat{\boldsymbol{\mu}}_{\boldsymbol{\theta}} \hat{\boldsymbol{\mu}}_{\boldsymbol{\theta}}^T - \hat{\boldsymbol{\mu}}_{\boldsymbol{\theta}} \mathbb{E}[\boldsymbol{\theta}_s^T] - \mathbb{E}[\boldsymbol{\theta}_s] \hat{\boldsymbol{\mu}}_{\boldsymbol{\theta}}^T \right]$
17: **Update** *Conv* based on Eq. (24)
18: } **End While**

## 4. Theoretical insights and comparison with existing methods

Most methods developed for inferring the structural parameters based on the modal data lead to the minimization of a quadratic loss function [15]. Such loss functions can generally be written as

$$L(\boldsymbol{\theta}) = \sum_{s=1}^{N_D} \sum_{i=1}^{N_m} w_i \left( \varpi_i^2(\boldsymbol{\theta}) - \hat{\omega}_{i,s}^2 \right)^2 + \sum_{s=1}^{N_D} \sum_{i=1}^{N_m} w_{N_m+i} \left\| \chi_i \gamma_i \boldsymbol{\psi}_i(\boldsymbol{\theta}) - \left\| \hat{\boldsymbol{\phi}}_{i,s} \right\|^{-1} \hat{\boldsymbol{\phi}}_{i,s} \right\|^2 \quad (32)$$

where $w_i$ is the weighting coefficient of the *i*th modal feature, governing its contribution to the loss function. Appropriate selection of these weights is important as they directly affect the optimal values of the parameters. In this paper, such a relationship can be obtained based on Eq. (18). The maximization of Eq. (18) for the underlying parameters yields the MPV, which can be achieved equivalently through the minimization of negative logarithm of the distribution, given by

$$L(\{\varpi^2(\boldsymbol{\theta}_s), \boldsymbol{\Psi}(\boldsymbol{\theta}_s), \boldsymbol{\theta}_s\}_{s=1}^{N_D}, \varphi) = \frac{1}{2}\sum_{s=1}^{N_D}\left[\ln\left|\hat{\boldsymbol{\Sigma}}_{\omega_s^2} + \boldsymbol{\Sigma}_{\omega^2}\right| + \ln\left|\hat{\boldsymbol{\Sigma}}_{\Phi_s} + \boldsymbol{\Sigma}_{\Phi}\right| + \ln\left|\boldsymbol{\Sigma}_{\theta}\right|\right]$$

$$+ \frac{1}{2}\sum_{s=1}^{N_D}\left[tr\left[\boldsymbol{\Sigma}_{\theta}^{-1}(\boldsymbol{\theta}_s - \boldsymbol{\mu}_{\theta})(\boldsymbol{\theta}_s - \boldsymbol{\mu}_{\theta})^T\right]\right]$$

$$+ \frac{1}{2}\sum_{s=1}^{N_D} tr\left[(\hat{\boldsymbol{\Sigma}}_{\omega_s^2} + \boldsymbol{\Sigma}_{\omega^2})^{-1}(\hat{\boldsymbol{\omega}}_s^2 - \varpi^2(\boldsymbol{\theta}_s))(\hat{\boldsymbol{\omega}}_s^2 - \varpi^2(\boldsymbol{\theta}_s))^T\right]$$

$$+ \frac{1}{2}\sum_{s=1}^{N_D} tr\left[(\hat{\boldsymbol{\Sigma}}_{\Phi_s} + \boldsymbol{\Sigma}_{\Phi})^{-1}(\hat{\boldsymbol{\Phi}}_s - \boldsymbol{\Gamma}_s\boldsymbol{\Psi}(\boldsymbol{\theta}_s))(\hat{\boldsymbol{\Phi}}_s - \boldsymbol{\Gamma}_s\boldsymbol{\Psi}(\boldsymbol{\theta}_s))^T\right] + cte.$$

(33)

where $L(\{\varpi^2(\boldsymbol{\theta}_s), \boldsymbol{\Psi}(\boldsymbol{\theta}_s), \boldsymbol{\theta}_s\}_{s=1}^{N_D}, \varphi) \doteq -\ln\left[p(\{\varpi^2(\boldsymbol{\theta}_s), \boldsymbol{\Psi}(\boldsymbol{\theta}_s), \boldsymbol{\theta}_s\}_{s=1}^{N_D}, \varphi \mid \mathbf{D}, \mathbb{M}_p)\right]$ is the negative logarithm of the marginal posterior distribution. Comparing Eq. (32) with Eq. (33), the following points can be highlighted:

- The uncertainty of the analytical modal frequency ($\varpi^2(\boldsymbol{\theta}_s)$) and the incomplete mode shape vector ($\boldsymbol{\Gamma}_s\boldsymbol{\Psi}(\boldsymbol{\theta}_s)$) are determined as $\hat{\boldsymbol{\Sigma}}_{\omega_s^2} + \boldsymbol{\Sigma}_{\omega^2}$ and $\hat{\boldsymbol{\Sigma}}_{\Phi_s} + \boldsymbol{\Sigma}_{\Phi}$, respectively. This aggregate uncertainty is the sum of the identification and prediction uncertainties.

- When minimizing this objective function, the modal parameters $\varpi^2(\boldsymbol{\theta}_s)$ and $\boldsymbol{\Gamma}_s\boldsymbol{\Psi}(\boldsymbol{\theta}_s)$ are weighed by $(\hat{\boldsymbol{\Sigma}}_{\omega_s^2} + \boldsymbol{\Sigma}_{\omega^2})^{-1}$ and $(\hat{\boldsymbol{\Sigma}}_{\Phi_s} + \boldsymbol{\Sigma}_{\Phi})^{-1}$, respectively. This weighting scheme makes intuitive sense as more uncertain parameters have smaller contributions and vice versa.

- Since the natural logarithm of a number is a monotonically increasing function, the expressions $\ln\left|\hat{\boldsymbol{\Sigma}}_{\omega_s^2} + \boldsymbol{\Sigma}_{\omega^2}\right|$ and $\ln\left|\hat{\boldsymbol{\Sigma}}_{\Phi_s} + \boldsymbol{\Sigma}_{\Phi}\right|$ in Eq. (33) penalize large uncertainties reflected through the covariance matrices.

- The expression $\ln\left|\boldsymbol{\Sigma}_{\theta}\right|$ acts as a penalty over large variances of the structural parameters, and the expression $\boldsymbol{\Sigma}_{\theta}^{-1}(\boldsymbol{\theta}_s - \boldsymbol{\mu}_{\theta})(\boldsymbol{\theta}_s - \boldsymbol{\mu}_{\theta})^T$ accounts for the fitting accuracy of the hyper-parameters.

It is also worth comparing the proposed framework with its Bayesian predecessors. The multi-objective framework proposed in [32] weighs the modal features according to the inverse sum of the modal residuals. Based on the notation used herein, this scheme considers the weights proportional to $\boldsymbol{\Sigma}_{\omega^2}^{-1}$ and $\boldsymbol{\Sigma}_{\Phi}^{-1}$. On the other hand, the Bayesian approach proposed by Au and Zhang [39] considers the weights equal to the inverse of the identification covariance matrices, i.e., $\hat{\mathbf{\Sigma}}_{\omega_s^2}^{-1}$ and $\hat{\mathbf{\Sigma}}_{\mathbf{\Phi}_s}^{-1}$. The proposed HBM framework unifies these two perspectives by characterizing the modal features' weights as equal to the inverse of the aggregate uncertainties, i.e., $(\hat{\mathbf{\Sigma}}_{\omega_s^2} + \mathbf{\Sigma}_{\omega^2})^{-1}$ and $(\hat{\mathbf{\Sigma}}_{\mathbf{\Phi}_s} + \mathbf{\Sigma}_{\mathbf{\Phi}})^{-1}$. This result also underpins that the above methods are only special cases of the proposed HBM framework.

The Bayesian approach presented in Goller et al. [34] maximizes the evidence term for providing optimal weights. The proposed HBM framework does a similar task through a more general and natural approach, identifying the weights as a part of the Bayesian optimization.

## 5. Illustrative examples

### 5.1. Numerical example with synthetic data

The two-story frame shown in Fig. 2 is used for the verification of the proposed methodology. The mass is lumped at the first and second floors, both considered 1kg, and the nominal value of the lateral stiffness of the columns is set to $k_1 = k_2 = 0.5\text{kN/m}$. The lateral stiffness of the second story is expressed as $2\theta k_2$, where $\theta$ is an unknown parameter. Three sets of experimental modal data ($N_D = 3$), each comprising two modal frequencies and shapes ($N_m = N_o = 2$), are simulated using perturbed values $\theta_1 = 0.98$, $\theta_2 = 1.00$, and $\theta_3 = 1.02$, which creates variability in the modal features. No additive noise is imposed on the experimental modal features, and their identification uncertainty is characterized using 0.001 coefficients of variation. We intend to apply the proposed probabilistic model to this simulated modal data and update the unknown parameters, collected into $\{\theta_1, \theta_2, \theta_3, \mu_\theta, \sigma_\theta^2, \tau_{\omega_1^2}^2, \tau_{\omega_2^2}^2, \sigma_{\phi_1}^2, \sigma_{\phi_2}^2\}$. For the sake of simplicity, the covariance matrices $\mathbf{\Sigma}_{\phi_1}$ and $\mathbf{\Sigma}_{\phi_2}$ are reparametrized as $\mathbf{\Sigma}_{\phi_1} = \sigma_{\phi_1}^2 \mathbf{I}$ and $\mathbf{\Sigma}_{\phi_2} = \sigma_{\phi_2}^2 \mathbf{I}$, respectively. The actual values of the hyper parameters are calculated as $\mu_\theta = 1$ and $\sigma_\theta^2 = 2.7 \times 10^{-4}$. Since the simulated modal features are free of additive noise, the actual values of the variances $\{\tau_{\omega_1^2}^2, \tau_{\omega_2^2}^2, \sigma_{\phi_1}^2, \sigma_{\phi_2}^2\}$ are all zero.

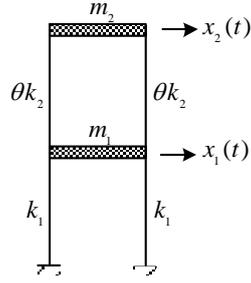

**Fig. 2,** Two-story structure considered for validation of the proposed method

Having made the above assumptions, the proposed computational algorithms are used for inferring the unknown parameters. Table 1 compares the MPV of the parameters with the actual values. The MPV of $\{\theta_1,\theta_2,\theta_3,\mu_\theta,\sigma_\theta^2\}$ matches almost exactly with the actual values, and the variances $\{\tau_{\omega_1^2}^2,\tau_{\omega_2^2}^2,\sigma_{\phi_1}^2,\sigma_{\phi_2}^2\}$ appear to be very close to zero. This result confirms the validity of the EM algorithms.

**Table 1.** Identification of the unknown parameters and comparison with actual values

| Model Class | Computational Method | Unknown parameters | | | | | | | | |
|---|---|---|---|---|---|---|---|---|---|---|
| | | $\theta_1$ | $\theta_2$ | $\theta_3$ | $\mu_\theta$ | $\sigma_\theta^2$ | $\tau_{\omega_1^2}^2$ | $\tau_{\omega_2^2}^2$ | $\sigma_{\phi_1}^2$ | $\sigma_{\phi_2}^2$ |
| | Exact Values | 0.98 | 1.00 | 1.02 | 1.00 | 2.7e-4 | 0 | 0 | 0 | 0 |
| $M_{P,1}$ [1] | EM Algorithms [4] | 0.981 | 1.000 | 1.020 | 1.000 | 2.7e-4 | 9e-12 | 4e-12 | 9e-7 | 8e-7 |
| | Sampling [5] | 0.981 | 1.001 | 1.020 | 1.001 | 0.0004 | 4.4e-4 | 5.3e-4 | 2.3e-4 | 2.4e-4 |
| $M_{P,2}$ [2] | Optimization [6] | ---- | ---- | ---- | 0.999 | 0.0023 | 4e-5 | 4e-5 | 4e-6 | 4e-6 |
| $M_{P,3}$ [3] | Optimization [6] | ---- | ---- | ---- | 1.001 | 0.0012 | 2e-5 | 2e-5 | 5e-6 | 5e-6 |

[1] Proposed HBM framework
[2] Vanik et al. [27] (involves no individual weighting of the modal features)
[3] Goller et al. [34] (uses a scalar to weight modal frequencies and shapes differently)
[4] MPV Calculated based on Algorithms 1 and 2
[5] Mean of posterior samples
[6] Calculated by Laplace Approximation

Additionally, the results are compared with a full sampling strategy whose details are provided in Appendix (E). The prior distribution of $\{\theta_1,\theta_2,\theta_3,\mu_\theta\}$ is considered uniform varying within the interval $U(0.75,1.25)$; the prior distribution of the variances $\{\sigma_\theta^2,\tau_{\omega_1^2}^2,\tau_{\omega_2^2}^2,\sigma_{\phi_1}^2,\sigma_{\phi_2}^2\}$ is selected to be uniform within the interval $U(0,0.01)$. The sample size is considered as large as $N_{sa}=100000$, and an MCMC sampling algorithm is used for drawing the samples from the posterior distribution [71]. The results of sampling are visualized in a matrix plot format in Fig. 3. Histogram plots are shown on the main diagonal, and the smoothed joint distributions are shown in the lower-diagonal plots. The mean of the samples is presented in Table 1, indicating remarkable accuracy of the sampling approach. However, the variances obtained from the sampling approach are slightly larger than those of the EM algorithms and the actual values (zero), which is attributed to the fact that the sampling approach also incorporates the uncertainty in the hyper-parameters whereas the proposed EM algorithms ignore it.

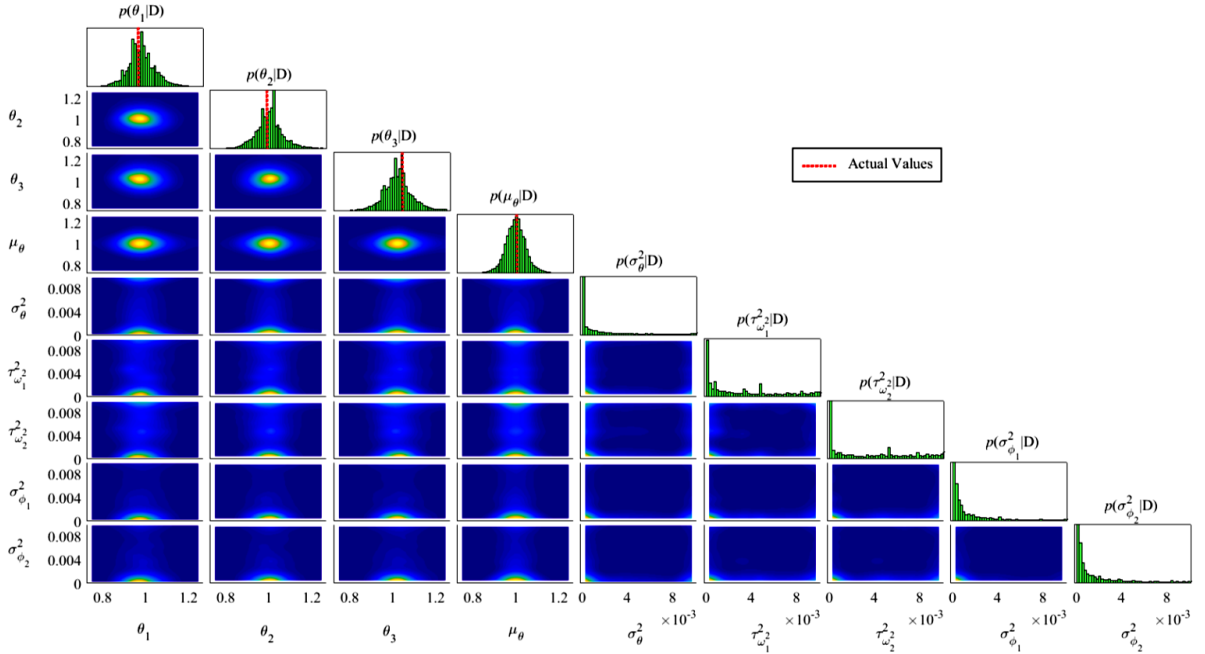

**Fig. 3,** Visualization of the joint posterior distribution and its histogram plots using MCMC sampling

In terms of computation time, the MCMC sampling was completed in 2186s whereas the EM algorithms converged only in 26s. Note that both simulations were performed on the same computer having an Intel Core i7 2.8GHz processor. Although this comparison can suggest the computational savings of the proposed EM algorithms, care must be taken when interpreting the time. Specifically, the computational cost of the EM algorithm depends on the number of iterations, which can be adjusted through the predefined tolerance and the stopping criteria. On the other hand, the number of samples governs the computational cost of the sampling approach. Consequently, an unrealistic tolerance or an excessively large number of samples can lead to an indefinitely large time of computation. In this example, we balanced the stopping criteria and the sample size to obtain a similar level of accuracy by both methods such that the time of computation becomes a measure of computational efficiency.

In Table 1, the results of two Bayesian formulations are also provided for the sake of comparison. These probabilistic models appear as $M_{P,2}$ and $M_{P,3}$, which refer to Vanik et al. [27] and Goller et al. [34], respectively. Although these models perform well, the proposed probabilistic model ($M_{P,3}$) yields the most accurate results in this example. This conclusion is strengthened by performing a Bayesian model class selection [11,72]. In Table 2, the Bayesian Information Criterion (BIC) and the logarithm of the evidence term are presented. Note that the BIC is calculated from $\text{BIC} = \ln P(\mathbf{D}|\hat{\Theta}, M_{P,j}) - (N_\Theta / 2)\ln\left(N_m(N_o + 1)\right)$ [11,72,73], where $N_\Theta$ is the total number of parameters; $\ln P(\mathbf{D}|\hat{\Theta}, M_{P,j})$ is the log-likelihood function evaluated at the MPV ($\hat{\Theta}$), reflecting the fitting accuracy of the model; the expression $(N_\Theta / 2)\ln\left(N_m(N_o + 1)\right)$ penalizes parameterization. Based on this table, it can be concluded that the proposed method ($M_{P,1}$) corresponds to larger values of the BIC and the evidence. Additionally, when equal prior probabilities are attributed to each model class, i.e., $P(M_{P,j}|\mathbb{M}_\mathbb{P}) = 1/3$; $\forall j = 1, 2, 3$, the posterior probabilities $P(M_{P,j}|\mathbf{D}, \mathbb{M}_\mathbb{P})$ can be calculated through the Bayes' rule. Based on the results in Table 2, it can be confirmed that the proposed probabilistic model ($M_{P,1}$) is much more likely than the other two formulations. It should be noticed that, although the number of parameters is comparatively large, the gain in the fitting accuracy seems to compensate for the loss of BIC score due to having additional parameterization.

**Table 2.** Bayesian model class selection

| Selection Criterion | Probabilistic Model Class | | |
|---|---|---|---|
| | $M_{P,1}$ [1] | $M_{P,2}$ [2] | $M_{P,3}$ [3] |
| BIC $= \ln P(\mathbf{D}|\hat{\Theta}, M_{P,j}) - (N_\Theta/2)\ln(N_m(N_o+1))$ | 131.56 | 66.82 | 75.43 |
| ln (Evidence): $\ln p(\mathbf{D}|M_P)$ | 144.57 | 69.71 | 79.76 |
| Posterior Probability: $P(M_P|\mathbf{D}, \mathbb{M}_\mathbb{P})$ | ≈1.0 | 3e-33 | 7e-29 |

[1] Proposed HBM framework
[2] Vanik et al. [27] (involves no individual weighting of the modal features)
[3] Goller et al. [34] (uses a scalar to weight modal frequencies and shapes differently)

## 5.2. Small-scale shear frame

A three-story shear frame shown in Fig. 4(a) is used for demonstrating the proposed framework. The acceleration time-history responses of all three floors were measured while the structure was subjected to $N_D = 19$ independent GWN base excitations. Each data set is 120s long and sampled at 0.005s intervals. The data sets are collected while the temperature, excitation properties, and base fixity were the same throughout the experimentation. Thus, any variability in the realizations would be attributed to modeling errors.

The lateral stiffness of the columns along the y-axis is much larger than the x-axis. Thus, vibrations along the y-axis can be ignored, which allows modeling the structure by a 2D shear frame shown in Fig. 4(b). The mass matrix is considered to be known and lumped on each floor, given as $m_1$ = 5.63kg, $m_2$ = 6.03kg, and $m_3$ = 4.66kg [50]. On the contrary, the stiffness parameters should be identified using Algorithms 1-2. The stiffness matrix of this model is parameterized by multiplying the nominal stiffness of the three stories by the unknown parameters $\theta_1$, $\theta_2$, and $\theta_3$. The nominal stiffness of the stories was previously reported as $k_1$ = 20.88kN/m, $k_2$ = 22.37kN/m, and $k_3$ = 24.21kN/m [50].

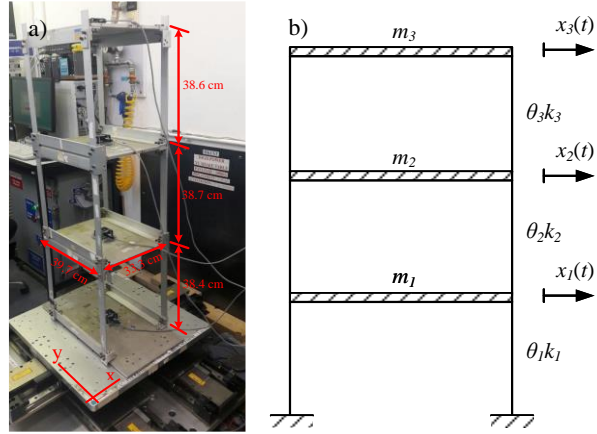

**Fig. 4,** a) Small-scale frame tested on a shaking table under GWN base excitation along the x-axis b) Shear frame model considered to simulate dynamical responses

The averaged singular value spectrum is obtained and shown in Fig. 5. Three resonant bands are considered as [3.2–5.2Hz], [12.0–14.0Hz], and [17.5–19.5Hz]. The FFT-based modal identification approach gives the MPV of the modal parameters and their identification uncertainties. Fig. 6 shows a schematic view of the three dynamical modes. For each data set, the modal frequencies and mode shape vectors are obtained along with their identification uncertainty. Table 3 presents the modal identification results, including the MPV and identification uncertainty of all modal parameters. For comparison purposes, the ensemble mean of the MPV and standard deviations is presented in this table. This outcome is in good agreement with those reported in [50], which confirms the validity of the results.

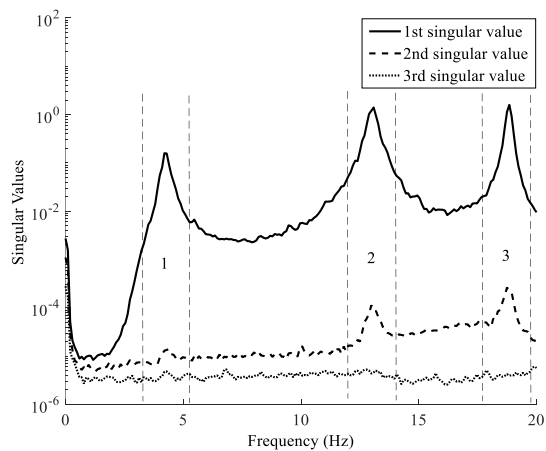

**Fig. 5,** Averaged singular value spectrum

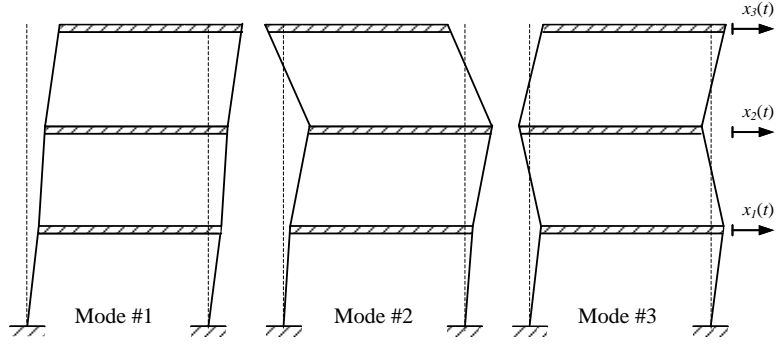

**Fig. 6,** A schematic view of the dynamical modes

**Table 3.** MPV and identification uncertainty of the experimental modal parameters

| Experimental Modal Parameters | Mode #1 | | Mode #2 | | Mode #3 | |
|---|---|---|---|---|---|---|
| | Mean[*] | S.D.[**] | Mean[*] | S.D.[**] | Mean[*] | S.D.[**] |
| $\omega_i/(2\pi)$ (Hz) | 4.245 | 0.0032 | 12.937 | 0.0018 | 18.852 | 0.0011 |
| $\xi_i$ (%) | 0.0150 | 0.2330 | 0.0126 | 0.1683 | 0.0065 | 0.1832 |
| $S_i$ | 0.0012 | 0.1029 | 6.6e-5 | 0.1192 | 5.2e-6 | 0.1163 |
| $S_{ei}$ | 0.0002 | 0.0644 | 2.6e-5 | 0.0643 | 2.0e-5 | 0.0644 |
| $\phi_{1i}$ | -0.3746 | 0.0021 | -0.6946 | 0.0016 | -0.4954 | 0.0034 |
| $\phi_{2i}$ | -0.5873 | 0.0019 | -0.1767 | 0.0022 | 0.7202 | 0.0028 |
| $\phi_{3i}$ | -0.7174 | 0.0016 | 0.6973 | 0.0016 | -0.4857 | 0.0034 |

[*] Averaged over the MPV obtained from different data sets
[**] Identification uncertainty averaged over all data sets (S.D. stands for standard deviation)

In the formulation presented earlier, the correlation between the modal frequencies and mode shape vectors was ignored. This assumption can be verified by calculating the cross-correlations based on the data sets. For instance, the correlation coefficients obtained from the data set ($D_1$) are calculated based on the non-diagonal elements of the Hessian inverse, giving:

$$\begin{cases} \rho_{\omega_{1,1}^2,\phi_{11,1}} = 4.9\times 10^{-6} \\ \rho_{\omega_{1,1}^2,\phi_{21,1}} = 3.2\times 10^{-7} \\ \rho_{\omega_{1,1}^2,\phi_{31,1}} = -3.7\times 10^{-6} \end{cases} ; \begin{cases} \rho_{\omega_{2,1}^2,\phi_{12,1}} = 1.5\times 10^{-6} \\ \rho_{\omega_{2,1}^2,\phi_{22,1}} = -5.7\times 10^{-6} \\ \rho_{\omega_{2,1}^2,\phi_{32,1}} = -5.2\times 10^{-7} \end{cases} ; \begin{cases} \rho_{\omega_{3,1}^2,\phi_{13,1}} = -9.7\times 10^{-6} \\ \rho_{\omega_{3,1}^2,\phi_{23,1}} = -4.2\times 10^{-7} \\ \rho_{\omega_{3,1}^2,\phi_{33,1}} = 4.9\times 10^{-6} \end{cases} \quad (34)$$

This result indicates that the cross-correlation between the modal frequencies and the components of the mode shape vector is indeed negligible, confirming the assumption made earlier in Eq. (10).

The experimental modal parameters obtained for each data set can next be used for updating the model parameters. Table 4 presents the MPV and the standard deviations of the analytical modal parameters. The standard deviations account for both the identification uncertainty of experimental modal parameters and the variability over different data sets. Comparing Tables 3 and 4, the matching accuracy of the modal frequencies appears to be good, whereas the elements of the mode shape vectors are in fair agreement. This conclusion can be reinforced by looking at the MPV of the prediction error variances presented in Table 5. These parameters govern the mismatch between the experimental and modal parameters, revealing that the uncertainty of the elements of mode shape vectors is mainly greater than those of the modal frequencies, except for the 1st modal frequency.

**Table 4.** MPV and identification uncertainty of the analytical modal parameters

| Analytical Modal Parameters | Mode #1 | | Mode #2 | | Mode #3 | |
|---|---|---|---|---|---|---|
| | Mean* | S.D.** | Mean* | S.D.** | Mean* | S.D.** |
| $\varpi_i/(2\pi)$ (Hz) | 4.3683 | 0.3174 | 12.956 | 0.0774 | 18.852 | 7.2e-5 |
| $\psi_{1i}$ | -0.3976 | 0.0192 | -0.7216 | 0.0276 | -0.5186 | 0.0303 |
| $\psi_{2i}$ | -0.5981 | 0.0148 | -0.1580 | 0.0164 | 0.7267 | 0.0053 |
| $\psi_{3i}$ | -0.6958 | 0.0227 | 0.6740 | 0.0245 | -0.4504 | 0.0268 |

* Averaged over the MPV obtained from different data sets
** Identification uncertainty averaged over all data sets

**Table 5.** Standard deviation of the modal parameters reflecting the mismatch between the modal features

| Analytical Modal Parameters | Mode #1 | Mode #2 | Mode #3 |
|---|---|---|---|
| $\tau_{\omega_i^2}$ | 0.0596 | 0.0048 | 5.1e-7 |
| $\sigma_{\phi_{1i}}$ | 0.0191 | 0.0275 | 0.0301 |
| $\sigma_{\phi_{2i}}$ | 0.0146 | 0.0163 | 0.0045 |
| $\sigma_{\phi_{3i}}$ | 0.0226 | 0.0245 | 0.0266 |

Fig. 7 depicts the posterior distribution of the stiffness parameters in a matrix plot format. The plots on the main diagonal correspond to the marginal posterior distribution $\theta_i \sim N(\hat{\mu}_{\theta_i}, \hat{\sigma}_{\theta_i}^2)$, where $\hat{\mu}_{\theta_i}$ and $\hat{\sigma}_{\theta_i}^2$ are calculated based on the MPV of $\mathbf{\mu_\theta}$ and $\mathbf{\Sigma_\theta}$, respectively. The upper triangular plots correspond to the joint distribution given as

$$\begin{bmatrix} \theta_i \\ \theta_j \end{bmatrix} \sim N\left( \begin{bmatrix} \hat{\mu}_{\theta_i} \\ \hat{\mu}_{\theta_j} \end{bmatrix}, \begin{bmatrix} \hat{\sigma}_{\theta_i}^2 & \hat{\rho}_{\theta_i \theta_j} \hat{\sigma}_{\theta_i} \hat{\sigma}_{\theta_j} \\ \hat{\rho}_{\theta_i \theta_j} \hat{\sigma}_{\theta_i} \hat{\sigma}_{\theta_j} & \hat{\sigma}_{\theta_j}^2 \end{bmatrix} \right) \tag{35}$$

where $\hat{\rho}_{\theta_i \theta_j}$ is the correlation between $\theta_i$ and $\theta_j$. The error bars on the lower triangular plots exhibit the realizations of each pair of the stiffness parameters along with their identification uncertainties (±3S.D.). Unlike the identification uncertainties that appear to be relatively small, the variability over data sets turns out to be much larger. When the environmental conditions and excitation levels are almost constant, as is the case in this example, the HBM framework provides insights into the variability attributed to modeling errors. On this plot, it is also evident that the correlation of the stiffness parameters is well accounted for by the hierarchical model.

Table 6 provides the estimations of the stiffness parameters. The validity of the results can be assured by comparing them with the results reported in [50], where the time-domain HBM is applied to the same experimental data sets. The ensemble mean and standard deviation agree well with the MPV of the hyper-parameters. This result indicates that they can be used as initial estimations. The identification uncertainty of the stiffness parameters obtained using the Laplace approximation is much smaller than the standard deviations $\hat{\sigma}_{\theta_i}$'s. This finding also indicates to what extent the identification uncertainty can be smaller than the test-to-test variability and highlights how well the proposed framework can quantify both the identification uncertainty and variability. Since the hyper covariance matrix is estimated in full, the correlation coefficients can also be realized, as tabulated in Table 6.

Given the uncertainties, it is straightforward to propagate them for the virtual sensing of unobserved response quantities. This step is skipped herein but is well discussed in our pervious works [50].

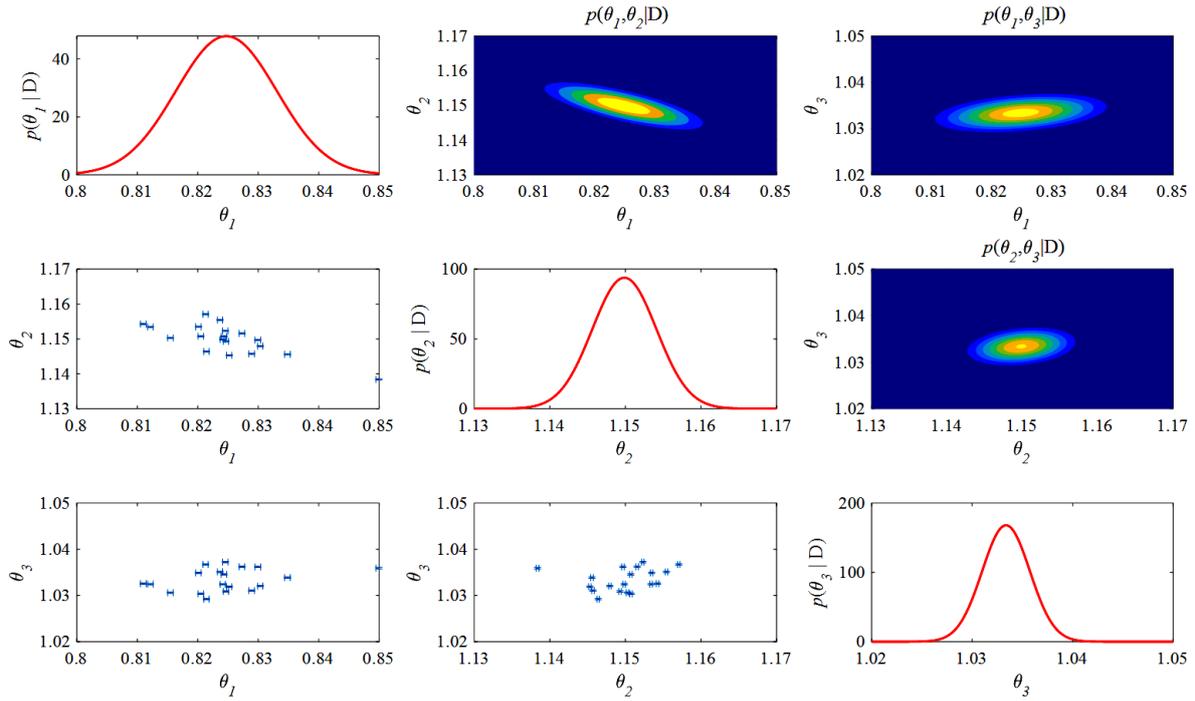

**Fig. 7,** Uncertainty Quantification of the stiffness parameters (Plots on the main diagonal present the marginal posterior distributions; The upper triangular plots show the joint posterior distribution of each pair of parameters; The lower triangular plots show the variation of each pair of the parameters over 19 data sets).

Table 6. MPV of stiffness hyper-parameters

| Stiffness parameters | Nominal Values [50] | Ensemble Mean | Ensemble Variability (S.D.) | Identification Uncertainty (S.D.) | Hyper-parameters of Stiffness | | | | |
|---|---|---|---|---|---|---|---|---|---|
| | | | | | $\hat{\mu}_{\theta_i}$ * | $\hat{\sigma}_{\theta_i}$ ** | | $\hat{\rho}_{\theta_i\theta_j}$ *** | |
| $\theta_1$ | 0.838 | 0.8247 | 0.0086 | 1.5e-4 | 0.8247 | 0.0083 | 1 | -0.7577 | 0.3373 |
| $\theta_2$ | 1.141 | 1.1499 | 0.0044 | 5.5e-5 | 1.1499 | 0.0043 | -0.7577 | 1 | 0.2448 |
| $\theta_3$ | 1.033 | 1.0334 | 0.0024 | 4.7e-5 | 1.0334 | 0.0024 | 0.3373 | 0.2448 | 1 |

* MPV of hyper mean vector
** MPV of the square-root of the diagonal elements of the hyper covariance
*** MPV of the correlation coefficients (based on the off-diagonal elements of the hyper covariance matrix)

## 5.3. Single-tower cable footbridge

The ambient vibration response of a footbridge located at the Hong Kong University of Science and Technology (HKUST) is a practical case for further investigation of the accuracy of the proposed framework. Fig. 8(a) shows the Northern view of the bridge. The structural drawings are used for creating a detailed FE model in CSI SAP 2000 [74] depicted in Fig. 8(b). This model consists of 31 frame elements, 6 shell elements, and 4 cable elements, creating 386 DOFs in total. The primary objective is to update the mechanical properties of this structure using acceleration responses provided by six triaxial Imote2 wireless sensors [75]. The dimensions of the bridge and the sensor placement are shown in Fig. 8(c). The experimental data includes 19 sets of acceleration time-history responses, where each data set is 300s long, measured at 0.01s intervals. The data sets were collected on a cloudy afternoon in March 2019 between 13:00-17:00 while the structure was under ambient excitations and pedestrian loadings. Further information about the experimental setup and data sets is available in [56].

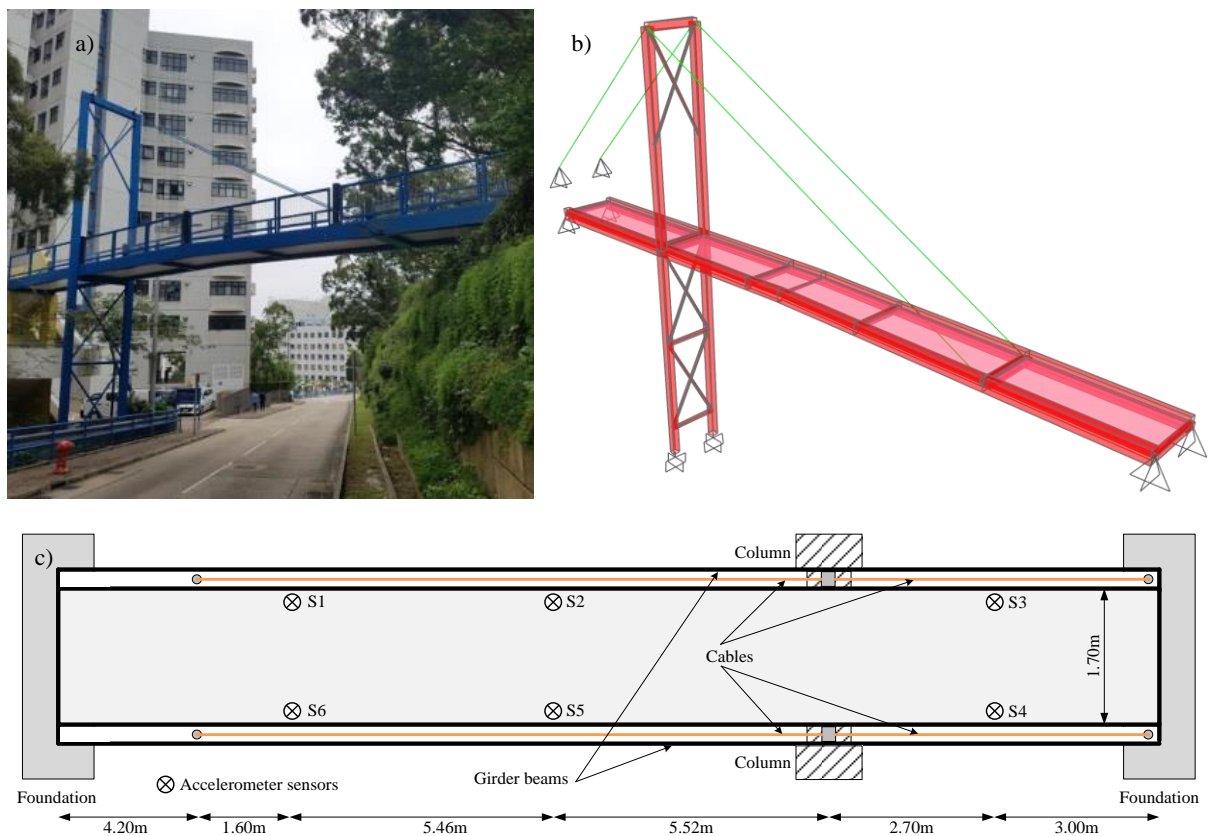

**Fig. 8,** a) The footbridge selected for collecting ambient vibration data b) FE model developed in SAP 2000 c) Placement of six accelerometers on the concrete deck for collecting ambient vibration data

Fig. 9 shows the singular value spectrum obtained using the experimental data. Four modes are well-excited, which allows selecting the resonant bands as [3.3–4.3Hz], [9.7–10.7Hz], [11.2–12.3Hz], and [19.5–20.5Hz]. Given these intervals, the FFT-based modal identification approach provides the experimental modal parameters, including the modal frequencies, modal damping ratios, the PSD of the input, the PSD of the prediction errors, and the elements of the mode shape vectors.

The MPV and the identification uncertainty are obtained for each data set, and their average over data sets is presented in Table 7.

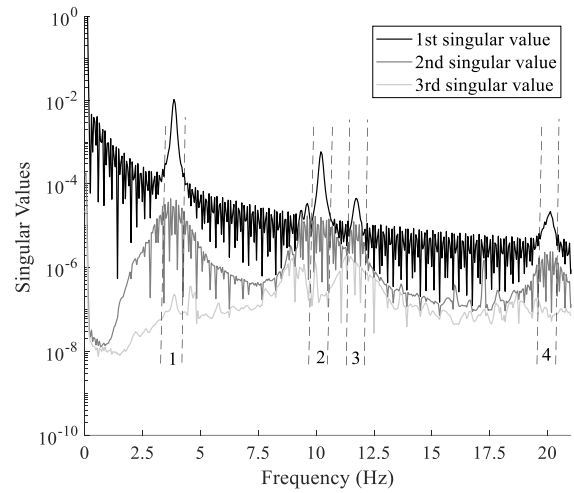

**Fig. 9,** Averaged singular value spectrum of the ambient vibration response used for selecting four well-separated dynamical modes

**Table 7.** MPV and identification uncertainty of the experimental modal parameters

| Modal Parameters | Mode #1 | | Mode #2 | | Mode #3 | | Mode #4 | |
|---|---|---|---|---|---|---|---|---|
| | Mean* | S.D.** | Mean* | S.D.** | Mean* | S.D.** | Mean* | S.D.** |
| $\omega_i/(2\pi)$ (Hz) | 3.8483 | 0.0014 | 10.2152 | 0.0006 | 11.7235 | 0.0007 | 20.0531 | 0.0007 |
| $\xi_i$ (%) | 0.0121 | 0.1274 | 0.0053 | 0.1204 | 0.0078 | 0.1053 | 0.0091 | 0.1035 |
| $S_i$ | 0.7998 | 0.0258 | 0.0084 | 0.0258 | 0.0013 | 0.0245 | 0.0007 | 0.0258 |
| $S_{ei}$ | 0.0773 | 0.0257 | 0.1149 | 0.0257 | 0.3224 | 0.0245 | 0.0525 | 0.0255 |
| $\phi_{1i}$ | -0.5161 | 0.0002 | -0.4308 | 0.0012 | -0.5252 | 0.0058 | -0.1902 | 0.0032 |
| $\phi_{2i}$ | -0.4787 | 0.0002 | 0.5559 | 0.0011 | 0.4335 | 0.0063 | 0.1063 | 0.0032 |
| $\phi_{3i}$ | -0.4701 | 0.0002 | 0.5466 | 0.0011 | -0.4879 | 0.0062 | 0.0961 | 0.0032 |
| $\phi_{4i}$ | 0.0496 | 0.0003 | -0.1177 | 0.0013 | -0.0248 | 0.0067 | -0.6555 | 0.0025 |
| $\phi_{5i}$ | 0.0491 | 0.0003 | -0.1145 | 0.0013 | 0.0456 | 0.0066 | -0.6959 | 0.0023 |
| $\phi_{6i}$ | -0.5278 | 0.0002 | -0.4236 | 0.0012 | 0.5430 | 0.0057 | -0.1682 | 0.0032 |

* Averaged MPV
** Averaged identification uncertainty

Fig. 10 visualizes the mode shapes of the FE model. The dynamical modes of the initial FE model are in good agreement with the experimental ones documented in [56]. The next step is to consider these modal properties for updating the FE model. Three substructures are considered to parameterize the FE model: the steel girders, columns, and bracings whose modulus of elasticity is to be updated ($200 \times \theta_1$ GPa); the four cables having an identical unknown modulus of elasticity ($200 \times \theta_2$ GPa); and finally, the reinforced concrete deck having an unknown modulus of elasticity ($21.5 \times \theta_3$

MPa). The mass matrix is estimated according to the dimensions and the material properties of the members.

The proposed framework is applied to this example. Table 8 presents the analytical modal parameters. The mean values correspond to the average of the MPV obtained from different data sets, and the standard deviations represent the aggregate uncertainty. The analytical modal parameters are in fair agreement with the experimental ones, and their mismatch can be realized based on the MPV of the hyper covariance matrices provided in Table 9. The analytical modal frequencies of the dynamical modes #1 and #2 do not fit well into the experimental ones, while the mode shape vector components match well. On the contrary, the modal properties of the dynamical modes #3 and #4 are in close agreement with the experimental ones.

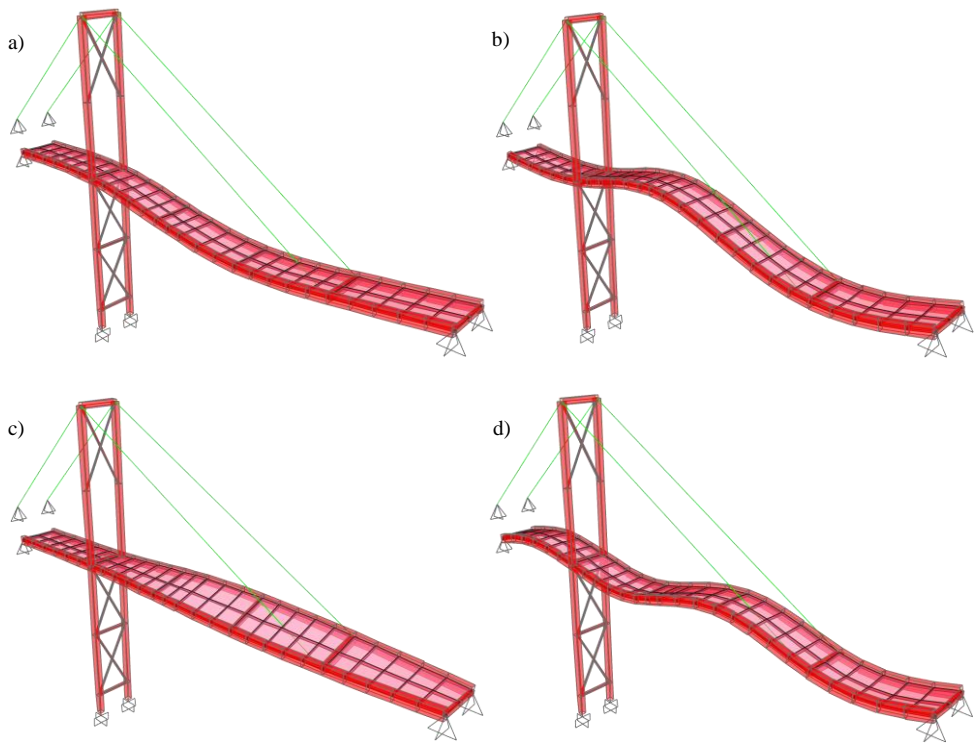

**Fig. 10,** Dynamical mode shapes of the FE model a) mode shape #1 b) mode shape #2 c) mode shape #3 d) mode shape #4

**Table 8.** MPV and identification uncertainty of modal parameters

| Modal Parameters | Mode #1 | | Mode #2 | | Mode #3 | | Mode #4 | |
|---|---|---|---|---|---|---|---|---|
| | Mean* | S.D.** | Mean* | S.D.** | Mean* | S.D.** | Mean* | S.D.** |
| $\varpi_i/(2\pi)$ (Hz) | 4.680 | 2.313 | 11.043 | 2.162 | 11.724 | 0.0001 | 20.053 | 0.0015 |
| $\psi_{1i}$*** | -0.5005 | 0.0169 | -0.4682 | 0.0351 | -0.5186 | 0.0112 | 0.1616 | 0.0262 |
| $\psi_{2i}$ | -0.4982 | 0.0192 | 0.5206 | 0.0315 | 0.4800 | 0.0423 | -0.1438 | 0.0418 |
| $\psi_{3i}$ | -0.4982 | 0.0284 | 0.5206 | 0.0261 | -0.4800 | 0.0159 | -0.1438 | 0.0497 |
| $\psi_{4i}$ | 0.0353 | 0.0140 | -0.0991 | 0.0250 | -0.0242 | 0.0171 | 0.6732 | 0.0232 |
| $\psi_{5i}$ | 0.0353 | 0.0098 | -0.0991 | 0.0177 | 0.0242 | 0.0198 | 0.6732 | 0.0261 |
| $\psi_{6i}$ | -0.5005 | 0.0263 | -0.4682 | 0.0417 | 0.5186 | 0.0308 | 0.1616 | 0.0165 |

* MPV averaged over all data sets
** Total uncertainty averaged over all data sets (S.D. stands for standard deviation)
*** The observed elements of the analytical mode shape

**Table 9.** Standard deviation of the modal parameters reflecting the mismatch between the modal parameters

| Analytical Modal Parameters | Mode #1 | Mode #2 | Mode #3 | Mode #4 |
|---|---|---|---|---|
| $\tau_{\omega_i^2}$ | 0.2299 | 0.0285 | 1.5e-13 | 3.3e-9 |
| $\sigma_{\phi_{1i}}$ | 0.0169 | 0.0350 | 0.0093 | 0.0260 |
| $\sigma_{\phi_{2i}}$ | 0.0192 | 0.0315 | 0.0418 | 0.0416 |
| $\sigma_{\phi_{3i}}$ | 0.0284 | 0.0261 | 0.0144 | 0.0496 |
| $\sigma_{\phi_{4i}}$ | 0.0140 | 0.0249 | 0.0156 | 0.0231 |
| $\sigma_{\phi_{5i}}$ | 0.0098 | 0.0176 | 0.0186 | 0.0260 |
| $\sigma_{\phi_{6i}}$ | 0.0263 | 0.0416 | 0.0302 | 0.0162 |

Based on Algorithm 1, the stiffness parameters are identified from each data set. Then, the identification uncertainty is calculated using the inverse of the Hessian matrix evaluated at the MPV. Similar to Fig. 7, the lower diagonal plots in Fig. 10 depict the variation in the MPV, and the identification uncertainties (±3S.D.) shown by the error bars. The diagonal plots show the marginal posterior distribution of the stiffness parameters obtained by combining the realizations from multiple data sets, and the upper triangular plots show the joint posterior distributions of each pair of parameters (similar to Fig. 7). As shown in Fig. 11, the correlation pattern and the variability of realizations obtained by the HBM framework agree with the variation of the MPV shown in the lower diagonal plots. Table 10 provides the ensemble mean, the ensemble variance, the identification uncertainties averaged over data sets, and the MPV of the stiffness hyper-parameters. The correlation coefficient of the stiffness parameters is identified and reported in Table 10, which agrees well with the results shown in Fig. 10.

The variability due to modeling errors turned out to be the dominant source of uncertainty, although the environmental parameters were constant. Moreover, the ensemble mean and covariance turn out to be good approximations for the MPV of the hyper-parameters. From a practical point of view, one could use these approximations rather than a thorough Bayesian analysis. However, such an approach would be unable to perform uncertainty propagation, response predictions, model class selection, and damage detection. Although these features are not demonstrated in this study, they constitute key merits of following such a Bayesian framework.

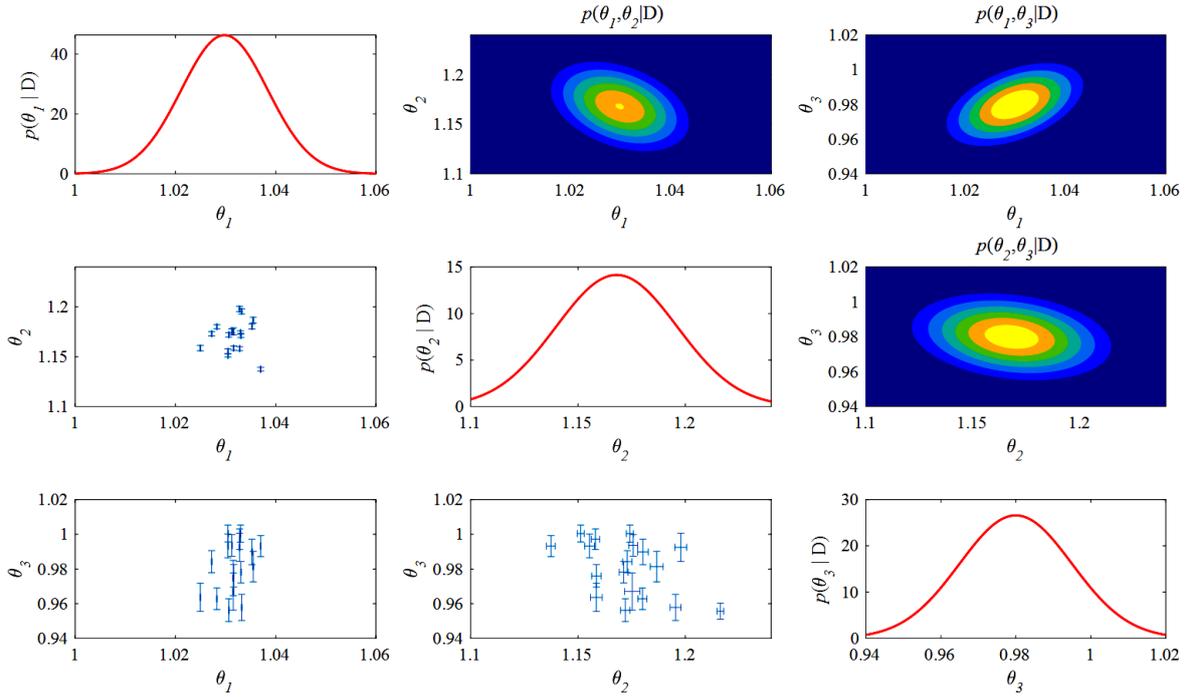

**Fig. 11,** Visualization of the identification of stiffness parameters (Plots on the main diagonal present marginal posterior distributions; The upper triangular plots show the joint posterior distribution of each pair of parameters; The lower triangular plots show the variation of each pair of the parameters over 19 data sets).

**Table 10.** Probabilistic identification of the stiffness parameters

| Stiffness Parameters | Ensemble Mean | Ensemble Variability (S.D.) | Identification Uncertainty (S.D.) | Hyper-parameters of Stiffness Parameters | | | | |
|---|---|---|---|---|---|---|---|---|
| | | | | $\hat{\mu}_{\theta_i}$ * | $\hat{\sigma}_{\theta_i}$ ** | $\hat{\rho}_{\theta_i\theta_j}$ *** | | |
| $\theta_1$ | 1.0298 | 0.0089 | 2.6e-5 | 1.0298 | 0.0086 | 1 | -0.3671 | 0.4685 |
| $\theta_2$ | 1.1680 | 0.0290 | 8.0e-4 | 1.1680 | 0.0282 | -0.3671 | 1 | -0.2234 |
| $\theta_3$ | 0.9799 | 0.0156 | 2.4e-3 | 0.9800 | 0.0150 | 0.4685 | -0.2234 | 1 |

* MPV of hyper mean vector
** MPV of the diagonal elements of the hyper covariance
*** MPV of the correlation coefficients (elements of correlation matrix)

The proposed framework is demonstrated using two experimental examples, wherein the number of structural parameters is relatively low. However, it has the potential to be applied to more involved problems with a large number of structural parameters. Although, in this case, the computational cost could be large, the proposed formulation and computational algorithms remain valid.

## 6. Summary and conclusions

A novel HBM framework is developed to update FE models of structures. This framework is coherent and consistent in using the Bayesian probability logic, tailored for updating the parameters of FE models based on modal information. In this framework, the misfit between the modal parameters is described using a zero-mean Gaussian discrepancy model, whose covariance matrix should be updated based on the data. This discrepancy model is required since a perfect matching between the analytical and experimental modal parameters is not possible in most practical cases. Additionally, the variability of structural parameters over data sets is considered using a parameterized probability distribution. This distribution is assumed to be Gaussian with unknown parameters, which should be calibrated by fusing multiple data sets. It is notable the proposed framework is general and flexible in terms of the structure chosen for the hyper covariance matrices.

Although the proposed Bayesian formulation is general, the computation method is heuristic, addressed via two EM algorithms associated with Laplace approximations. In this algorithm, the unit-norm constraint of the mode shape vectors is satisfied through an elegant analytical manner (see Appendix B), and the modal features are weighed optimally based on their aggregate uncertainty. A broad scope of uncertainty quantification is exercised by considering the identification uncertainties and the variability of parameters over data sets.

When this framework is used in conjunction with multiple data sets corresponding to different environmental and operational conditions, it delivers additional robustness by considering the variability of estimations. When this variability is large, it is necessary to reduce it by considering the correlation with structure temperature and loading intensity, similar to [65]. Nevertheless, when the data sets correspond to fixed operational and environmental conditions, the variability could be attributed to modeling errors, e.g., systematic nonlinearities and model inadequacies. The latter is the case in the two experimental examples demonstrated herein. Although in these examples the operational and environmental variabilities are kept almost similar for all data sets, the variability of realizations is the dominant source of uncertainty and much larger than identification uncertainties.

**Acknowledgements**

Financial support from the Hong Kong Research Grants Council under grants 16212918 and 16211019 is gratefully acknowledged.

**Credit Author Statement**

**Omid Sedehi:** Conceptualization, Methodology, Formal Analysis, Investigation, Validation, Visualization, Data curation, Software, Writing - Original Draft, Writing - Review & Editing.

**Costas Papadimitriou:** Conceptualization, Methodology, Writing - Review & Editing.

**Lambros S. Katafygiotis:** Conceptualization, Funding Acquisition, Project Administration, Writing - Review & Editing.

**Appendix (A). Some properties of multivariate Gaussian distributions**

**Theorem 1.** Let $\mathbf{X}$ be a random vector described by a multivariate Gaussian distribution $N(\mathbf{X}|\boldsymbol{\mu},\boldsymbol{\Sigma})$, having the mean $\boldsymbol{\mu}$ and the covariance matrix $\boldsymbol{\Sigma}$. The mean vector $\boldsymbol{\mu}$ is described by $N(\boldsymbol{\mu}|\boldsymbol{\mu}_0,\boldsymbol{\Sigma}_0)$; the covariance matrices $\boldsymbol{\Sigma}$ and $\boldsymbol{\Sigma}_0$, as well as the vector $\boldsymbol{\mu}_0$ are all known. In this case, the following equations hold true:

$$p(\mathbf{X}) = \int_{\boldsymbol{\mu}} N(\mathbf{X}|\boldsymbol{\mu},\boldsymbol{\Sigma})N(\boldsymbol{\mu}|\boldsymbol{\mu}_0,\boldsymbol{\Sigma}_0)d\boldsymbol{\mu} = N(\mathbf{X}|\boldsymbol{\mu}_0,\boldsymbol{\Sigma}+\boldsymbol{\Sigma}_0) \tag{A1}$$

$$p(\boldsymbol{\mu}|\mathbf{X}) = N(\boldsymbol{\mu}|(\boldsymbol{\Sigma}^{-1}+\boldsymbol{\Sigma}_0^{-1})^{-1}(\boldsymbol{\Sigma}^{-1}\mathbf{X}+\boldsymbol{\Sigma}_0^{-1}\boldsymbol{\mu}_0),(\boldsymbol{\Sigma}^{-1}+\boldsymbol{\Sigma}_0^{-1})^{-1}) \tag{A2}$$

A proof of this result can be found in [56].

**Appendix (B). Derivation of the ECM algorithm proposed in Eq. (22)**

*B.1. Derivation of the E-step*

In this section, the proposed ECM algorithm is investigated mathematically to derive explicit formulations. For ease of computation, the cross-correlation between the parameters of different modes in the matrices $\hat{\Sigma}_{\omega_s^2}$ and $\hat{\Sigma}_{\Phi_s}$ is ignored. We also introduce a beneficial normalization of the modal frequencies' variance, considering $\sigma_{\omega_i^2}^2 = \hat{\omega}_{i,s}^4 \tau_{\omega_i^2}^2$ such that $\tau_{\omega_i^2}^2$ denotes the dimensionless variance of the modal frequencies. By considering this assumption and rewriting Eq. (21) in terms of individual modal parameters, one finds:

$$p(\mathbf{D}|\beta, \Sigma_{\omega^2}, \Sigma_{\Phi}, \mathbb{M}_p)$$
$$= \prod_{s=1}^{N_D} \prod_{i=1}^{N_m} N(\hat{\omega}_{i,s}^2 | \omega_{i,s}^2, \hat{\sigma}_{\omega_{i,s}^2}^2) N(\omega_{i,s}^2 | \varpi_{i,s}^2, \hat{\omega}_{i,s}^4 \tau_{\omega_i^2}^2) N(\hat{\boldsymbol{\phi}}_{i,s} | \|\boldsymbol{\phi}_{i,s}\|^{-1} \boldsymbol{\phi}_{i,s}, \hat{\Sigma}_{\boldsymbol{\phi}_{i,s}}) N(\|\boldsymbol{\phi}_{i,s}\|^{-1} \boldsymbol{\phi}_{i,s} | \chi_{i,s} \gamma_{i,s} \psi_{i,s}, \Sigma_{\boldsymbol{\phi}}) \quad (B1)$$

where $\hat{\omega}_{i,s}^2$ and $\hat{\boldsymbol{\phi}}_{i,s}$ are the *i*th modal frequency and mode shape vector extracted from their augmented counterparts $\hat{\boldsymbol{\omega}}_s^2$ and $\hat{\boldsymbol{\Phi}}_s$, respectively. This simplification reads the objective function in Eq. (23) as

$$L(\{\boldsymbol{\varpi}^2(\boldsymbol{\theta}_s), \boldsymbol{\Psi}(\boldsymbol{\theta}_s), \boldsymbol{\theta}_s\}_{s=1}^{N_D}, \Sigma_{\omega^2}, \Sigma_{\Phi}) = -\frac{N_D}{2} \sum_{i=1}^{N_m} \left[ \ln(\hat{\omega}_{i,s}^{-4} \tau_{\omega_i^2}^2) + \ln|\Sigma_{\boldsymbol{\phi}}| \right]$$
$$-\frac{1}{2} \sum_{s=1}^{N_D} \sum_{i=1}^{N_m} \left[ \hat{\sigma}_{\omega_{i,s}^2}^{-2} \left[ \mathbb{E}[\omega_{i,s}^4] - 2\hat{\omega}_{i,s}^2 \mathbb{E}[\omega_{i,s}^2] + \hat{\omega}_{i,s}^4 \right] + \hat{\omega}_{i,s}^{-4} \tau_{\omega_i^2}^{-2} \left[ \mathbb{E}[\omega_{i,s}^4] - 2\varpi_{i,s}^2 \mathbb{E}[\omega_{i,s}^2] + \varpi_{i,s}^4 \right] \right]$$
$$-\frac{1}{2} \sum_{s=1}^{N_D} \sum_{i=1}^{N_m} tr\left[ \hat{\Sigma}_{\boldsymbol{\phi}_{i,s}}^{-1} \left[ \mathbb{E}[\|\boldsymbol{\phi}_{i,s}\|^{-2} \boldsymbol{\phi}_{i,s} \boldsymbol{\phi}_{i,s}^T] - \mathbb{E}[\|\boldsymbol{\phi}_{i,s}\|^{-1} \boldsymbol{\phi}_{i,s}] \hat{\boldsymbol{\phi}}_{i,s}^T - \hat{\boldsymbol{\phi}}_{i,s} \mathbb{E}[\|\boldsymbol{\phi}_{i,s}\|^{-1} \boldsymbol{\phi}_{i,s}^T] + \hat{\boldsymbol{\phi}}_{i,s} \hat{\boldsymbol{\phi}}_{i,s}^T \right] \right]$$
$$-\frac{1}{2} \sum_{s=1}^{N_D} \sum_{i=1}^{N_m} tr\left[ \Sigma_{\boldsymbol{\phi}}^{-1} \left[ \mathbb{E}[\|\boldsymbol{\phi}_{i,s}\|^{-2} \boldsymbol{\phi}_{i,s} \boldsymbol{\phi}_{i,s}^T] - \chi_{i,s} \mathbb{E}[\|\boldsymbol{\phi}_{i,s}\|^{-1} \boldsymbol{\phi}_{i,s}] \psi_{i,s}^T \gamma_{i,s}^T - \chi_{i,s} \gamma_{i,s} \psi_{i,s} \mathbb{E}[\|\boldsymbol{\phi}_{i,s}\|^{-1} \boldsymbol{\phi}_{i,s}^T] + \chi_{i,s}^2 \gamma_{i,s} \psi_{i,s} \psi_{i,s}^T \gamma_{i,s}^T \right] \right] + cte.$$
(B2)

Calculating this objective function would require the second-moment information of the experimental modal parameters $(\boldsymbol{\omega}_s^2, \boldsymbol{\Phi}_s)$. This task constitutes the E-Step of the algorithm, which can be performed independently for the parameters of each data set. Specifically, the independence of the modal frequencies and mode shapes of each specific dynamical mode implies:

$$p(\omega_{i,s}^2 | \varpi_{i,s}^2, \tau_{\omega_i^2}^2, \mathbf{D}) \propto N(\hat{\omega}_{i,s}^2 | \omega_{i,s}^2, \hat{\sigma}_{\omega_{i,s}^2}^2) N(\omega_{i,s}^2 | \varpi_{i,s}^2, \hat{\omega}_{i,s}^4 \tau_{\omega_i^2}^2) \quad (B3)$$

$$p\left(\boldsymbol{\phi}_{i,s} \mid \boldsymbol{\psi}_{i,s}, \boldsymbol{\Sigma}_{\boldsymbol{\phi}}, \mathbf{D}\right) \propto N\left(\hat{\boldsymbol{\phi}}_{i,s} \mid \|\boldsymbol{\phi}_{i,s}\|^{-1} \boldsymbol{\phi}_{i,s}, \hat{\boldsymbol{\Sigma}}_{\boldsymbol{\phi}_{i,s}}\right) N\left(\|\boldsymbol{\phi}_{i,s}\|^{-1} \boldsymbol{\phi}_{i,s} \mid \chi_{i,s} \gamma_{i,s} \boldsymbol{\psi}_{i,s}, \boldsymbol{\Sigma}_{\boldsymbol{\phi}}\right) \tag{B4}$$

where $p(\omega_{i,s}^2 \mid \varpi_{i,s}^2, \tau_{\omega_i^2}^2, \mathbf{D})$ and $p(\boldsymbol{\phi}_{i,s} \mid \boldsymbol{\psi}_{i,s}, \boldsymbol{\Sigma}_{\boldsymbol{\phi}}, \mathbf{D})$ are the marginal distribution of experimental modal frequencies and mode shape vectors, respectively. In these equations, the dependence on irrelevant parameters is dropped for brevity. By virtue of Theorem 1 of Appendix (A), the marginal posterior distribution of modal frequencies can be simplified into

$$p(\omega_{i,s}^2 \mid \varpi_{i,s}^2, \tau_{\omega_i^2}^2, \mathbf{D}) = N(\omega_{i,s}^2 \mid (\hat{\sigma}_{\omega_{i,s}^2}^{-2} + \hat{\omega}_{i,s}^{-4} \tau_{\omega_i^2}^{-2})^{-1}(\hat{\sigma}_{\omega_{i,s}^2}^{-2} \hat{\omega}_{i,s}^2 + \hat{\omega}_{i,s}^{-4} \tau_{\omega_i^2}^{-2} \varpi_{i,s}^2), (\hat{\omega}_{i,s}^{-4} \tau_{\omega_i^2}^{-2} + \hat{\sigma}_{\omega_{i,s}^2}^{-2})^{-1}) \tag{B5}$$

Thus, the second-moment statistics of individual modal frequencies can be calculated as

$$\mathbb{E}[\omega_{i,s}^2] = (\hat{\sigma}_{\omega_{i,s}^2}^{-2} + \hat{\omega}_{i,s}^{-4} \tau_{\omega_i^2}^{-2})^{-1}(\hat{\sigma}_{\omega_{i,s}^2}^{-2} \hat{\omega}_{i,s}^2 + \hat{\omega}_{i,s}^{-4} \tau_{\omega_i^2}^{-2} \varpi_{i,s}^2) \tag{B6}$$

$$\mathbb{E}[\omega_{i,s}^4] = (\hat{\sigma}_{\omega_{i,s}^2}^{-2} + \hat{\omega}_{i,s}^{-4} \tau_{\omega_i^2}^{-2})^{-1} + (\mathbb{E}[\omega_{i,s}^2])^2 \tag{B7}$$

Unlike the modal frequencies, the statistical moments of the mode shape vectors are non-trivial to be calculated due to the scaling by the norm of $\boldsymbol{\phi}_{i,s}$. To circumvent this problem, a Laplace approximation is proposed, simplifying $p(\boldsymbol{\phi}_{i,s} \mid \boldsymbol{\psi}_{i,s}, \boldsymbol{\Sigma}_{\boldsymbol{\phi}}, \mathbf{D})$ by a Gaussian distribution. For this purpose, the following function should be minimized:

$$\begin{aligned}L(\boldsymbol{\phi}_{i,s}) = &\frac{1}{2}\left[\left(\hat{\boldsymbol{\phi}}_{i,s} - \|\boldsymbol{\phi}_{i,s}\|^{-1} \boldsymbol{\phi}_{i,s}\right)^T \hat{\boldsymbol{\Sigma}}_{\boldsymbol{\phi}_{i,s}}^{-1} \left(\hat{\boldsymbol{\phi}}_{i,s} - \|\boldsymbol{\phi}_{i,s}\|^{-1} \boldsymbol{\phi}_{i,s}\right)\right] \\ &+ \frac{1}{2}\left[\left(\|\boldsymbol{\phi}_{i,s}\|^{-1} \boldsymbol{\phi}_{i,s} - \chi_{i,s} \gamma_{i,s} \boldsymbol{\psi}_{i,s}\right)^T \boldsymbol{\Sigma}_{\boldsymbol{\phi}}^{-1} \left(\|\boldsymbol{\phi}_{i,s}\|^{-1} \boldsymbol{\phi}_{i,s} - \chi_{i,s} \gamma_{i,s} \boldsymbol{\psi}_{i,s}\right)\right] + cte.\end{aligned} \tag{B8}$$

where $L(\boldsymbol{\phi}_{i,s}) = -\ln p(\boldsymbol{\phi}_{i,s} \mid \boldsymbol{\psi}_{i,s}, \boldsymbol{\Sigma}_{\boldsymbol{\phi}}, \mathbf{D})$ is determined from Eq. (B4). Since the minimization of this objective function for $\boldsymbol{\phi}_{i,s}$ could be difficult, a Lagrange multiplier approach is adopted herein, which requires the minimization of the unconstrained objective function:

$$\begin{aligned}L(\boldsymbol{\phi}_{i,s}, \delta_{i,s}) = &\frac{1}{2}\left[\left(\hat{\boldsymbol{\phi}}_{i,s} - \boldsymbol{\phi}_{i,s}\right)^T \hat{\boldsymbol{\Sigma}}_{\boldsymbol{\phi}_{i,s}}^{-1} \left(\hat{\boldsymbol{\phi}}_{i,s} - \boldsymbol{\phi}_{i,s}\right)\right] \\ &+ \frac{1}{2}\left[\left(\boldsymbol{\phi}_{i,s} - \chi_{i,s} \gamma_{i,s} \boldsymbol{\psi}_{i,s}\right)^T \boldsymbol{\Sigma}_{\boldsymbol{\phi}}^{-1} \left(\boldsymbol{\phi}_{i,s} - \chi_{i,s} \gamma_{i,s} \boldsymbol{\psi}_{i,s}\right) + \delta_{i,s}\left(1 - \boldsymbol{\phi}_{i,s}^T \boldsymbol{\phi}_{i,s}\right)\right]\end{aligned} \tag{B9}$$

where $\delta_{i,s}$ is a multiplier included for satisfying the unit norm constraint $\boldsymbol{\phi}_{i,s}^T \boldsymbol{\phi}_{i,s} = 1$. The first derivatives of $L(\boldsymbol{\phi}_{i,s}, \delta_{i,s})$ with respect to $\boldsymbol{\phi}_{i,s}$ and $\delta_{i,s}$ are given as

$$\frac{\partial L(\boldsymbol{\phi}_{i,s},\delta_{i,s})}{\partial \boldsymbol{\phi}_{i,s}} = \hat{\boldsymbol{\Sigma}}_{\boldsymbol{\phi}_{i,s}}^{-1}(\boldsymbol{\phi}_{i,s} - \hat{\boldsymbol{\phi}}_{i,s}) + \boldsymbol{\Sigma}_{\boldsymbol{\phi}}^{-1}(\boldsymbol{\phi}_{i,s} - \chi_{i,s}\boldsymbol{\gamma}_{i,s}\boldsymbol{\psi}_{i,s}) - \delta_{i,s}\boldsymbol{\phi}_{i,s} \tag{B10}$$

$$\frac{\partial L(\boldsymbol{\phi}_{i,s},\delta_{i,s})}{\partial \delta_{i,s}} = 1 - \boldsymbol{\phi}_{i,s}^T \boldsymbol{\phi}_{i,s} \tag{B11}$$

These equations should be set to zero and solved simultaneously, leading to the following system of equations:

$$\begin{cases} (\hat{\boldsymbol{\Sigma}}_{\boldsymbol{\phi}_{i,s}}^{-1} + \boldsymbol{\Sigma}_{\boldsymbol{\phi}}^{-1})\boldsymbol{\phi}_{i,s} + \mathbf{b}_{i,s} = \delta_{i,s}\boldsymbol{\phi}_{i,s} \\ \boldsymbol{\phi}_{i,s}^T \boldsymbol{\phi}_{i,s} = 1 \end{cases} \tag{B12}$$

where $\mathbf{b}_{i,s} = -\hat{\boldsymbol{\Sigma}}_{\boldsymbol{\phi}_{i,s}}^{-1}\hat{\boldsymbol{\phi}}_{i,s} - \chi_{i,s}\boldsymbol{\Sigma}_{\boldsymbol{\phi}}^{-1}\boldsymbol{\gamma}_{i,s}\boldsymbol{\psi}_{i,s}$ is a known vector having the same dimension as $\boldsymbol{\phi}_{i,s}$, whereas $\boldsymbol{\phi}_{i,s}$ and $\delta_{i,s}$ are the parameters to be determined. This set of equations implies that:

$$\boldsymbol{\phi}_{i,s} = -(\hat{\boldsymbol{\Sigma}}_{\boldsymbol{\phi}_{i,s}}^{-1} + \boldsymbol{\Sigma}_{\boldsymbol{\phi}}^{-1} - \delta_{i,s}\mathbf{I}_{N_o})^{-1}\mathbf{b}_{i,s} \tag{B13}$$

Substituting this result into the unit-norm equation readily provides:

$$\mathbf{b}_{i,s}^T \mathbf{z}_{i,s} = 1 \tag{B14}$$

where $\mathbf{z}_{i,s} = (\hat{\boldsymbol{\Sigma}}_{\boldsymbol{\phi}_{i,s}}^{-1} + \boldsymbol{\Sigma}_{\boldsymbol{\phi}}^{-1} - \delta_{i,s}\mathbf{I}_{N_o})^{-2}\mathbf{b}_{i,s}$ is an unknown vector, depending only on $\delta_{i,s}$. When this vector is rewritten in terms of $\boldsymbol{\phi}_{i,s}$, the following equation is obtained:

$$(\hat{\boldsymbol{\Sigma}}_{\boldsymbol{\phi}_{i,s}}^{-1} + \boldsymbol{\Sigma}_{\boldsymbol{\phi}}^{-1})\mathbf{z}_{i,s} + \boldsymbol{\phi}_{i,s} = \delta_{i,s}\mathbf{z}_{i,s} \tag{B15}$$

Pre-multiplying $\mathbf{b}_{i,s}$ to both sides of Eq. (B14) yields:

$$\mathbf{b}_{i,s}\mathbf{b}_{i,s}^T\mathbf{z}_{i,s} = \mathbf{b}_{i,s} \tag{B16}$$

Replacing this equation into Eq. (B12) gives:

$$(\hat{\boldsymbol{\Sigma}}_{\boldsymbol{\phi}_{i,s}}^{-1} + \boldsymbol{\Sigma}_{\boldsymbol{\phi}}^{-1})\boldsymbol{\phi}_{i,s} + \mathbf{b}_{i,s}\mathbf{b}_{i,s}^T\mathbf{z}_{i,s} = \delta_{i,s}\boldsymbol{\phi}_{i,s} \tag{B17}$$

Eqs. (B15) and (B17) should be solved simultaneously in the following matrix form [20]:

$$\begin{bmatrix} \hat{\boldsymbol{\Sigma}}_{\boldsymbol{\phi}_{i,s}}^{-1} + \boldsymbol{\Sigma}_{\boldsymbol{\phi}}^{-1} & \mathbf{b}_{i,s}\mathbf{b}_{i,s}^T \\ \mathbf{I}_{N_o} & \hat{\boldsymbol{\Sigma}}_{\boldsymbol{\phi}_{i,s}}^{-1} + \boldsymbol{\Sigma}_{\boldsymbol{\phi}}^{-1} \end{bmatrix} \begin{bmatrix} \boldsymbol{\phi}_{i,s} \\ \mathbf{z}_{i,s} \end{bmatrix} = \delta_{i,s} \begin{bmatrix} \boldsymbol{\phi}_{i,s} \\ \mathbf{z}_{i,s} \end{bmatrix} \tag{B18}$$

This eigenvalue problem yields the optimal values of the unknown parameters, denoted by $\tilde{\delta}_{i,s}$ and $\tilde{\boldsymbol{\phi}}_{i,s}$. To find out which Eigen-pair we should pick, it is necessary to replace these optimal values into the loss function in Eq. (B9). For this purpose, by expanding the multiplications, Eq. (B9) can be simplified into

$$L(\boldsymbol{\phi}_{i,s}, \delta_{i,s}) = \frac{1}{2}\left[ \hat{\boldsymbol{\phi}}_{i,s}^T \hat{\boldsymbol{\Sigma}}_{\boldsymbol{\phi}_{i,s}}^{-1} \hat{\boldsymbol{\phi}}_{i,s} + \chi_{i,s}^2 \boldsymbol{\psi}_{i,s}^T \boldsymbol{\gamma}_{i,s}^T \boldsymbol{\Sigma}_{\boldsymbol{\phi}}^{-1} \boldsymbol{\gamma}_{i,s} \boldsymbol{\psi}_{i,s} + \boldsymbol{\phi}_{i,s}^T (\boldsymbol{\Sigma}_{\boldsymbol{\phi}}^{-1} + \hat{\boldsymbol{\Sigma}}_{\boldsymbol{\phi}_{i,s}}^{-1}) \boldsymbol{\phi}_{i,s} + 2\boldsymbol{\phi}_{i,s}^T \mathbf{b}_{i,s} + \delta_{i,s}(1 - \boldsymbol{\phi}_{i,s}^T \boldsymbol{\phi}_{i,s}) \right] \quad \text{(B19)}$$

Combining this equation with Eq. (B12) leads to

$$L(\tilde{\boldsymbol{\phi}}_{i,s}, \tilde{\delta}_{i,s}) = \frac{1}{2} \hat{\boldsymbol{\phi}}_{i,s}^T \hat{\boldsymbol{\Sigma}}_{\boldsymbol{\phi}_{i,s}}^{-1} \hat{\boldsymbol{\phi}}_{i,s} + \frac{1}{2} \chi_{i,s}^2 \boldsymbol{\psi}_{i,s}^T \boldsymbol{\gamma}_{i,s}^T \boldsymbol{\Sigma}_{\boldsymbol{\phi}}^{-1} \boldsymbol{\gamma}_{i,s} \boldsymbol{\psi}_{i,s} + \frac{1}{2} \boldsymbol{\phi}_{i,s}^T \mathbf{b}_{i,s} + \frac{1}{2} \tilde{\delta}_{i,s} \quad \text{(B20)}$$

where $L(\tilde{\boldsymbol{\phi}}_{i,s}, \tilde{\delta}_{i,s})$ is the objective function evaluated at the optimal values. According to this equation, the smallest eigenvalue of Eq. (B18) gives the optimal value $\tilde{\delta}_{i,s}$, and the first half of the corresponding eigenvector contains $\tilde{\boldsymbol{\phi}}_{i,s}$.

To estimate the uncertainty, the inverse of the Hessian matrix of the constrained objective function should be evaluated at the optimal values. To do so, the constrained objective function is given as

$$L(\boldsymbol{\phi}_{i,s}) = \frac{1}{2}\left[ (\hat{\boldsymbol{\phi}}_{i,s} - \boldsymbol{\phi}_{i,s})^T \hat{\boldsymbol{\Sigma}}_{\boldsymbol{\phi}_{i,s}}^{-1} (\hat{\boldsymbol{\phi}}_{i,s} - \boldsymbol{\phi}_{i,s}) + (\boldsymbol{\phi}_{i,s} - \chi_{i,s} \boldsymbol{\gamma}_{i,s} \boldsymbol{\psi}_{i,s})^T \boldsymbol{\Sigma}_{\boldsymbol{\phi}}^{-1} (\boldsymbol{\phi}_{i,s} - \chi_{i,s} \boldsymbol{\gamma}_{i,s} \boldsymbol{\psi}_{i,s}) \right] \quad \text{(B21)}$$

This function should be minimized subjected to $G_c(\boldsymbol{\phi}_{i,s}) = 0$, where $G_c(\boldsymbol{\phi}_{i,s}) = 1 - \boldsymbol{\phi}_{i,s}^T \boldsymbol{\phi}_{i,s}$ ensures that $\boldsymbol{\phi}_{i,s}$ has unit norm. Thus, the unconstrained objective function is rewritten as

$$L(\boldsymbol{\phi}_{i,s}, \delta_{i,s}) = L(\boldsymbol{\phi}_{i,s}) + \frac{1}{2} \delta_{i,s} G_c(\boldsymbol{\phi}_{i,s}) \quad \text{(B22)}$$

Let $\mathbf{v}_c(\boldsymbol{\phi}_{i,s}) = \boldsymbol{\phi}_{i,s} / \|\boldsymbol{\phi}_{i,s}\|$ be the unit-norm mode shape vector. The calculation of the Hessian matrix of $L(\mathbf{v}_c(\boldsymbol{\phi}_{i,s}), \delta_{i,s})$ at $\tilde{\boldsymbol{\phi}}_{i,s}$ and $\tilde{\delta}_{i,s}$ is desired, while satisfying the unit-norm constraint. However, the direct derivation of the Hessian matrix of $L(\mathbf{v}_c(\boldsymbol{\phi}_{i,s}), \delta_{i,s})$ entails considerable mathematical effort. As proven in [20,76], the chain rule is a convenient substitute approach to calculate the Hessian matrix of $L(\mathbf{v}_c(\boldsymbol{\phi}_{i,s}), \delta_{i,s})$ in terms of the derivatives of $L(\boldsymbol{\phi}_{i,s}, \delta_{i,s})$ and $\mathbf{v}_c(\boldsymbol{\phi}_{i,s})$, giving:

$$\frac{\partial^2 L(\mathbf{v}_c(\boldsymbol{\phi}_{i,s}),\delta_{i,s})}{\partial \boldsymbol{\phi}_{i,s} \partial \boldsymbol{\phi}_{i,s}^T} = \frac{\partial \mathbf{v}_c}{\partial \boldsymbol{\phi}_{i,s}^T} \frac{\partial^2 L(\boldsymbol{\phi}_{i,s})}{\partial \boldsymbol{\phi}_{i,s} \partial \boldsymbol{\phi}_{i,s}^T} \frac{\partial \mathbf{v}_c}{\partial \boldsymbol{\phi}_{i,s}} + \left( \mathbf{I}_{N_o} \otimes \frac{\partial L(\boldsymbol{\phi}_{i,s})}{\partial \boldsymbol{\phi}_{i,s}} \right) \frac{\partial^2 \mathbf{v}_c}{\partial \boldsymbol{\phi}_{i,s} \partial \boldsymbol{\phi}_{i,s}^T} \tag{B23}$$

Evaluation of this matrix at the optimal values of $\boldsymbol{\phi}_{i,s}$ and $\delta_{i,s}$ provides:

$$\left. \frac{\partial^2 L(\mathbf{v}_c(\hat{\boldsymbol{\phi}}_{i,s}),\hat{\delta}_{i,s})}{\partial \boldsymbol{\phi}_{i,s} \partial \boldsymbol{\phi}_{i,s}^T} \right|_{\substack{\boldsymbol{\phi}_{i,s}=\tilde{\boldsymbol{\phi}}_{i,s} \\ \delta_{i,s}=\tilde{\delta}_{i,s}}} = \left[ \frac{\partial \mathbf{v}_c^T}{\partial \boldsymbol{\phi}_{i,s}} \frac{\partial^2 L(\boldsymbol{\phi}_{i,s},\delta_{i,s})}{\partial \boldsymbol{\phi}_{i,s} \partial \boldsymbol{\phi}_{i,s}^T} \frac{\partial \mathbf{v}_c}{\partial \boldsymbol{\phi}_{i,s}} \right]_{\substack{\boldsymbol{\phi}_{i,s}=\tilde{\boldsymbol{\phi}}_{i,s} \\ \delta_{i,s}=\tilde{\delta}_{i,s}}} \tag{B24}$$

where the derivatives are given as

$$\frac{\partial^2 L(\boldsymbol{\phi}_{i,s},\delta_{i,s})}{\partial \boldsymbol{\phi}_{i,s} \partial \boldsymbol{\phi}_{i,s}^T} = \hat{\boldsymbol{\Sigma}}_{\boldsymbol{\phi}_{i,s}}^{-1} + \boldsymbol{\Sigma}_{\boldsymbol{\phi}}^{-1} - \delta_{i,s} \mathbf{I}_{N_o} \tag{B25}$$

$$\frac{\partial \mathbf{v}_c}{\partial \boldsymbol{\phi}_{i,s}} = \|\boldsymbol{\phi}_{i,s}\|^{-1} \mathbf{I}_{N_o} - \|\boldsymbol{\phi}_{i,s}\|^{-3} \boldsymbol{\phi}_{i,s} \boldsymbol{\phi}_{i,s}^T \tag{B26}$$

Having obtained the Hessian matrix, the covariance matrix of $\boldsymbol{\phi}_{i,s}$ can be computed using the Laplace asymptotic approximation, which gives [20,76]:

$$\mathbf{H}^{-1}(\boldsymbol{\phi}_{i,s}) = \left. \frac{\partial \mathbf{v}_c}{\partial \boldsymbol{\phi}_{i,s}} \left[ \frac{\partial \mathbf{v}_c^T}{\partial \boldsymbol{\phi}_{i,s}} \frac{\partial^2 L(\boldsymbol{\phi}_{i,s},\delta_{i,s})}{\partial \boldsymbol{\phi}_{i,s} \partial \boldsymbol{\phi}_{i,s}^T} \frac{\partial \mathbf{v}_c}{\partial \boldsymbol{\phi}_{i,s}} \right]^+ \frac{\partial \mathbf{v}_c^T}{\partial \boldsymbol{\phi}_{i,s}} \right|_{\substack{\boldsymbol{\phi}_{i,s}=\tilde{\boldsymbol{\phi}}_{i,s} \\ \delta_{i,s}=\tilde{\delta}_{i,s}}} \tag{B27}$$

where $\mathbf{H}(\boldsymbol{\phi}_{i,s})$ is the Hessian matrix evaluated at $\tilde{\boldsymbol{\phi}}_{i,s}$ and $\tilde{\delta}_{i,s}$. Replacing the derivatives from Eqs. (B25-B26) into the latest equation yields the second moment of $\boldsymbol{\phi}_{i,s}$ as follows:

$$\mathbf{H}^{-1}(\boldsymbol{\phi}_{i,s}) = (\mathbf{I}_{N_o} - \tilde{\boldsymbol{\phi}}_{i,s}\tilde{\boldsymbol{\phi}}_{i,s}^T)\left((\mathbf{I}_{N_o} - \tilde{\boldsymbol{\phi}}_{i,s}\tilde{\boldsymbol{\phi}}_{i,s}^T)^T(\hat{\boldsymbol{\Sigma}}_{\boldsymbol{\phi}_{i,s}}^{-1} + \boldsymbol{\Sigma}_{\boldsymbol{\phi}}^{-1} - \hat{\delta}_{i,s}\mathbf{I}_{N_o})(\mathbf{I}_{N_o} - \tilde{\boldsymbol{\phi}}_{i,s}\tilde{\boldsymbol{\phi}}_{i,s}^T)\right)^+ (\mathbf{I}_{N_o} - \tilde{\boldsymbol{\phi}}_{i,s}\tilde{\boldsymbol{\phi}}_{i,s}^T)^T \tag{B28}$$

This equation gives the second moment of the mode shape vectors as

$$\mathbb{E}\left[\boldsymbol{\phi}_{i,s}\boldsymbol{\phi}_{i,s}^T\right] = \mathbf{H}^{-1}(\boldsymbol{\phi}_{i,s})\Big|_{\boldsymbol{\phi}_{i,s}=\tilde{\boldsymbol{\phi}}_{i,s}} + \tilde{\boldsymbol{\phi}}_{i,s}\tilde{\boldsymbol{\phi}}_{i,s}^T \tag{B29}$$

*B.2. Derivation of the M-Step 1*

In the M-Step, the statistical moments obtained earlier should be substituted into Eq. (B2). Since the relationship between $\boldsymbol{\theta}_s$ and $\{\boldsymbol{\varpi}^2(\boldsymbol{\theta}_s), \boldsymbol{\Psi}(\boldsymbol{\theta}_s)\}$ is deterministic (referring to Eq. (2)), the maximization of $L(\{\boldsymbol{\varpi}^2(\boldsymbol{\theta}_s), \boldsymbol{\Psi}(\boldsymbol{\theta}_s), \boldsymbol{\theta}_s\}_{s=1}^{N_D}, \boldsymbol{\Sigma}_{\boldsymbol{\omega}^2}, \boldsymbol{\Sigma}_{\boldsymbol{\Phi}})$ for $\boldsymbol{\theta}_s$'s would be sufficient to infer the analytical modal parameters $\{\varpi^2(\boldsymbol{\theta}_s), \boldsymbol{\Psi}(\boldsymbol{\theta}_s)\}$. Additionally, given $\boldsymbol{\Sigma}_{\boldsymbol{\omega}^2}$ and $\boldsymbol{\Sigma}_{\boldsymbol{\Phi}}$, the maximization of the objective function in Eq. (B2) can be performed independently for each $\boldsymbol{\theta}_s$, based on the following objective function:

$$L(\boldsymbol{\theta}_s) = -\frac{N_D}{2}\sum_{i=1}^{N_m}\left[\ln(\hat{\omega}_{i,s}^{-4}\tau_{\omega_i^2}^2) + \ln|\boldsymbol{\Sigma}_{\boldsymbol{\phi}_i}|\right] - \frac{1}{2}\sum_{i=1}^{N_m}\left[\hat{\omega}_{i,s}^{-4}\tau_{\omega_i^2}^{-2}\left[\mathbb{E}\left[\omega_{i,s}^4\right] - 2\varpi_{i,s}^2\mathbb{E}\left[\omega_{i,s}^2\right] + \varpi_{i,s}^4\right]\right]$$

$$-\frac{1}{2}\sum_{i=1}^{N_m}tr\left[\boldsymbol{\Sigma}_{\boldsymbol{\phi}_i}^{-1}\left[\mathbb{E}\left[\|\boldsymbol{\phi}_{i,s}\|^{-2}\boldsymbol{\phi}_{i,s}\boldsymbol{\phi}_{i,s}^T\right] - \chi_{i,s}\mathbb{E}\left[\|\boldsymbol{\phi}_{i,s}\|^{-1}\boldsymbol{\phi}_{i,s}\right]\boldsymbol{\psi}_{i,s}^T\boldsymbol{\gamma}_{i,s}^T - \chi_{i,s}\boldsymbol{\gamma}_{i,s}\boldsymbol{\psi}_{i,s}\mathbb{E}\left[\|\boldsymbol{\phi}_{i,s}\|^{-1}\boldsymbol{\phi}_{i,s}^T\right] + \chi_{i,s}^2\boldsymbol{\gamma}_{i,s}\boldsymbol{\psi}_{i,s}\boldsymbol{\psi}_{i,s}^T\boldsymbol{\gamma}_{i,s}^T\right]\right] + cte.$$
(B30)

where $L(\boldsymbol{\theta}_s)$ embeds only the relevant part of Eq. (B2) that governs the identification of $\boldsymbol{\theta}_s$. In this equation, the expectations should be replaced from Eqs. (B6-B7,B18,B29). In Appendix (D), the analytical derivatives of this objective function are provided, aiming to help obtain reliable estimations.

*B.3. Derivation of the M-Step 2*

Having estimated $\boldsymbol{\theta}_s$'s, it is straightforward to estimate $\boldsymbol{\Sigma}_{\boldsymbol{\omega}^2}$ and $\boldsymbol{\Sigma}_{\boldsymbol{\Phi}}$ by calculating the derivatives of $L(\{\varpi^2(\boldsymbol{\theta}_s), \boldsymbol{\Psi}(\boldsymbol{\theta}_s), \boldsymbol{\theta}_s\}_{s=1}^{N_D}, \boldsymbol{\Sigma}_{\boldsymbol{\omega}^2}, \boldsymbol{\Sigma}_{\boldsymbol{\Phi}})$ and setting them to zero. By doing so, we obtain:

$$\hat{\tau}_{\omega_i^2}^2 = \frac{1}{N_D}\sum_{s=1}^{N_D}\hat{\omega}_{i,s}^{-4}\left[\mathbb{E}\left[\omega_{i,s}^4\right] - 2\varpi_{i,s}^2\mathbb{E}\left[\omega_{i,s}^2\right] + \varpi_{i,s}^4\right] \tag{B31}$$

$$\hat{\boldsymbol{\Sigma}}_{\boldsymbol{\phi}_i} = \frac{1}{N_D}\sum_{s=1}^{N_D}\left[\mathbb{E}\left[\|\boldsymbol{\phi}_{i,s}\|^{-2}\boldsymbol{\phi}_{i,s}\boldsymbol{\phi}_{i,s}^T\right] - \chi_{i,s}\mathbb{E}\left[\|\boldsymbol{\phi}_{i,s}\|^{-1}\boldsymbol{\phi}_{i,s}\right]\boldsymbol{\psi}_{i,s}^T\boldsymbol{\gamma}_{i,s}^T - \chi_{i,s}\boldsymbol{\gamma}_{i,s}\boldsymbol{\psi}_{i,s}\mathbb{E}\left[\|\boldsymbol{\phi}_{i,s}\|^{-1}\boldsymbol{\phi}_{i,s}^T\right] + \chi_{i,s}^2\boldsymbol{\gamma}_{i,s}\boldsymbol{\psi}_{i,s}\boldsymbol{\psi}_{i,s}^T\boldsymbol{\gamma}_{i,s}^T\right]$$
(B32)

where $\hat{\tau}_{\omega_i^2}^2$ and $\hat{\boldsymbol{\Sigma}}_{\boldsymbol{\phi}_i}$ denote the estimations of the hyper covariance matrices.

The ECM algorithm requires initial estimations of $\boldsymbol{\theta}_s$, $\boldsymbol{\Sigma}_{\boldsymbol{\omega}^2}$ and $\boldsymbol{\Sigma}_{\boldsymbol{\Phi}}$. For this purpose, Algorithm 1 recommends estimating the structural parameters ($\boldsymbol{\theta}_s$) by applying a least-squares approach. Then, the initial estimations $\{\varpi_{i,s}^2|_0, \boldsymbol{\psi}_{i,s}|_0, \chi_{i,s}|_0\}$ can be calculated using Eq. (2), giving:

$$\hat{\tau}_{\omega_i^2}^2|_0 = \frac{1}{N_D}\sum_{s=1}^{N_D}\hat{\omega}_{i,s}^{-4}\left[(\varpi_{i,s}^2|_0 - \hat{\omega}_{i,s}^2)^2\right] + \frac{1}{N_D}\sum_{s=1}^{N_D}\hat{\omega}_{i,s}^{-4}\hat{\sigma}_{\omega_{i,s}^2}^2 \tag{B33}$$

$$\hat{\boldsymbol{\Sigma}}_{\boldsymbol{\phi}}\mid_0 = \frac{1}{N_D}\sum_{s=1}^{N_D}\left[(\boldsymbol{\chi}_{i,s}\mid_0 \boldsymbol{\gamma}_{i,s}\boldsymbol{\psi}_{i,s}\mid_0 -\hat{\boldsymbol{\phi}}_{i,s})(\boldsymbol{\chi}_{i,s}\mid_0 \boldsymbol{\gamma}_{i,s}\boldsymbol{\psi}_{i,s}\mid_0 -\hat{\boldsymbol{\phi}}_{i,s})^T\right] + \frac{1}{N_D}\sum_{s=1}^{N_D}\hat{\boldsymbol{\Sigma}}_{\boldsymbol{\phi},s} \qquad (B34)$$

where the index $.\mid_0$ indicates the initial estimation of the parameters. Due to these approximations, the prediction error covariance matrices are determined as the sum of ensemble variability and the mean of identification uncertainties.

**Appendix (C). Derivation of the EM algorithm in Eq. (30)**

*C.1. Derivation of the E-Step*

In the E-Step, the second-moment statistical information of $\boldsymbol{\theta}_s$'s should be computed while considering $\boldsymbol{\mu}_{\boldsymbol{\theta}}$ and $\boldsymbol{\Sigma}_{\boldsymbol{\theta}}$ as given. From Eq. (29), it is realized that $\boldsymbol{\theta}_s$'s are statistically independent when both $\boldsymbol{\mu}_{\boldsymbol{\theta}}$ and $\boldsymbol{\Sigma}_{\boldsymbol{\theta}}$ are given. Therefore, one can write:

$$p(\boldsymbol{\theta}_s \mid \boldsymbol{\mu}_{\boldsymbol{\theta}}, \boldsymbol{\Sigma}_{\boldsymbol{\theta}}, \mathbf{D}) \propto N\left(\hat{\boldsymbol{\theta}}_s \mid \boldsymbol{\theta}_s, \hat{\boldsymbol{\Sigma}}_{\boldsymbol{\theta}_s}\right) N\left(\boldsymbol{\theta}_s \mid \boldsymbol{\mu}_{\boldsymbol{\theta}}, \boldsymbol{\Sigma}_{\boldsymbol{\theta}}\right) \qquad (C1)$$

Due to Theorem 1 of Appendix A, this Gaussian multiplication can be simplified into

$$p(\boldsymbol{\theta}_s \mid \boldsymbol{\mu}_{\boldsymbol{\theta}}, \boldsymbol{\Sigma}_{\boldsymbol{\theta}}, \mathbf{D}) = N\left(\boldsymbol{\theta}_s \mid (\hat{\boldsymbol{\Sigma}}_{\boldsymbol{\theta}_s}^{-1} + \boldsymbol{\Sigma}_{\boldsymbol{\theta}}^{-1})^{-1}(\hat{\boldsymbol{\Sigma}}_{\boldsymbol{\theta}_s}^{-1}\hat{\boldsymbol{\theta}}_s + \boldsymbol{\Sigma}_{\boldsymbol{\theta}}^{-1}\boldsymbol{\mu}_{\boldsymbol{\theta}}), (\boldsymbol{\Sigma}_{\boldsymbol{\theta}}^{-1} + \hat{\boldsymbol{\Sigma}}_{\boldsymbol{\theta}_s}^{-1})^{-1}\right) \qquad (C2)$$

Thus, the second-moment information of $\boldsymbol{\theta}_s$ is given as

$$\mathbb{E}\left[\boldsymbol{\theta}_s\right] = (\hat{\boldsymbol{\Sigma}}_{\boldsymbol{\theta}_s}^{-1} + \boldsymbol{\Sigma}_{\boldsymbol{\theta}}^{-1})^{-1}(\hat{\boldsymbol{\Sigma}}_{\boldsymbol{\theta}_s}^{-1}\hat{\boldsymbol{\theta}}_s + \boldsymbol{\Sigma}_{\boldsymbol{\theta}}^{-1}\boldsymbol{\mu}_{\boldsymbol{\theta}}) \qquad (C3)$$

$$\mathbb{E}\left[\boldsymbol{\theta}_s\boldsymbol{\theta}_s^T\right] = (\boldsymbol{\Sigma}_{\boldsymbol{\theta}}^{-1} + \hat{\boldsymbol{\Sigma}}_{\boldsymbol{\theta}_s}^{-1})^{-1} + \mathbb{E}\left[\boldsymbol{\theta}_s\right]\mathbb{E}\left[\boldsymbol{\theta}_s\right]^T \qquad (C4)$$

*C.2. Derivation of the E-Step*

In the M-Step, the hyper-parameters $\boldsymbol{\mu}_{\boldsymbol{\theta}}$ and $\boldsymbol{\Sigma}_{\boldsymbol{\theta}}$ should be updated. Given the second moment statistics of $\boldsymbol{\theta}_s$'s, the derivatives of $L(\boldsymbol{\mu}_{\boldsymbol{\theta}}, \boldsymbol{\Sigma}_{\boldsymbol{\theta}})$ with respect to the hyper-parameters should be calculated and set to zero, which leads to

$$\hat{\boldsymbol{\mu}}_{\boldsymbol{\theta}} = \frac{1}{N_D}\sum_{s=1}^{N_D}\mathbb{E}\left[\boldsymbol{\theta}_s\right] \qquad (C5)$$

$$\hat{\Sigma}_{\theta} = \frac{1}{N_D}\sum_{s=1}^{N_D}\left[\mathbb{E}\left[\theta_s\theta_s^T\right]+\hat{\mu}_{\theta}\hat{\mu}_{\theta}^T-\hat{\mu}_{\theta}\mathbb{E}\left[\theta_s^T\right]-\mathbb{E}\left[\theta_s\right]\hat{\mu}_{\theta}^T\right] \tag{C6}$$

where $\hat{\mu}_{\theta}$ and $\hat{\Sigma}_{\theta}$ are the updated estimations of the hyper-parameters. Several iterations of Eqs. (C3-C6) converge to the MPV of the parameters. However, to start Algorithm 2, the initial estimates of $\mu_{\theta}$ and $\Sigma_{\theta}$ are required. A good estimate can be obtained from Eqs. (C5-C6) by replacing $\mathbb{E}[\theta_s]$ and $\mathbb{E}[\theta_s\theta_s^T]$ by $\hat{\theta}_s$ and $\hat{\theta}_s\hat{\theta}_s^T + \hat{\Sigma}_{\theta_s}$, respectively, which provides:

$$\hat{\mu}_{\theta}\big|_0 = \frac{1}{N_D}\sum_{s=1}^{N_D}\hat{\theta}_s \tag{C7}$$

$$\hat{\Sigma}_{\theta}\big|_0 = \frac{1}{N_D}\sum_{s=1}^{N_D}(\hat{\theta}_s-\hat{\mu}_{\theta}\big|_0)(\hat{\theta}_s-\hat{\mu}_{\theta}\big|_0)^T + \frac{1}{N_D}\sum_{s=1}^{N_D}\hat{\Sigma}_{\theta_s} \tag{C8}$$

where $\hat{\mu}_{\theta}\big|_0$ and $\hat{\Sigma}_{\theta}\big|_0$ are the initial estimates of the hyper-parameter to be used in the beginning. According to these equations, the hyper mean vector is estimated as the mean of optimal values ($\hat{\theta}_s$'s), and the hyper covariance matrix is approximated as the sum of ensemble variability and the mean of identification uncertainty.

**Appendix (D). Analytical derivatives of Eqs. (27) and (B30)**

In this appendix, we provide the analytical derivatives of $J(\theta_s)$ and $L(\theta_s)$ expressed earlier in Eqs. (27) and (B30), respectively. These derivatives are required for finding the MPV. Let $\theta_{s,p}$ be the $p$th element of $\theta_s$. The gradient vector of $L(\theta_s)$ can be calculated in a component-wise manner as follows:

$$\frac{\partial L(\theta_s)}{\partial \theta_{s,p}} = -\sum_{i=1}^{N_m}\hat{\omega}_{i,s}^{-4}\tau_{\omega_i^2}^{-2}\left[\varpi_{i,s}^2-\mathbb{E}[\omega_{i,s}^2]\right]\frac{\partial \varpi_{i,s}^2}{\partial \theta_{s,p}} - \sum_{i=1}^{N_m}\left[\chi_{i,s}\psi_{i,s}^T\gamma_{i,s}^T-\mathbb{E}[\phi_{i,s}^T]\right]\Sigma_{\phi_i}^{-1}\frac{\partial(\chi_{i,s}\gamma_{i,s}\psi_{i,s})}{\partial \theta_{s,p}} \tag{D1}$$

Remember that $\chi_{i,s} = \text{sign}(\mathbb{E}[\phi_{i,s}^T]\gamma_{i,s}\psi_{i,s})/\|\gamma_{i,s}\psi_{i,s}\|$ is also a function of $\psi_{i,s}$. Thus, the derivatives of $\chi_{i,s}\psi_{i,s}$ with respect to $\theta_{s,p}$ is calculated from

$$\frac{\partial(\chi_{i,s}\boldsymbol{\gamma}_{i,s}\boldsymbol{\psi}_{i,s})}{\partial\theta_{s,p}} = \text{sign}(\mathbb{E}[\boldsymbol{\phi}_{i,s}]^T \boldsymbol{\gamma}_{i,s}\boldsymbol{\psi}_{i,s})\boldsymbol{\gamma}_{i,s}\frac{\partial(\boldsymbol{\psi}_{i,s}/\|\boldsymbol{\gamma}_{i,s}\boldsymbol{\psi}_{i,s}\|)}{\partial\theta_{s,p}} \tag{D2}$$

$$\frac{\partial(\boldsymbol{\psi}_{i,s}/\|\boldsymbol{\gamma}_{i,s}\boldsymbol{\psi}_{i,s}\|)}{\partial\theta_{s,p}} = \|\boldsymbol{\gamma}_{i,s}\boldsymbol{\psi}_{i,s}\|^{-1}\frac{\partial\boldsymbol{\psi}_{i,s}}{\partial\theta_{s,p}} - \|\boldsymbol{\gamma}_{i,s}\boldsymbol{\psi}_{i,s}\|^{-3}(\boldsymbol{\psi}_{i,s}^T\boldsymbol{\gamma}_{i,s}^T\boldsymbol{\gamma}_{i,s}\frac{\partial\boldsymbol{\psi}_{i,s}}{\partial\theta_{s,p}})\boldsymbol{\psi}_{i,s} \tag{D3}$$

Note that the derivatives of sign(.) is zero. Similarly, the entries of the gradient vector $J(\boldsymbol{\theta}_s)$ are calculated as

$$\frac{\partial J(\boldsymbol{\theta}_s)}{\partial\theta_{s,p}} = \sum_{i=1}^{N_m}\left[(\hat{\sigma}_{\varpi_{i,s}^2}^2 + \hat{\omega}_{i,s}^4\hat{\tau}_{\omega_i^2}^2)^{-1}(\varpi_{i,s}^2 - \hat{\omega}_{i,s}^2)\frac{\partial\varpi_{i,s}^2}{\partial\theta_{s,p}} + (\chi_{i,s}\boldsymbol{\gamma}_{i,s}\boldsymbol{\psi}_{i,s} - \hat{\boldsymbol{\phi}}_{i,s})^T(\hat{\boldsymbol{\Sigma}}_{\boldsymbol{\phi}_{i,s}} + \hat{\boldsymbol{\Sigma}}_{\boldsymbol{\phi}})^{-1}\frac{\partial(\chi_{i,s}\boldsymbol{\gamma}_{i,s}\boldsymbol{\psi}_{i,s})}{\partial\theta_{s,p}}\right] \tag{D4}$$

In Algorithm 2, the Hessian matrix of $J(\boldsymbol{\theta}_s)$ is also required, whose elements are determined from

$$\begin{aligned}\frac{\partial^2 J(\boldsymbol{\theta}_s)}{\partial\theta_{s,p}\partial\theta_{s,q}} &= \sum_{i=1}^{N_m}(\hat{\sigma}_{\varpi_{i,s}^2}^2 + \hat{\omega}_{i,s}^4\hat{\tau}_{\omega_i^2}^2)^{-1}\left[(\varpi_{i,s}^2 - \hat{\omega}_{i,s}^2)\frac{\partial^2\varpi_{i,s}^2}{\partial\theta_{s,p}\partial\theta_{s,q}} + \frac{\partial\varpi_{i,s}^2}{\partial\theta_{s,q}}\frac{\partial\varpi_{i,s}^2}{\partial\theta_{s,p}}\right] \\ &+ \sum_{i=1}^{N_m}\left[(\chi_{i,s}\boldsymbol{\gamma}_{i,s}\boldsymbol{\psi}_{i,s} - \hat{\boldsymbol{\phi}}_{i,s})^T(\hat{\boldsymbol{\Sigma}}_{\boldsymbol{\phi}_{i,s}} + \hat{\boldsymbol{\Sigma}}_{\boldsymbol{\phi}})^{-1}\frac{\partial^2(\chi_{i,s}\boldsymbol{\gamma}_{i,s}\boldsymbol{\psi}_{i,s})}{\partial\theta_{s,p}\partial\theta_{s,q}} + \frac{\partial(\chi_{i,s}\boldsymbol{\psi}_{i,s}^T\boldsymbol{\gamma}_{i,s}^T)}{\partial\theta_{s,q}}(\hat{\boldsymbol{\Sigma}}_{\boldsymbol{\phi}_{i,s}} + \hat{\boldsymbol{\Sigma}}_{\boldsymbol{\phi}})^{-1}\frac{\partial(\chi_{i,s}\boldsymbol{\gamma}_{i,s}\boldsymbol{\psi}_{i,s})}{\partial\theta_{s,p}}\right]\end{aligned} \tag{D5}$$

where $\partial^2(\chi_{i,s}\boldsymbol{\psi}_{i,s})/\partial\theta_{s,p}\partial\theta_{s,q}$ is calculated from

$$\frac{\partial^2(\chi_{i,s}\boldsymbol{\gamma}_{i,s}\boldsymbol{\psi}_{i,s})}{\partial\theta_{s,p}\partial\theta_{s,q}} = \text{sign}(\hat{\boldsymbol{\phi}}_{i,s}^T\boldsymbol{\gamma}_{i,s}\boldsymbol{\psi}_{i,s})\frac{\partial^2(\boldsymbol{\gamma}_{i,s}\boldsymbol{\psi}_{i,s}/\|\boldsymbol{\gamma}_{i,s}\boldsymbol{\psi}_{i,s}\|)}{\partial\theta_{s,p}\partial\theta_{s,q}} \tag{D6}$$

and

$$\begin{aligned}\frac{\partial^2(\boldsymbol{\psi}_{i,s}/\|\boldsymbol{\gamma}_{i,s}\boldsymbol{\psi}_{i,s}\|)}{\partial\theta_{s,p}\partial\theta_{s,q}} &= \|\boldsymbol{\gamma}_{i,s}\boldsymbol{\psi}_{i,s}\|^{-1}\frac{\partial^2\boldsymbol{\psi}_{i,s}}{\partial\theta_{s,p}\partial\theta_{s,q}} - \|\boldsymbol{\gamma}_{i,s}\boldsymbol{\psi}_{i,s}\|^{-3}(\boldsymbol{\psi}_{i,s}^T\boldsymbol{\gamma}_{i,s}^T\boldsymbol{\gamma}_{i,s}\frac{\partial\boldsymbol{\psi}_{i,s}}{\partial\theta_{s,q}})\frac{\partial\boldsymbol{\psi}_{i,s}}{\partial\theta_{s,p}} \\ &- \|\boldsymbol{\gamma}_{i,s}\boldsymbol{\psi}_{i,s}\|^{-3}(\boldsymbol{\psi}_{i,s}^T\boldsymbol{\gamma}_{i,s}^T\boldsymbol{\gamma}_{i,s}\frac{\partial\boldsymbol{\psi}_{i,s}}{\partial\theta_{s,p}})\frac{\partial\boldsymbol{\psi}_{i,s}}{\partial\theta_{s,q}} + 3\|\boldsymbol{\gamma}_{i,s}\boldsymbol{\psi}_{i,s}\|^{-5}(\boldsymbol{\psi}_{i,s}^T\boldsymbol{\gamma}_{i,s}^T\boldsymbol{\gamma}_{i,s}\frac{\partial\boldsymbol{\psi}_{i,s}}{\partial\theta_{s,q}})(\boldsymbol{\psi}_{i,s}^T\boldsymbol{\gamma}_{i,s}^T\boldsymbol{\gamma}_{i,s}\frac{\partial\boldsymbol{\psi}_{i,s}}{\partial\theta_{s,p}})\boldsymbol{\psi}_{i,s} \\ &- \|\boldsymbol{\gamma}_{i,s}\boldsymbol{\psi}_{i,s}\|^{-3}(\boldsymbol{\psi}_{i,s}^T\boldsymbol{\gamma}_{i,s}^T\boldsymbol{\gamma}_{i,s}\frac{\partial^2\boldsymbol{\psi}_{i,s}}{\partial\theta_{s,p}\partial\theta_{s,q}})\boldsymbol{\psi}_{i,s} - \|\boldsymbol{\gamma}_{i,s}\boldsymbol{\psi}_{i,s}\|^{-3}(\frac{\partial\boldsymbol{\psi}_{i,s}^T}{\partial\theta_{s,q}}\boldsymbol{\gamma}_{i,s}^T\boldsymbol{\gamma}_{i,s}\frac{\partial\boldsymbol{\psi}_{i,s}}{\partial\theta_{s,p}})\boldsymbol{\psi}_{i,s}\end{aligned} \tag{D7}$$

In these equations, the derivatives of the analytical modal parameters with respect to the structural parameters are still required, which can be calculated from the following equation:

$$\mathbf{K}(\boldsymbol{\theta}_s)\boldsymbol{\psi}_{i,s} = \varpi_{i,s}^2\mathbf{M}(\boldsymbol{\theta}_s)\boldsymbol{\psi}_{i,s} \tag{D8}$$

Taking the derivatives with respect to the entries $\theta_{s,p}$ leads to

$$\frac{\partial \mathbf{K}(\boldsymbol{\theta}_s)}{\partial \theta_{s,p}} \boldsymbol{\psi}_{i,s} + \mathbf{K}(\boldsymbol{\theta}_s) \frac{\partial \boldsymbol{\psi}_{i,s}}{\partial \theta_{s,p}} = \frac{\partial \varpi_{i,s}^2}{\partial \theta_{s,p}} \mathbf{M}(\boldsymbol{\theta}_s) \boldsymbol{\psi}_{i,s} + \varpi_{i,s}^2 \frac{\partial \mathbf{M}(\boldsymbol{\theta}_s)}{\partial \theta_{s,p}} \boldsymbol{\psi}_{i,s} + \varpi_{i,s}^2 \mathbf{M}(\boldsymbol{\theta}_s) \frac{\partial \boldsymbol{\psi}_{i,s}}{\partial \theta_{s,p}} \tag{D9}$$

Pre-multiplying this equation by $\boldsymbol{\psi}_{i,s}^T$ gives:

$$\frac{\partial \varpi_{i,s}^2}{\partial \theta_{s,p}} = \boldsymbol{\psi}_{i,s}^T \left( \frac{\partial \mathbf{K}(\boldsymbol{\theta}_s)}{\partial \theta_{s,p}} - \varpi_{i,s}^2 \frac{\partial \mathbf{M}(\boldsymbol{\theta}_s)}{\partial \theta_{s,p}} \right) \boldsymbol{\psi}_{i,s} \,/\, \boldsymbol{\psi}_{i,s}^T \mathbf{M}(\boldsymbol{\theta}_s) \boldsymbol{\psi}_{i,s} \tag{D10}$$

This equation provides the derivatives of analytical modal frequencies with respect to individual elements of $\boldsymbol{\theta}_s$. Rearranging the expressions in Eq. (D9) gives the derivatives of the analytical mode shape vector with respect to $\theta_{s,p}$, which leads to

$$\frac{\partial \boldsymbol{\psi}_{i,s}}{\partial \theta_{s,p}} = \left( \mathbf{K}(\boldsymbol{\theta}_s) - \varpi_{i,s}^2 \mathbf{M}(\boldsymbol{\theta}_s) \right)^{+} \left( \frac{\partial \varpi_{i,s}^2}{\partial \theta_{s,p}} \mathbf{M}(\boldsymbol{\theta}_s) + \varpi_{i,s}^2 \frac{\partial \mathbf{M}(\boldsymbol{\theta}_s)}{\partial \theta_{s,p}} - \frac{\partial \mathbf{K}(\boldsymbol{\theta}_s)}{\partial \theta_{s,p}} \right) \boldsymbol{\psi}_{i,s} \tag{D11}$$

For the calculation of the second derivatives, we take the derivatives of Eq. (D9) with respect to another arbitrary element like $\theta_{s,p}$, which yields:

$$\begin{aligned}
& \frac{\partial^2 \mathbf{K}(\boldsymbol{\theta}_s)}{\partial \theta_{s,q} \partial \theta_{s,p}} \boldsymbol{\psi}_{i,s} + \frac{\partial \mathbf{K}(\boldsymbol{\theta}_s)}{\partial \theta_{s,p}} \frac{\partial \boldsymbol{\psi}_{i,s}}{\partial \theta_{s,q}} + \frac{\partial \mathbf{K}(\boldsymbol{\theta}_s)}{\partial \theta_{s,q}} \frac{\partial \boldsymbol{\psi}_{i,s}}{\partial \theta_{s,p}} + \mathbf{K}(\boldsymbol{\theta}_s) \frac{\partial^2 \boldsymbol{\psi}_{i,s}}{\partial \theta_{s,q} \partial \theta_{s,p}} \\
& = \frac{\partial^2 \varpi_{i,s}^2}{\partial \theta_{s,q} \partial \theta_{s,p}} \mathbf{M}(\boldsymbol{\theta}_s) \boldsymbol{\psi}_{i,s} + \left[ \frac{\partial \varpi_{i,s}^2}{\partial \theta_{s,p}} \frac{\partial \mathbf{M}(\boldsymbol{\theta}_s)}{\partial \theta_{s,q}} + \frac{\partial \varpi_{i,s}^2}{\partial \theta_{s,q}} \frac{\partial \mathbf{M}(\boldsymbol{\theta}_s)}{\partial \theta_{s,p}} + \varpi_{i,s}^2 \frac{\partial^2 \mathbf{M}(\boldsymbol{\theta}_s)}{\partial \theta_{s,q} \partial \theta_{s,p}} \right] \boldsymbol{\psi}_{i,s} \\
& + \varpi_{i,s}^2 \left[ \frac{\partial \mathbf{M}(\boldsymbol{\theta}_s)}{\partial \theta_{s,q}} \frac{\partial \boldsymbol{\psi}_{i,s}}{\partial \theta_{s,p}} + \frac{\partial \mathbf{M}(\boldsymbol{\theta}_s)}{\partial \theta_{s,q}} \frac{\partial \boldsymbol{\psi}_{i,s}}{\partial \theta_{s,p}} \right] + \mathbf{M}(\boldsymbol{\theta}_s) \left[ \frac{\partial \varpi_{i,s}^2}{\partial \theta_{s,p}} \frac{\partial \boldsymbol{\psi}_{i,s}}{\partial \theta_{s,q}} + \frac{\partial \varpi_{i,s}^2}{\partial \theta_{s,q}} \frac{\partial \boldsymbol{\psi}_{i,s}}{\partial \theta_{s,p}} + \varpi_{i,s}^2 \frac{\partial^2 \boldsymbol{\psi}_{i,s}}{\partial \theta_{s,q} \partial \theta_{s,p}} \right]
\end{aligned} \tag{D12}$$

Pre-multiplying this equation by $\boldsymbol{\psi}_{i,s}^T$ and simplifying the expressions readily provide the second-order derivatives of $\varpi_{i,s}^2$ as

$$\begin{aligned}
\frac{\partial^2 \varpi_{i,s}^2}{\partial \theta_{s,q} \partial \theta_{s,p}} = & \frac{1}{\boldsymbol{\psi}_{i,s}^T \mathbf{M}(\boldsymbol{\theta}_s) \boldsymbol{\psi}_{i,s}} \left\{ \boldsymbol{\psi}_{i,s}^T \frac{\partial^2 \mathbf{K}(\boldsymbol{\theta}_s)}{\partial \theta_{s,q} \partial \theta_{s,p}} \boldsymbol{\psi}_{i,s} + \boldsymbol{\psi}_{i,s}^T \frac{\partial \mathbf{K}(\boldsymbol{\theta}_s)}{\partial \theta_{s,p}} \frac{\partial \boldsymbol{\psi}_{i,s}}{\partial \theta_{s,q}} + \boldsymbol{\psi}_{i,s}^T \frac{\partial \mathbf{K}(\boldsymbol{\theta}_s)}{\partial \theta_{s,q}} \frac{\partial \boldsymbol{\psi}_{i,s}}{\partial \theta_{s,p}} \right. \\
& - \frac{\partial \varpi_{i,s}^2}{\partial \theta_{s,p}} \left[ \boldsymbol{\psi}_{i,s}^T \frac{\partial \mathbf{M}(\boldsymbol{\theta}_s)}{\partial \theta_{s,q}} \boldsymbol{\psi}_{i,s} + \boldsymbol{\psi}_{i,s}^T \mathbf{M}(\boldsymbol{\theta}_s) \frac{\partial \boldsymbol{\psi}_{i,s}}{\partial \theta_{s,q}} \right] - \frac{\partial \varpi_{i,s}^2}{\partial \theta_{s,q}} \left[ \boldsymbol{\psi}_{i,s}^T \frac{\partial \mathbf{M}(\boldsymbol{\theta}_s)}{\partial \theta_{s,p}} \boldsymbol{\psi}_{i,s} + \boldsymbol{\psi}_{i,s}^T \mathbf{M}(\boldsymbol{\theta}_s) \frac{\partial \boldsymbol{\psi}_{i,s}}{\partial \theta_{s,p}} \right] \\
& \left. - \varpi_{i,s}^2 \left[ \boldsymbol{\psi}_{i,s}^T \frac{\partial^2 \mathbf{M}(\boldsymbol{\theta}_s)}{\partial \theta_{s,q} \partial \theta_{s,p}} \boldsymbol{\psi}_{i,s} + \boldsymbol{\psi}_{i,s}^T \frac{\partial \mathbf{M}(\boldsymbol{\theta}_s)}{\partial \theta_{s,p}} \frac{\partial \boldsymbol{\psi}_{i,s}}{\partial \theta_{s,q}} + \boldsymbol{\psi}_{i,s}^T \frac{\partial \mathbf{M}(\boldsymbol{\theta}_s)}{\partial \theta_{s,q}} \frac{\partial \boldsymbol{\psi}_{i,s}}{\partial \theta_{s,p}} \right] \right\}
\end{aligned} \tag{D13}$$

Replacing this result into Eq. (D12) gives the second derivatives with respect to the mode shape vector:

$$\begin{aligned}
\frac{\partial^2 \mathbf{\psi}_{i,s}}{\partial \theta_{s,q} \partial \theta_{s,p}} &= \left(\mathbf{K}(\mathbf{\theta}_s) - \varpi_{i,s}^2 \mathbf{M}(\mathbf{\theta}_s)\right)^+ \left\{ \frac{\partial^2 \varpi_{i,s}^2}{\partial \theta_{s,q} \partial \theta_{s,p}} \mathbf{M}(\mathbf{\theta}_s) \mathbf{\psi}_{i,s} + \frac{\partial \varpi_{i,s}^2}{\partial \theta_{s,p}} \left[ \frac{\partial \mathbf{M}(\mathbf{\theta}_s)}{\partial \theta_{s,q}} \mathbf{\psi}_{i,s} + \mathbf{M}(\mathbf{\theta}_s) \frac{\partial \mathbf{\psi}_{i,s}}{\partial \theta_{s,q}} \right] \right. \\
&+ \frac{\partial \varpi_{i,s}^2}{\partial \theta_{s,q}} \left[ \frac{\partial \mathbf{M}(\mathbf{\theta}_s)}{\partial \theta_{s,p}} \mathbf{\psi}_{i,s} + \mathbf{M}(\mathbf{\theta}_s) \frac{\partial \mathbf{\psi}_{i,s}}{\partial \theta_{s,p}} \right] + \varpi_{i,s}^2 \left[ \frac{\partial^2 \mathbf{M}(\mathbf{\theta}_s)}{\partial \theta_{s,q} \partial \theta_{s,p}} \mathbf{\psi}_{i,s} + \frac{\partial \mathbf{M}(\mathbf{\theta}_s)}{\partial \theta_{s,p}} \frac{\partial \mathbf{\psi}_{i,s}}{\partial \theta_{s,q}} + \frac{\partial \mathbf{M}(\mathbf{\theta}_s)}{\partial \theta_{s,q}} \frac{\partial \mathbf{\psi}_{i,s}}{\partial \theta_{s,p}} \right] \quad \text{(D14)}\\
&\left. - \frac{\partial^2 \mathbf{K}(\mathbf{\theta}_s)}{\partial \theta_{s,q} \partial \theta_{s,p}} \mathbf{\psi}_{i,s} - \frac{\partial \mathbf{K}(\mathbf{\theta}_s)}{\partial \theta_{s,p}} \frac{\partial \mathbf{\psi}_{i,s}}{\partial \theta_{s,q}} - \frac{\partial \mathbf{K}(\mathbf{\theta}_s)}{\partial \theta_{s,q}} \frac{\partial \mathbf{\psi}_{i,s}}{\partial \theta_{s,p}} \right\}
\end{aligned}$$

**Appendix (E). TMCMC sampling algorithm**

**Algorithm 3.** Sampling method for estimating the posterior distribution

1: **Set** prior distributions of $\{\mathbf{\theta}_s\}_{s=1}^{N_D}$, $\mathbf{\Sigma}_{\omega^2}$, and $\mathbf{\Sigma}_{\Phi}$

2: **Set** prior distributions of $\mathbf{\mu}_{\theta}$ and $\mathbf{\Sigma}_{\theta}$

3: **Draw** samples of $\{\mathbf{\theta}_s\}_{s=1}^{N_D}$, $\mathbf{\Sigma}_{\omega^2}$, and $\mathbf{\Sigma}_{\Phi}$ from $\prod_{s=1}^{N_D} N(\hat{\mathbf{\omega}}_s^2 | \mathbf{\varpi}^2(\mathbf{\theta}_s), \hat{\mathbf{\Sigma}}_{\omega_s^2} + \mathbf{\Sigma}_{\omega^2}) N(\hat{\mathbf{\Phi}}_s | \mathbf{\Gamma}_s \mathbf{\Psi}(\mathbf{\theta}_s), \hat{\mathbf{\Sigma}}_{\Phi_s} + \mathbf{\Sigma}_{\Phi})$ using TMCMC [71]

4: **Draw** samples of $\mathbf{\mu}_{\theta}$ and $\mathbf{\Sigma}_{\theta}$ from $\prod_{s=1}^{N_D} \sum_{m=1}^{N_{sa}} N(\mathbf{\theta}_s^{(m)} | \mathbf{\mu}_{\theta}, \mathbf{\Sigma}_{\theta})$, where $\mathbf{\theta}_s^{(m)}$ are the posterior samples of $\mathbf{\theta}_s$ using TMCMC [71]